\documentclass[11pt,english]{article}

\renewcommand{\Im}{{\rm Im \, }}

\newcommand{\CO}{{\cal O}}
\newcommand{\CD}{{\cal D}}
\newcommand{\CL}{{\cal L}}
\newcommand{\SL}{{\text{SL}}}
\newcommand{\CC}{{\cal C}}
\newcommand{\CK}{{\cal K}}
\newcommand{\CW}{{\cal W}}
\newcommand{\CU}{{\cal U}}
\newcommand{\CV}{{\cal V}}
\newcommand{\mL}{{\mathfrak L}}
\newcommand{\CF}{{\cal F}}
\newcommand{\CP}{{\cal P}}

\newcommand{\mY}{{\mathbb Y}}
\newcommand{\CS}{{\cal S}}
\newcommand{\tS}{{\text S}}
\newcommand{\tK}{{\text K}}
\newcommand{\tN}{{\text N}}
\newcommand{\tM}{{\text M}}
\newcommand{\tA}{{\text A}}
\newcommand{\ttN}{{\widetilde{\text{N}}}}
\newcommand{\so}{{\mathfrak{so}}}
\newcommand{\SO}{{\text{SO}}}

\newcommand{\CM}{{\cal M}}
\usepackage{amsthm}
\newtheorem{theorem}{Theorem}[section]
\newtheorem{remark}[theorem]{Remark}
\newtheorem{proposition}[theorem]{Proposition}
\newtheorem{lemma}[theorem]{Lemma}
\newtheorem{definition}[theorem]{Definition}
\newtheorem{claim}{Claim}

\makeatletter
\newcommand*{\rom}[1]{\expandafter\@slowromancap\romannumeral #1@}
\makeatother

\usepackage[latin9]{inputenc}
\usepackage{geometry}
\usepackage{verbatim}
\usepackage{prettyref}
\usepackage[numbers,sort&compress]{natbib}
\usepackage{amsmath}
\usepackage{amssymb}
\usepackage{tikz-cd}
\tikzset{commutative diagrams/row sep/huge=4cm}
\tikzset{commutative diagrams/column sep/huge=4cm}
\tikzcdset{scale cd/.style={every label/.append style={scale=#1},
    cells={nodes={scale=#1}}}}
\usepackage{lmodern}
\usepackage{tcolorbox}
\usepackage{cases}
\usepackage{bbold}
\usepackage{ytableau}
\usepackage{mathtools}
\usepackage{graphicx}
\usepackage[nottoc]{tocbibind}
\usepackage{mathrsfs}
\usepackage{caption}
\usepackage{subcaption}
\usepackage{cancel}
\usepackage{float}
\usepackage{bm}
\usepackage{tikz}
\usepackage{lipsum}
\usepackage{adjustbox}
\usepackage{mathtools}

\DeclarePairedDelimiter\floor{\lfloor}{\rfloor}

 \usepackage{slashed}
\usepackage{esvect}
\usepackage{dsfont}

\usepackage{color}
\definecolor{darkgreen}{rgb}{0,0.5,0}
\definecolor{darkblue}{rgb}{0,0,0.6}
\definecolor{purple}{rgb}{0.4,.2,0.7}
\usepackage[colorlinks=true,citecolor=darkgreen,linkcolor=purple,urlcolor=purple]{hyperref}

\numberwithin{equation}{section}
\numberwithin{figure}{section}
\numberwithin{table}{section}

\def\CH{{\cal H}}

\def\CU{{\cal U}}
\def\CM{{\cal M}}
\def\CN{{\cal N}}
\def\CD{{\cal D}}

\DeclareMathOperator{\Tr}{Tr}

\DeclareFontShape{OT1}{cmr}{mx}{n}{<->cmr10}{}

\begin{document}

\fontseries{mx}\selectfont

\begin{center}
\LARGE \bf A note on the  representations of $\SO(1, d+1)$
\end{center}

\vskip1cm

\begin{center}
Zimo Sun$^1$
\vskip5mm
{
\it{\footnotesize $^1$ Princeton Gravity Initiative, Princeton University}\\
}
\end{center}

\vskip2cm

\vskip0mm

\begin{abstract}
$\SO(1, d+1)$ is the isometry group of $(d+1)$-dimensional de Sitter spacetime ($\text{dS}_{d+1}$) and the conformal group of $\mathbb{R}^{d}$.
This note gives a pedagogical introduction to the representation theory of $\SO(1, d+1)$, from the perspective of de Sitter quantum field theory and using tools from conformal field theory.
Topics include (1) the construction and classification of all unitary irreducible representations (UIRs) of $\SO(1,2)$ and $\SL(2,\mathbb R)$,
(2) the construction and classification of all UIRs of $\SO(1,d+1)$ that describe integer-spin fields in $\text{dS}_{d+1}$,
 (3) a physical framework for understanding these UIRs,
(4) the definition and  derivation of Harish-Chandra group characters of $\SO(1,d+1)$, and (5) a comparison between UIRs of $\SO(1, d+1)$ and $\SO(2,d)$.

\end{abstract}

\newpage

\tableofcontents

\section{Introduction}
According to Wigner's principle of classification \cite{Bargmann:1948ck}, particles in $\text{dS}_{d+1}$ are in one-to-one correspondence with UIRs of $\SO(1, d+1)$, a generalized Lorentz group. Because of the noncompactness of $\SO(1,d+1)$, these UIR are infinite dimensional. The original study of infinite irreducible representations of the Lorentz group, i.e. $\SO(1,3)$, can be dated back to Dirac \cite{Dirac:1945cm}, who inspired  three independent work classifying (unitary) irreducible representations of $\SO(1,3)$, one by Harish-Chandra \cite{10.2307/97833},  another by  Bargmann \cite{10.2307/1969129} and a third by Gel'fand and Naimark \cite{GelNai47}. In \cite{10.2307/1969129}, the (unitary) irreducible representations of $\SO(1,2)$ (and its double covering $\SL(2,\mathbb R)$) were also classified. The generalization to $\SO(1, 4)$ was solved in  \cite{ThomasSO,NewtonSO} by Thomas and Newton, where some heuristic reasonings were later made rigorous by Dixmier \cite{BSMF_1961__89__9_0}.  In particular, Dixmier proved  that the representations obtained in \cite{ThomasSO,NewtonSO} were square integrable.  The classification problem for any higher dimensional $\SO(1,d+1)$ was studied by Hirai \cite{10.3792/pja/1195523460,10.3792/pja/1195523378}  using the method of infinitesimal operators and by Takahashi \cite{BSMF_1963__91__289_0} using the induced representation method. 

Due to the dual role of $\SO(1, d+1)$  as a Euclidean conformal group and the de Sitter isometry group, it has remarkable applications in numerous physical settings,  including but certainly not limited to conformal field theories and de Sitter quantum field theories. On the CFT side, the modern application of $\SO(1, d+1)$ representation theory, in particular the harmonic analysis on $\SO(1,d+1)$, originated from Mack's work on conformal partial wave expansion \cite{Mack:2009mi}, which was then followed by many profound results in analytical conformal bootstrap like conformal Regge theory \cite{Costa:2012cb}, recursion relations for conformal blocks \cite{Penedones:2015aga} and the Lorentzian OPE inversion formula \cite{Caron-Huot:2017vep,Simmons-Duffin:2017nub}. On the dS side, the representation theory has been used to resolve ambiguities associated with the definition of  mass terms \cite{Garidi:2003bg, Garidi} and to develop an algebraic approach towards de Sitter quantum field theory \cite{Bros:1994dn,Bros:1995js,Bros:1998ik,Joung:2006gj,Joung:2007je}, since the 90s. More recently, $\SO(1, d+1)$ representations are used as a powerful tool for computing late-time cosmological correlators on a rigid de Sitter background \cite{Sleight:2020obc,Hogervorst:2021uvp, DiPietro:2021sjt,Sleight:2021plv}. In \cite{Sleight:2020obc,DiPietro:2021sjt,Sleight:2021plv}, the late-time four-point functions are computed perturbatively by making a connection between Feynman diagrams in dS and Euclidean AdS,  and in \cite{Hogervorst:2021uvp}, a nonperturbative program is proposed for bootstrapping late-time four-point functions.  In addition to these bootstrap-type developments,  it has also been noticed that the Euclidean one-loop partition function for every effective field theory of quantum gravity in dS  can be written as an integral of $\SO(1, d+1)$ Harish-Chandra characters, up to edge corrections \cite{Anninos:2020hfj, Law:2020cpj} and these characters encode quasinormal modes of dS static patch horizon \cite{Sun:2020sgn}. The Harish-Chandra character, invented by Harish-Chandra \cite{bams/1183520006,bams/1183525024},  is a central notion in the representation theory of real semi-simple Lie groups ($\SO(1,d+1)$ is an example of such groups of rank 1), closely related to the construction of discrete series \cite{10.1007/BF02391779, 10.1007/BF02392813} and to harmonic analysis on groups \cite{HARISHCHANDRA1975104,HC2,10.2307/1971058}. The Harish-Chandra characters of $\SO(1,d+1)$ were first computed in \cite{10.3792/pja/1195522333}, and were rederived recently in \cite{Basile:2016aen} using Bernstein-Gel'fand-Gel'fand resolutions in order to understand mixed-symmetry fields in dS. We will discuss $\SO(1, d+1)$ Harish-Chandra characters in section \ref{HCchar}. It is also worth mentioning other important applications of $\SO(1, d+1)$ in some lower dimensions. For example,  $\SL(2,\mathbb R)$ governs the low energy behaviors of SYK model \cite{1993, Kitaevtalk,Maldacena:2016hyu}, and $\SO(1,3)$ (or its covering group $\SL(2,\mathbb C)$) relates 4D scattering amplitudes in asymptotically spacetimes and 2D celestial amplitudes \cite{20171,20172,2020,raclariu2021lectures,pasterski2021lectures}.

The main goal of this note is to provide a pedagogical introduction to the UIRs of $\SO(1,d+1)$, their Harish-Chandra characters and their relations with quantum fields in de Sitter spacetime. For this purpose, we will show explicitly the construction of  these UIRs from scratch, by using a CFT-type language that is more familiar to physicists rather than the standard induced representation approach in math literature. Building on the construction, we will be able to compute Harish-Chandra characters for these UIRs, without turning to advanced tools like Bernstein-Gel'fand-Gel'fand resolutions. By quantizing fields of different masses and spins in $\text{dS}_{d+1}$, we will also identify their single-particle Hilbert spaces as UIRs of $\SO(1,d+1)$.

Although we have tried to make this note as self-contained as possible, we are not going to present all the mathematical proofs, in particular those technical and lengthy ones that are not directly relevant to our purpose. Our strategy is that we will  refer the readers to literature that contains the details of the proofs and then illustrate the underlying ideas of the proofs by working out some examples. For instance,  the irreducibility of $\CF_{\Delta,  s}$  (c.f. section \ref{d+1}) was proved by Hirai \cite{10.3792/pja/1195523460}. Instead of repeating his steps, we give a simpler and more straightforward proof  for the scalar case, i.e. $s=0$ (c.f. appendix \ref{irrCF}), in the same spirit  as Hirai's.

The note is organized as follows:
\begin{enumerate}
\item In section \ref{preandconv}, we briefly review the Lie algebra $\so(1, d+1)$ and fix our conventions, e.g. reality conditions and the quadratic Casimir operator.
\item In section \ref{SO12}, we give two different but equivalent constructions of all UIRs of $\SO(1,2)$, one akin to the quantization of angular momentum operators in quantum mechanics  and the other akin to conformal field theory. The UIRs are classified, i.e. principle series, complementary series and discrete series, and shadow transformations are constructed, using each method. The realization of these UIRs in $\text{dS}_2$ quantum field theory is also discussed.

\item The CFT-type construction used for $\SO(1,2)$ is generalized to higher dimensional $\SO(1, d+1)$ in section \ref{d+1}. The UIRs describing integer-spin fields in $\text{dS}_{d+1}$ are classified. They fall into principal series, complementary series and exceptional series \footnote{When $d=3$, the exceptional series is the same as discrete series. We will comment more on this in section \ref{SO14}.}. The bulk mass range for each series is spelled out explicitly and the Higuchi bound \cite{Higuchi:1986py,2019Lust} is recovered as a result of unitarity (in the representation sense). In search of exceptional series, a chain of intertwining maps including shadow transformations are constructed. We also decompose these UIRs into $\SO(d+1)$ irreducible representations.

\item In section \ref{HCchar}, we introduce the notion of Harish-Chandra characters as the analogue of usual Weyl characters of compact Lie group  and explain why they exist for the representations constructed in the previous sections. We also give a simple and straightforward (though not very rigorous) derivation of  $\SO(1,d+1)$ Harish-Chandra characters based on the constructions in section \ref{d+1}.

\item In section \ref{versus}, we briefly comment on the distinctions between the physically relevant UIRs of $\SO(1,d+1)$  and $\SO(2,d)$. We also compare $\SO(1,d+1)$ Harish-Chandra characters and $\so(2,d)$ characters \footnote{It will be clear in section \ref{versus} why we call them $\so(2,d)$ characters rather than $\SO(2,d)$ characters.}.

\end{enumerate}

\noindent{}\textbf{Omissions}:  The beautiful topic regarding harmonic analysis on $\SO(1,d+1)$ is not covered in this note. An excellent and comprehensive review on this topic is \cite{Dobrev:1977qv}. A more condensed version can be found in \cite{Karateev:2018oml}, summarizing and generalizing some key results in \cite{Dobrev:1977qv}.

\section{Preliminaries and conventions}\label{preandconv}
$(d+1)$-dimensional de Sitter spacetime can be represented as a hypersurface in a $(d+2)$-dimensional embedding space
\begin{align}\label{dSemb}
\text{dS}_{d+1}:\,\,\,\,\, -X_0^2+X_1^2+\cdots + X_{d+1}^2=1
\end{align}
where the de Sitter radius is chosen to be 1. Its isometry group is  $\SO(1,d+1)$,  isomorphic to the conformal conformal group of $\mathbb R^d$. A standard basis for the Lie algebra $\so(1, d+1)$ is $L_{AB}=-L_{BA}, A=0, 1,\cdots d+1$, with commutation relations 
\begin{align}\label{defiso}
[L_{AB}, L_{CD}]=\eta_{BC} L_{AD}-\eta_{AC} L_{BD}+\eta_{AD} L_{BC}-\eta_{BD} L_{AC}
\end{align} 
where 
\begin{align}
\eta_{AB}=\text{diag}(-1,1,\cdots, 1)
\end{align}
The isomorphism between $\so(1, d+1)$ and the $d$-dimensional Euclidean conformal algebra is realized as
\begin{align}\label{defconf}
L_{ij}=M_{ij}, \,\,\,\,\, L_{0, d+1}=D, \,\,\,\,\, L_{d+1, i}=\frac{1}{2}(P_i+K_i), \,\,\,\,\, L_{0, i}=\frac{1}{2}(P_i-K_i)
\end{align}
where $D$ is the dilatation, $P_i$ ($i=1, 2,\cdots d$) are translations, $K_i$ are special conformal transformations and $M_{ij}=-M_{ji}$ are spatial rotations.
The commutation relations of conformal algebra following from (\ref{defiso}) and (\ref{defconf}) are
\begin{align}\label{confalg}
&[D, P_i]=P_i, \,\,\,\,\, [D, K_i]=-K_i, \,\,\,\,\, [K_i, P_j]=2\delta_{ij}D-2M_{ij}\nonumber\\
&[M_{ij}, P_k]=\delta_{jk} P_i-\delta_{ik}P_j, \,\,\,\,\,[M_{ij}, K_k]=\delta_{jk} K_i-\delta_{ik}K_j\nonumber\\
&[M_{ij}, M_{k\ell}]=\delta_{jk} M_{i\ell}-\delta_{ik} M_{j\ell}+\delta_{i\ell} M_{jk}-\delta_{j\ell} M_{ik}
\end{align}

\noindent{}The generators $L_{AB}$  exponentiate to group elements in $\SO(1, d+1)$
\begin{align}
U(\theta)= \exp\left(\frac{1}{2}\theta^{AB} L_{AB}\right)
\end{align}
where $\theta^{AB}=-\theta^{BA}$ are real parameters. Some important subgroups of $\SO(1, d+1)$ that will be used later are 
\begin{align}\label{subgroups}
&\text{K}= \,\SO(d+1), \,\,\,\,\, \text{M}=\left\{e^{\frac{1}{2}\omega^{ij}M_{ij}}, \omega^{ij}\in\mathbb R\right\}=\SO(d)\nonumber \\
&\text{N}=\left\{e^{b\cdot K}, b^i\in\mathbb R\right\},\,\,\,\,\, \widetilde{\text{N}}=\left\{e^{x\cdot P}, x^i\in\mathbb R\right\}, \,\,\,\,\, \text{A}=\left\{e^{\lambda D}, \lambda\in\mathbb R\right\}
\end{align}
where $\text{K}$ is the {\it maximal compact} subgroup of $\SO(1, d+1)$. 
 
To get a unitary representation of $\SO(1, d+1)$, the Lie algebra generators must be realized
as \textbf{anti-hermitian} operators on some Hilbert space, i.e.
\begin{align}\label{dSreality}
L_{AB}^\dagger = - L_{AB}
\end{align}
This is the reality condition relevant to $(d+1)$-dimensional  unitary quantum field theories on a  fixed dS background. Notice it is different from the reality conditions relevant to $d$-dimensional  unitary  CFTs or  $(d+1)$-dimensional unitary quantum field theories on a  fixed  (E)AdS background. The latter corresponds to the reality condition of the $\so(2,d)$ algebra obtained
by Wick-rotating the $X^{d+1}$ direction. This gives for example $P_i^\dagger = K_i $ whereas for $\so(1, d+1)$ we have $P_i^\dagger=- P_i$. Throughout the paper, we will call (\ref{dSreality}) the {\it dS reality}.

The quadratic Casimir, which commutes with all $L_{AB}$, is chosen to be 
\begin{align}\label{generalcas}
\CC_2&=\frac{1}{2}L_{AB}L^{AB}=-D^2+\frac{1}{4}(P_i+K_i)^2-\frac{1}{4}(P_i-K_i)^2+\frac{1}{2} M_{ij}M^{ij}\nonumber\\
&=-D^2+\frac{1}{2}(P_iK_i+K_i P_i)+\frac{1}{2} M_{ij}^2\nonumber\\
&=D(d-D)+P_i K_i+\frac{1}{2} M_{ij}^2
\end{align}
Here $\frac{1}{2}M_{ij}^2\equiv \frac{1}{2} M_{ij}M^{ij}$ is the quadratic Casimir of $\SO(d)$ and it is negative-definite for a unitary representation since $M_{ij}$ are anti-hermitian. For example, for a spin-$s$ representation of $\SO(d)$, it takes the value of $-s(s+d-2)$.

\section{UIRs of $\SO(1,2)$}\label{SO12}
When $d=1$, the Lie algebra $\so(1, 2)$ is generated by $\{P, D,K\}$ satisfying the following commutation conditions
\begin{align}\label{PDKcom}
[D, P]=P, \,\,\,\,\, [D, K]=-K, \,\,\,\,\, [K, P]=2D
\end{align}
The quadratic Casimir is $
\CC_2=D(1-D)+P K$. Before delving into the detailed construction of UIRs, we want to mention that $\so(1,2)$ is also the Lie algebra of $\SL(2,\mathbb R)$, the double covering group of $\SO(1,2)$. Some excellent mathematical textbooks on the representation theory of $\SL(2,\mathbb R)$ are \cite{SL2,10.2307/j.ctt1bpm9sn}. Kitaev's note \cite{Kitaev:2017hnr} about the representations of the universal covering group of $\SL(2,\mathbb R)$ is also a good resource on this subject. We will review the UIRs of $\SL(2,\mathbb R)$ in appendix \ref{SLrev}.

\subsection{A direct construction}\label{direct}
Define $L_0=i L_{12}=-\frac{i}{2}(P+K)$ to be the {\it Hermitian} operator that generates the maximal compact subgroup $\SO(2)$. It takes value in integers for an  $\SO(1,2)$ group representation since $e^{2\pi i L_0}$ is the identity operator in $\SO(1, 2)$. As in the $SU(2)$ case, we define ladder operators
\begin{align}
L_\pm=-\frac{i}{2}(P-K)\mp D
\end{align}
and they satisfy 
\begin{align}\label{comher}
[L_0, L_\pm]=\pm L_\pm, \,\,\,\,\, [L_-, L_+]=2L_0, \,\,\,\,\, L_-^\dagger=L_+
\end{align}
Thus $L_+$ raises the eigenvalue of $L_0$ by 1 while $L_-$ lowers it by 1. In terms of the basis $\{L_0, L_\pm\}$, the quadratic Casimir operator is expressed as 
\begin{align}\label{casimir2}
\CC_2=L_+L_-+L_0(1-L_0)
\end{align}

To construct the whole representation space of certain UIR $R$ of $\SO(1, 2)$, we start from an eigenstate of $L_0$ in $R$, i.e. a state $|N\rangle$ satisfying $L_0|N\rangle=N|N\rangle$ where $N\in\mathbb Z$. By acting $L_\pm$ on $|N\rangle$ repeatedly, we obtain a  tower of states: $\{L_+^{p_+} |N\rangle, |N\rangle, L_-^{p_-}|N\rangle\}_{p_\pm \in\mathbb Z_+}$. There exist three different scenarios: (i) the $L_-$ action truncates, (ii) the $L_+$ action truncates, and (iii) neither $L_+$ or $L_-$ truncates. In scenarios (i),  $L_0$ is bounded from below. Let $p\in\mathbb Z$ be the minimal eigenvalue of $L_0$. The spectrum of $L_0$ is $\{p, p+1,p+2,\cdots\}$ and the quadratic Casimir is equal to $p(1-p)$. In scenarios (ii), $L_0$ is bounded from above. Let $q-1\in\mathbb Z$ be the maximal eigenvalue of $L_0$. The spectrum of $L_0$ is $\{q-1, q-2, q-3,\cdots\}$ and  the quadratic Casimir is equal to $q(1-q)$.  In scenario (iii), $L_0$  takes value in all integers. The three scenarios can be described in a uniform way as follows. First, we fix the quadratic Casimir $\CC_2=\Delta(1-\Delta)$, where $\Delta$ is an arbitrary complex number. Then by choosing proper normalization for the eigenstates of $L_0$, we also fix the action of $L_-$
\begin{align}\label{L-action}
L_-|n\rangle=(n-\Delta)|n-1\rangle
\end{align}
where $|n\rangle$ is the eigenstate  of $L_0$ with eigenvalue $n$. The action of $L_+$ follows from (\ref{L-action}) and the Casimir condition, i.e. $\CC_2|n\rangle=\Delta(1-\Delta)|n\rangle$
\begin{align}\label{L+action}
L_+|n\rangle=(n+\Delta)|n+1\rangle
\end{align}
It is clear that the truncations of $L_\pm$ discussed above happen when $\Delta$ hits integers.

With the action being known, we proceed to analyze the constraint of unitarity,  which consists of two parts: (a) the positivity of  $\langle n|n\rangle$,  and (b) the reality condition $L^\dagger_+=L_-$, i.e. $
\langle n|L_-|n+1\rangle =\langle n+1|L_+ |n\rangle^*$.
The latter implies
\begin{align}
\frac{\langle n+1|n+1\rangle}{\langle n|n\rangle}=\frac{n+\bar\Delta}{n+\Delta^*}\equiv \lambda_n
\end{align}
where $\bar\Delta=1-\Delta$ and $\Delta^*$ is the complex conjugate of $\Delta$. The reality of $\lambda_n$ yields two possible families of solutions for $\Delta$:
\begin{align}
(i) \, \Delta=\frac{1}{2}+i\mu, \,\,\mu\in\mathbb R, \,\,\,\,\, (ii)\, \Delta\in\mathbb R
\end{align}
The positivity of $\lambda_n$ requires separate discussions for different values of $\Delta$:
\begin{itemize}
\item $\lambda_n$ is identically equal to 1 when $\Delta=\frac{1}{2}+i\mu$. We can consistently choose $\langle n|n\rangle=1$ for any $n\in\mathbb Z$ and hence the resulting representation, denoted by $\CP_{\Delta}$, is unitary. $\CP_{\Delta}$ is a (unitary) {\it principal series} representation.
\item When $\Delta\in\mathbb R$ but $\Delta\notin\mathbb Z$, $n$ takes value in all integers and we need 
\begin{align}
\lambda_n=\frac{n+\bar\Delta}{n+\Delta}=\frac{(n+\frac{1}{2})^2-(\Delta-\frac{1}{2})^2}{(n+\Delta)^2}>0
\end{align}
to hold for all $n\in\mathbb Z$. The most stringent constraint clearly comes from $n=0$, which implies $0<\Delta<1$. For $\Delta$ in this range, we choose the following normalization
\begin{align}\label{normc}
\langle n|n\rangle=\frac{\Gamma(n+\bar\Delta)}{\Gamma(n+\Delta)}
\end{align}
 Altogether, we obtain a new continuous family of UIRs, denoted by $\CC_\Delta$ for each fixed $\Delta\in(0,1)$. They are called {\it complementary series} representations. At $\Delta=\frac{1}{2}$, the intersection point of principal series and complementary series, the norm  (\ref{normc}) is reduced to $\langle n|n\rangle=1$. Both principal and complementary series are irreducible as shown in the fig. (\ref{CP})
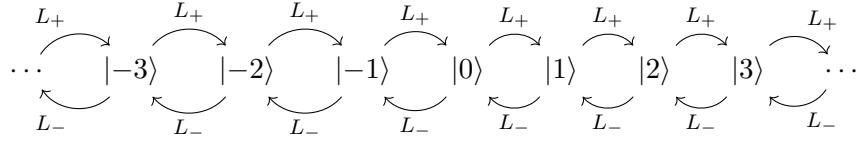
\begin{figure}[H]
\centering
\begin{tikzcd}[column sep=small]
\cdots \arrow[bend left=50]{r}{L_+}& \left|-3\right\rangle  \arrow[bend right=-50]{l}{L_-} \arrow[bend left=50]{r}{L_+} & \left|-2\right\rangle  \arrow[bend right=-50]{l}{L_-} \arrow[bend left=50]{r}{L_+} & \left|-1\right\rangle \arrow[bend right=-50]{l}{L_-} \arrow[bend left=50]{r}{L_+} & \left|0\right\rangle  \arrow[bend right=-50]{l}{L_-} \arrow[bend left=50]{r}{L_+} & \left|1\right\rangle \arrow[bend right=-50]{l}{L_-} \arrow[bend left=50]{r}{L_+} & \left|2\right\rangle  \arrow[bend right=-50]{l}{L_-} \arrow[bend left=50]{r}{L_+} & \left|3\right\rangle \arrow[bend right=-50]{l}{L_-} \arrow[bend left=50]{r}{L_+} &  \cdots  \arrow[bend right=-50]{l}{L_-} 
\end{tikzcd}
\caption{The action of $L_\pm$ in principal and complementary series. }
\label{CP}
\end{figure} 
 
\item When $\Delta\in\mathbb Z_+$, the vector space spanned by all $|n\rangle$ is reducible, c.f. diagram (\ref{Delta=1}), and contains two {\it irreducible} $\so(1, 2)$-invariant subspaces:
\begin{align}
&\{|n\rangle\}_{n\ge \Delta}: \text{carries a lowest-weight representation}\,\, \CD^{+}_\Delta\nonumber\\
&\{|n\rangle\}_{n\le -\Delta}: \text{carries a highest-weight representation}\,\, \CD^{-}_\Delta
\end{align}

\begin{figure}[H]
\centering
\captionsetup{width=.85\linewidth}
\begin{tikzcd}[column sep=small]
\cdots \arrow[bend left=50]{r}{L_+}& \left|-3\right\rangle  \arrow[bend right=-50]{l}{L_-} \arrow[bend left=50]{r}{L_+} & \left|-2\right\rangle  \arrow[bend right=-50]{l}{L_-} \arrow[bend left=50]{r}{L_+} & \left|-1\right\rangle \arrow[bend right=-50]{l}{L_-}  & \left|0\right\rangle  \arrow[bend right=-50]{l}{L_-} \arrow[bend left=50]{r}{L_+} & \left|1\right\rangle  \arrow[bend left=50]{r}{L_+} & \left|2\right\rangle  \arrow[bend right=-50]{l}{L_-} \arrow[bend left=50]{r}{L_+} & \left|3\right\rangle \arrow[bend right=-50]{l}{L_-} \arrow[bend left=50]{r}{L_+} &  \cdots  \arrow[bend right=-50]{l}{L_-} 
\end{tikzcd}
\caption{The action of $L_\pm$ for $\Delta=1$. $ \{|1\rangle, |2\rangle, \cdots\}$ furnishes an irreducible lowest-weight representation and $\{|-1\rangle, |-2\rangle, \cdots \}$ furnishes an irreducible highest-weight representation.}
\label{Delta=1}
\end{figure}
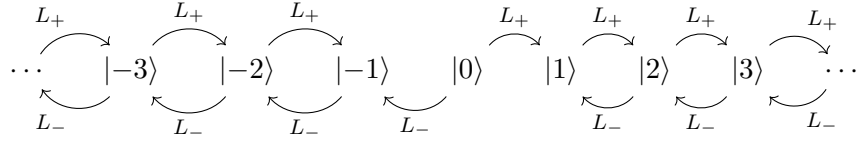
It is clear that all $\lambda_n$ are positive when restricted to these two subspaces. Thus $\CD^\pm_\Delta$ are also UIRs, called {\it discrete series} \footnote{A UIR $R$ of G is in the discrete series, if and only if  some matrix coefficient $\langle  v|R(g)|w\rangle$ is a square-integrable function on $G$. A proof about the existence of such matrix coefficients for $\CD^\pm_\Delta$ can be found in \cite{10.2307/j.ctt1bpm9sn}.  }.
In each $\CD^\pm_\Delta$, we choose the normalization to be 
\begin{align}\label{discketin}
\langle n|n\rangle_{\CD^\pm_\Delta}=\frac{\Gamma(\pm n+\bar\Delta)}{\Gamma(\pm n+\Delta)}
\end{align}
\item When $\Delta\in\mathbb Z_{-}$, the vector space spanned by all $|n\rangle$ contains only one {\it irreducible} $\so(1, 2)$-invariant subspace which furnishes a finite dimensional nonunitary representation. When $\Delta=0$, $|0\rangle$ itself furnishes the {\it trivial} representation. See fig. (\ref{Delta=-1}) for an example when $\Delta=-1$.
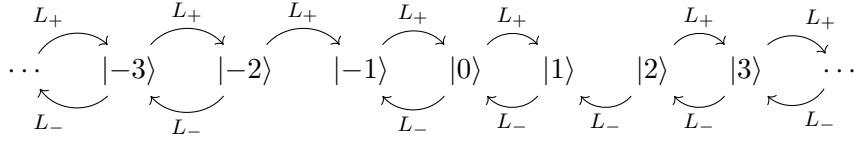
\begin{figure}[H]
\centering
\captionsetup{width=.85\linewidth}
\begin{tikzcd}[column sep=small]
\cdots \arrow[bend left=50]{r}{L_+}& \left|-3\right\rangle  \arrow[bend right=-50]{l}{L_-} \arrow[bend left=50]{r}{L_+} & \left|-2\right\rangle  \arrow[bend right=-50]{l}{L_-} \arrow[bend left=50]{r}{L_+} & \left|-1\right\rangle  \arrow[bend left=50]{r}{L_+} & \left|0\right\rangle  \arrow[bend right=-50]{l}{L_-} \arrow[bend left=50]{r}{L_+} & \left|1\right\rangle \arrow[bend right=-50]{l}{L_-}  & \left|2\right\rangle  \arrow[bend right=-50]{l}{L_-} \arrow[bend left=50]{r}{L_+} & \left|3\right\rangle \arrow[bend right=-50]{l}{L_-} \arrow[bend left=50]{r}{L_+} &  \cdots  \arrow[bend right=-50]{l}{L_-} 
\end{tikzcd}
\caption{The action of $L_\pm$ when $\Delta=-1$. In this case, $|\pm1\rangle$ and $|0\rangle$ span the spin-1 representation of $\SO(1,2)$.}
\label{Delta=-1}
\end{figure}
\end{itemize}
In conclusion, the conformal group $\SO(1, 2)$ admits four types of unitary irreducible representations (the irreducibility is automatically guaranteed by the construction given above): the principal series $\CP_\Delta$ for $\Delta\in\frac{1}{2}+i\mathbb R$, the complementary series $\CC_\Delta$ for $0<\Delta<1$, the discrete series  $\CD^\pm_\Delta$ for $\Delta\in\mathbb Z_+$ and the trivial representation.
In addition, for the principal and complementary series, there exists an isomorphism between $\CP_{\Delta}$ ($\CC_\Delta$) and $\CP_{\bar \Delta}$ ($\CC_{\bar\Delta}$),  which is realized by the following {\it invertible intertwining map}:
\begin{align}\label{Shadow1}
\CS_\Delta: |n\rangle_\Delta\longmapsto s_\Delta \frac{\Gamma(n+\bar\Delta)}{\Gamma(n+\Delta)} |n\rangle_{\bar\Delta}
\end{align}
where $|n\rangle_\Delta$ denotes the $|n\rangle$-basis in $\CP_{\Delta}$ ($\CC_\Delta$) and similarly for $|n\rangle_{\bar\Delta}$. Choose the normalization factor $s_\Delta$ such that $s_\Delta s_{\bar\Delta}=1$ and then $\CS_{\bar\Delta}\circ\CS_{\Delta}=1$, where $\circ$ means the composition of maps. In the CFT terminologies,  this intertwining map $\CS_\Delta$ corresponds to a  {\it shadow transformation} \cite{Ferrara:1972xe,Ferrara:1972ay,Ferrara:1972uq,Ferrara:1972kab}, which sends an operator with scaling dimension $\Delta$ to a nonlocal operator with scaling dimension $\bar\Delta$. We will derive the more common form of shadow formations in section \ref{St12} (for $\SO(1,2)$) and section \ref{shadownew} (for $\SO(1, d+1)$).

\subsection{A CFT-type construction}
In this section, we will show that all the UIRs of $\SO(1,2)$ constructed above can be realized in certain function spaces by borrowing ideas from conformal field theory. Compared to the previous construction, this method admits a straightforward generalization to the higher dimensions.

\subsubsection{Representation spaces}
We start with a {\it primary state of scaling dimension} $\bar\Delta= 1-\Delta$, i.e. a state $|\bar\Delta, 0\rangle$ satisfying:
\begin{align}
K|\bar\Delta, 0\rangle=0, \,\,\,\,\, D|\bar\Delta, 0\rangle=\bar\Delta|\bar\Delta, 0\rangle
\end{align}
However, unlike in usual unitary CFT, we do not require this state to be normalizable (indeed, as we shall see, it is not and hence does not belong to the Hilbert spaces we will construct). Acting by translations on this state produces a continuous family of states
\begin{align}
|\bar\Delta, x\rangle= e^{x P} |\bar\Delta, 0\rangle
\end{align}
Using this definition and the commutation relations (\ref{PDKcom}), we obtain (dropping the label $\bar\Delta$ for $|\bar\Delta, x\rangle$)
\begin{align}\label{onket}
P|x\rangle=\partial_x|x\rangle, \,\,\,\,\, D|x\rangle=(x\partial_x+\bar\Delta)|x\rangle,\,\,\,\,\,K|x\rangle=(x^2\partial_x+2\bar\Delta x)|x\rangle
\end{align}
Then a general state $|\psi\rangle$ can be expressed  as a linear combination
\begin{align}\label{generalstate}
|\psi\rangle\equiv \int_{\mathbb R}\, dx\, \psi(x)|x\rangle
\end{align}
At this point,  we allow $\psi(x)$ to be any smooth function. The action of the conformal generators on the wavefunctions $\psi(x)$ is obtained from (\ref{onket}) and integrating by part. For example, $P |\psi\rangle=\int dx\, \psi(x)\partial_x|x\rangle=\int dx\, (-\partial_x\psi(x))|x\rangle$. This gives 
\begin{align}\label{onfunction}
P\psi(x)=-\partial_x\psi(x), \,\,\,\,\, D\psi(x)=-(x\partial_x+\Delta)\psi(x),\,\,\,\,\,K\psi(x)=-(x^2\partial_x+2\Delta x)\psi(x)
\end{align}
The requirement of  lifting the Lie algebra action (\ref{onfunction}) to a group action imposes constraints on the wavefunctions. In particular, exponentiating the special conformal transformation yields
\begin{align}
\left(e^{b K}\psi\right)(x)=\left(\frac{1}{1+b x}\right)^{2\Delta}\psi\left(\frac{x}{1+b x}\right)
\end{align}
By taking the limit $x\to -\frac{1}{b}$, we can  read off the asymptotic behavior of $\psi(x)$
\begin{align}\label{asymp1}
\psi(x)\stackrel{|x|\to\infty}{\approx}\frac{1}{|x|^{2\Delta}}\left(c_\psi+\CO\left(\frac{1}{|x|}\right)\right)
\end{align}
where $c_\psi$ is some constant. Let $\CF_\Delta$ be the complex vector space of infinitely differentiable functions satisfying the asymptotic boundary condition (\ref{asymp1}) and it furnishes a representation of $\SO(1,2)$ whose infinitesimal version is given by (\ref{onfunction}).

\,

\,

\subsubsection{Consequences of unitarity}\label{cou}

Given the set of states/wavefunctions, we should define a positive definite inner product $\langle x|y\rangle$ that respects the dS reality condition. First notice that $P^\dagger=-P$ leads to the relation $\partial_y\langle x|y\rangle=\langle x|P|y\rangle=-\partial_x\langle x|y\rangle$, so $(\partial_x+\partial_y)\langle x|y\rangle=0$, i.e. $\langle x|y\rangle=f(x-y)$ for some function $f(x)$. Likewise the reality condition of dilatation and special conformal transformation requires
\begin{align}
&(x\partial_x+y\partial_y +\bar\Delta+\bar\Delta^*)f(x-y)=0\\
& (x^2\partial_x+y^2\partial_y+2\bar\Delta y+2\bar\Delta^* x)f(x-y)=0
\end{align}
where the first equation fixes the scaling property of $f(x)$ as $(x \partial_x+\bar\Delta+\bar\Delta^*)f(x)=0$, which together with the second equation implies
\begin{align}\label{fixing}
(\Delta-\Delta^*) x f(x)=0
\end{align}
The eq. (\ref{fixing}) holds when either $\Delta$ is real or $f(x)\propto \delta(x)$. The former further implies $f(x)\propto \frac{1}{|x|^{2\bar\Delta}}$ while the latter implies $\Delta+\bar\Delta=1$. Thus we arrive at the conclusion that the dS reality condition allows two qualitatively different branches of $\Delta$
\begin{align}\label{class12}
&\text{\rom{1}}:\Delta=\frac{1}{2}+i\mu, \,\, \mu\in\mathbb R,  \,\,\,\,\, \langle x| y\rangle=c\, \delta(x-y) \nonumber\\
&\text{\rom{2}}: \Delta\in\mathbb R,  \,\,\,\,\, \,\,\,\,\, \,\,\,\,\, \,\,\,\,\, \,\,\,\,\, \,\,\,\,\, \,\,\,\,\,\,\langle x| y\rangle=c\frac{2^{\bar\Delta}}{(x-y)^{2\bar\Delta}}
\end{align}
where $c$ is some constant that may depend on $\Delta$. Correspondingly the inner product of two states of the form (\ref{generalstate}) equals 
\begin{align}\label{innerproduct}
\langle \psi_1|\psi_2\rangle=\begin{cases}c \int_{\mathbb R} dx\, \psi_1^*(x)\psi_2(x)\,\,\,\,\, &\Delta=\frac{1}{2}+i\mu \\ 2^{\bar\Delta}c\int_{\mathbb R\times\mathbb R} dx\,dy\,  \frac{\psi_1^*(x)\psi_2(y)}{|x-y|^{2\bar\Delta}} &\Delta\in\mathbb R\end{cases}
\end{align}
In case \rom{1}, the inner product (\ref{innerproduct}) defines the usual $L^2$-norm for wavefunctions in the space $\CF_{\frac{1}{2}+i \mu}$ with $c$ chosen to be 1. This norm is clearly well-defined because for any $\psi\in\CF_{\frac{1}{2}+i\mu}$,  $|\psi|^2$ decays as $\frac{1}{|x|^2}$ near infinity. Therefore, $\psi\in\CF_{\frac{1}{2}+i\mu}$ equipped with the $L^2$-norm carries a unitary irreducible representation of $\SO(1,2)$, which as we shall see is isomorphic to $\CP_{\frac{1}{2}+i\mu}$. To analyze the positivity of (\ref{innerproduct}) in case \rom{2}, it is more convenient to  rewrite the kernel $\CK_\Delta(x,y)\equiv\frac{2^{\bar\Delta}\,c}{|x-y|^{2\bar\Delta}}$ in momentum space
\begin{align}\label{Kinp}
\CK_\Delta(p)&=\int_{-\infty}^\infty\,dx \frac{2^{\bar\Delta}\,c}{|x|^{2\bar\Delta}}e^{i p x}=\frac{2^{\bar\Delta}\,c}{\Gamma(\bar\Delta)}\int_0^\infty \frac{dt}{t} \, t^{\bar\Delta} \int_{-\infty}^\infty dx\, e^{-t x^2 +i px}\nonumber\\
&=\frac{2^{\bar\Delta}\,\sqrt{\pi} c}{\Gamma(\bar\Delta)}\int_0^\infty \frac{dt}{t} t^{\bar\Delta-\frac{1}{2}} e^{-\frac{p^2}{4t}}= \frac{c'}{p^{2\Delta-1}}, \,\,\,\,\,\,\, c'=2^{\Delta}\sqrt{\pi}\frac{\Gamma(\Delta-\frac{1}{2})}{\Gamma(\bar\Delta)} c
\end{align}
As long as we choose $c$ properly, the kernel $\CK_\Delta(p)$ is a  positive  function of $p$. However, the positivity of $\CK_\Delta(p)$ as a function does {\it not} necessarily imply that the inner product on $\CF_\Delta$ defined by $\CK_\Delta(p)$ is positive definite. We have to check that all the wavefunctions in $\CF_\Delta$ are normalizable with respect to this inner product, otherwise regularization can destroy the positivity. Since $\psi(x)\in\CF_{\Delta}$ is infinitely differentiable, its Fourier transformation decays rapidly as $|p|\to\infty$. It suffices to check the normalizability for small $p$. Let $\psi(p)$ be the Fourier transformation of $\psi(x)$. Its small $p$ behavior has two types of leading fall-offs: $c_1 \, p^0$ and $c_2\, p^{2\Delta-1}$ where $c_1, c_2$ are constants. The former comes from the regularity of $\psi(x)$ for finite $x$ \footnote{For example, for $\psi(x)=\frac{1}{x^{2\Delta}}$, which is singular at the origin,  we do not have the $p^0$ behavior.} and the latter arises from the large $x$ asymptotic behavior of $\psi(x)$. As an explicit example, let's take $\psi(x)=\frac{1}{(1+x^2)^\Delta}$, which clearly lives in the space $\CF_\Delta$. Its Fourier transformation can be performed analytically $\psi(p)\propto p^{\Delta-\frac{1}{2}}K_{\frac{1}{2}-\Delta}(p)$, where $K_{\frac{1}{2}-\Delta}(p)$ is a Bessel-$K$ function of order $\frac{1}{2}-\Delta$. Then the two leading fall-offs of $\psi(p)$ are a straightforward result of the following asymptotic property of Bessel $K$-function
\begin{align}
K_\alpha(z)\stackrel{z\to 0}{\approx} z^\alpha\left(2^{-1-\alpha}\Gamma(-\alpha) +\CO(z^2)\right)+z^{-\alpha}\left(2^{-1+\alpha}\Gamma(\alpha) +\CO(z^2)\right)
\end{align}
Plugging $\psi(p)\stackrel{p\to 0}{\approx}  c_1+c_2 p^{2\Delta-1}$ into the inner product $\int \frac{dp}{2\pi} \CK_\Delta(p)|\psi(p)|^2$ where $\CK_\Delta(p)=\frac{c'}{p^{2\Delta-1}}$, we conclude that $\psi(x)\in\CF_\Delta$ is normalizable if and only if $2\Delta-1>-1$ and $1-2\Delta>-1$, i.e. 
\begin{align}
0<\Delta<1
\end{align}
which is exactly the range of $\Delta$ for complementary series. 

\,

\,

\subsubsection{A discrete basis of $\CF_\Delta$}

Thus far we have shown that the function spaces $\CF_\Delta$ furnish unitary irreducible representations of $\SO(1,2)$ when $\Delta\in\frac{1}{2}+i\mathbb R$ and $0<\Delta<1$. Indeed they are the same as $\CP_\Delta$ and $\CC_\Delta$ respectively. To prove this claim, we first find the eigenbasis of $L_0=i \left(\frac{1+x^2}{2}\partial_x+\Delta x\right)$ in $\CF_\Delta$, i.e. wavefunctions $\psi^{(\Delta)}_n$ satisfying $L_0\psi^{(\Delta)}_n=n\psi^{(\Delta)}_n$:
\begin{align}\label{L0eigen}
\psi^{(\Delta)}_n(x)=\frac{1}{\sqrt{2\pi}}\left(\frac{1-i x}{1+i x}\right)^n\frac{2^\Delta}{(1+x^2)^\Delta}
\end{align}
In addition, it is straightforward to check 
\begin{align}
L_+\psi^{(\Delta)}_n=(n+\Delta)\psi^{(\Delta)}_{n+1}, \,\,\,\,\, L_-\psi^{(\Delta)}_n=(n-\Delta)\psi^{(\Delta)}_{n-1}
\end{align}
which take the same form as the action of $L_\pm$ on $|n\rangle$, c.f. (\ref{L-action}) and (\ref{L+action}). 

For  $\boxed{\Delta\in\frac{1}{2}+i\mathbb R}$, the $L^2(\mathbb R)$ inner product yields 
\begin{align}
(\psi^{(\Delta)}_n, \psi^{(\Delta)}_m)=c\int_{\mathbb R}\, dx\,  \psi^{(\Delta)}_n(x)^*\psi^{(\Delta)}_m(x)=c\, \delta_{n m}
\end{align}
By choosing $c=1$, we get to identify $\psi^{(\Delta)}_n(x)$ as the state $|n\rangle$ in $\CP_\Delta$. 

For $\boxed{0<\Delta<1}$, the inner product can be computed by expanding the kernel $\CK_\Delta(x_1\!-\!x_2)$ in terms of $\psi^{(\bar\Delta)}_n$. Define $x_i=\tan\frac{\theta_i}{2}$ and we rewrite $\CK_\Delta(x_1\!-\!x_2)$ as 
\begin{align}\label{kernelexp}
\CK_\Delta(x_{12})&=c\frac{2^{\bar\Delta}}{(1+x_1^2)^{\bar\Delta}}\frac{2^{\bar\Delta}}{(1+x^2_2)^{\bar\Delta}}\frac{1}{(1-\cos\theta_{12})^{\bar\Delta}}
\end{align}
where $x_{12}=x_1-x_2$ and $\theta_{12}\equiv \theta_1-\theta_2$. The desired expansion of $\CK_\Delta$ is equivalent to the harmonic expansion of $\frac{1}{(1-\cos\theta)^{\bar\Delta}}$, which can be done as follows. First use the Schwinger's trick to write 
\begin{align}
\frac{1}{(1-\cos\theta)^{\bar\Delta}}=\frac{1}{\Gamma(\bar\Delta)}\int_0^\infty \frac{ds}{s}\,s^{\bar\Delta} e^{-s+s\cos\theta}
\end{align}
Then use the plane wave harmonic expansion 
\begin{align}\label{harmexp}
e^{s\cos\theta}=\sum_{n\in\mathbb Z} I_n (s) \, e^{in\theta}
\end{align}
where  the coefficients are modified Bessel functions.
Using this in the above Schwinger integral we get
\begin{align}\label{harmexp2}
\frac{1}{(1-\cos\theta)^{\bar\Delta}}=\frac{1}{2^{\bar\Delta}\sqrt{\pi}}\frac{\Gamma(\Delta-\frac{1}{2})}{\Gamma(\bar\Delta)}\sum_{n\in\mathbb Z}\frac{\Gamma(\bar\Delta+n)}{\Gamma(\Delta+n)}\, e^{in\theta}
\end{align}
To be precise, convergence of Schwinger integral requires $\frac{1}{2}<\Delta<1$. For more general $\Delta$, the above result can be viewed as the integral renormalized by analytic continuation. Plugging  (\ref{harmexp2}) into (\ref{kernelexp}), we obtain the following expansion of $\CK_\Delta$ 
\begin{align}\label{expandCK}
\CK_\Delta(x_{12})&=\frac{c\, \Gamma(\Delta-\frac{1}{2})}{2^{\bar\Delta}\sqrt{\pi}\Gamma(\bar\Delta)}\frac{2^{\bar\Delta}}{(1+x_1^2)^{\bar\Delta}}\frac{2^{\bar\Delta}}{(1+x^2_2)^{\bar\Delta}}\, \sum_{n\in\mathbb Z}\frac{\Gamma(\bar\Delta+n)}{\Gamma(\Delta+n)}\, e^{in\theta_{12}}\nonumber\\
&=2^\Delta\sqrt{\pi}\, c\frac{ \Gamma(\Delta-\frac{1}{2})}{\, \Gamma(\bar\Delta)}\sum_{n\in\mathbb Z}\frac{\Gamma(\bar\Delta+n)}{\Gamma(\Delta+n)}\,\psi^{(\bar\Delta)}_n (x_1)^*\psi^{(\bar\Delta)}_n (x_2)
\end{align}
where in the last line we have used $e^{-i\theta_i}=\frac{1-i x_i}{1+i x_i}$. By choosing $c=\frac{\Gamma(\bar\Delta)}{2^\Delta\sqrt{\pi}\Gamma(\Delta-\frac{1}{2})}$, the basis $\psi^{(\Delta)}_n$ is normalized as 
\begin{align}
\left(\psi^{(\Delta)}_n,\psi^{(\Delta)}_m\right)=\int_{\mathbb R\times\mathbb R} dx_1\,dx_2\, \psi^{(\Delta)}_n(x_1)^*\CK_\Delta(x_{12}) \psi^{(\Delta)}_m(x_2)=\delta_{nm }\, \frac{\Gamma(\bar\Delta+n)}{\Gamma(\Delta+n)}
\end{align}
and hence we get to identify $\psi^{(\Delta)}_n$ as $|n\rangle$ in $\CC_\Delta$.

\subsubsection{Discrete series}\label{disd=1}
\begin{figure}
    \centering
    \includegraphics[width=0.7\textwidth]{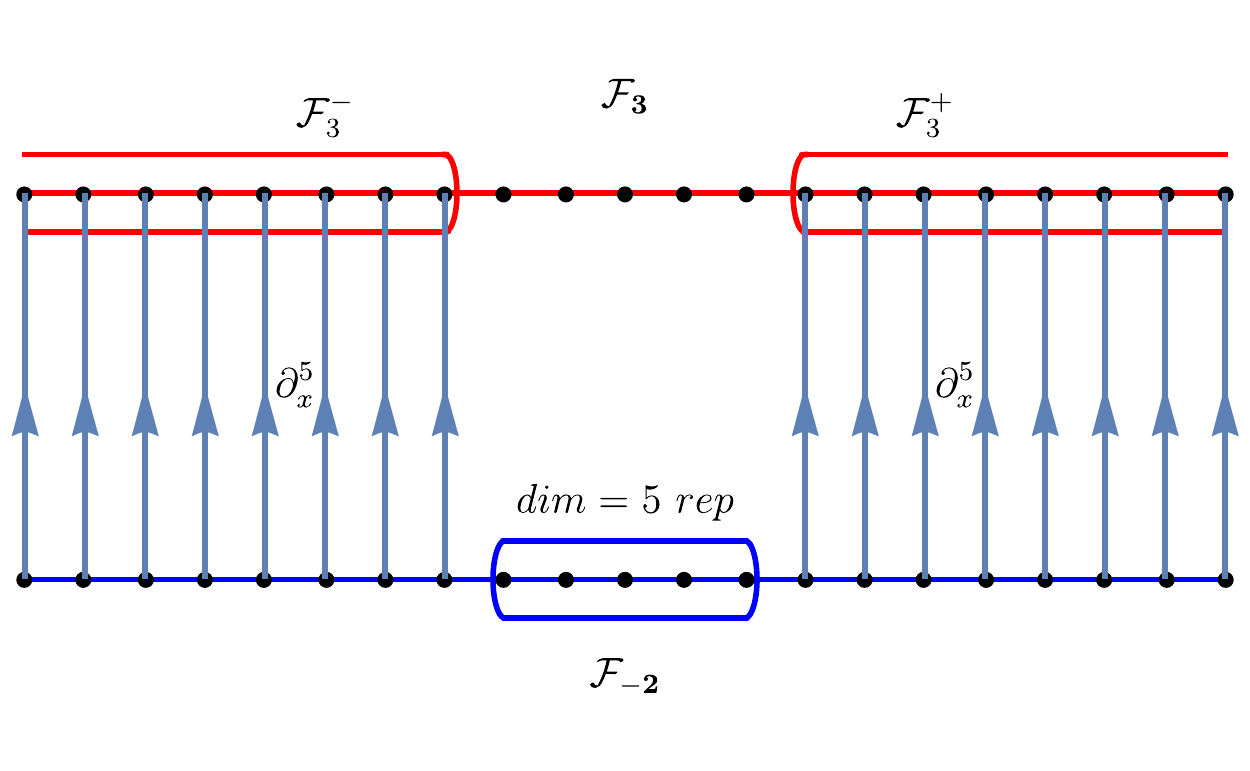}
    \caption{The relation between $\CF_3$ and its shadow $\CF_{-2}$. The upper red line represents the space $\CF_{3}$ and each dot denotes the basis $\psi^{(3)}_k$. It contains two invariant subspaces $\CF^\pm_3$ corresponding to the discrete series representations $\CD^\pm_3$. The lower blue line represents the space $\CF_{-2}$ and each dot denotes the basis $\psi^{(-2)}_k$. It contains a 5-dimensional invariant space $P_3$. The upward arrows denote the intertwining operators $\partial_x^5$. It annihilates  $P_3$ and maps $\psi_k^{(-2)}$ to $\psi_k^{(3)}$ for $|k|\ge 3$.}
        \label{fig:dS2}
\end{figure}

When $\Delta$ is an positive integer, say $\Delta=N\in\mathbb Z_+$, the function space $\CF_N$ has two invariant subspaces $\CF^+_N=\text{Span}\{\psi^{(N)}_n\}_{n\ge N}$ and $\CF^-_N=\text{Span}\{\psi^{(N)}_n\}_{n\le-N}$, which as we will show soon correspond to the discrete series $\CD^\pm_N$ respectively. Similarly, when $\Delta\equiv 1-N$ is a nonpositive integer, $\CF_{1-N}$ contains three invariant subspaces $\CF^+_{1-N}=\text{Span}\{\psi^{(N)}_n\}_{n\ge 1-N}$,  $\CF^-_{1-N}=\text{Span}\{\psi^{(N)}_n\}_{n\le N-1}$ and $P_N=\CF^+_{1-N}\cap\CF^-_{1-N}$, where $P_N$ is nothing but the space of polynomials in $x$ (with coefficients valued in $\mathbb C$) up to degree $2N-2$ and hence is annihilated by the differential operator $\partial_x^{2N-1}$. Furthermore, it is easy to check that $\partial_x^{2N-1}$ is an intertwining  map between $\CF_{1-N}$ and $\CF_N$, as shown in the commutative diagram (\ref{fig:comm1}). In particular, $\partial_x^{2N-1}$ maps $\psi^{(1-N)}_n(x)$ to $\psi^{(N)}_n(x)$ up to an overall constant for any $|n|\ge N$. 
\begin{figure}
\centering
\begin{tikzcd}[scale cd=1.5, column sep=huge, row sep=huge]
\CF_{1-N} \arrow[r, "\partial_x^{2N-1}"] \arrow[d, "L_{AB}"']
& \CF_N \arrow[d, "L_{AB}" ] \\
\CF_{1-N} \arrow[r,  "\partial_x^{2N-1}"' ]
& \CF_N
\end{tikzcd}
\caption{Intertwining operator $\partial_x^{2N-1}: \CF_{1-N}\to \CF_N$. $L_{AB}$ denote the action of $\so(1,2)$ on $\CF_N$ and $\CF_{1-N}$.}
\label{fig:comm1}
\end{figure}
\noindent{}Group theoretically, this claim follows from  $L_0 (\partial_x^{2N-1}\psi^{(1-N)}_n)= \partial_x^{2N-1} L_0\psi^{(1-N)}_n=n \,\partial_x^{2N-1}\psi^{(1-N)}_n$. On the other hand, it can also be proved by a direct computation, for example for $n\ge N$
\begin{align}\label{intertmap}
\partial_x^{2N-1}\psi^{(1-N)}_n(x)&=(-)^{N-1+n}\frac{2^{1-N}}{\sqrt{2\pi}}\partial_x^{2N-1}\frac{(1+ix -2)^{N-1+n}}{(1+i x)^{n+1-N}}\nonumber\\
&=(-)^{N-1+n}\frac{2^{1-N}}{\sqrt{2\pi}}\sum_{k=2N-1}^{N+n-1}(-2)^k \binom{N+n-1}{k}\partial_x^{2N-1}(1+ i x)^{2N-2-k}\nonumber\\
&=i (-)^{N} \frac{\Gamma(N+n)}{\Gamma(n-N+1)}\psi^{(N)}_n(x)
\end{align}
Taking  complex conjugate on both sides of (\ref{intertmap}) yields the result for $n\le -N$. Thus $\CF^\pm_N$ is isomorphic to the quotient space $\CF^\pm_{1-N}/P_N$.  Altogether, the relation between $\CF_N$ and $\CF_{1-N}$ is summarized in the diagram (\ref{fig:dS2})

To define an inner product on $\CF^+_N$ (and similarly on $\CF_N^-$), one would naively expect to use the kernel $\CK_N(x, y)=(x-y)^{2(N-1)}$ as in the complementary series case. However, this does not work because  $\CK_N$ is a delta function  with  derivatives in momentum space while the wavefunctions $\psi^{(N)}_n(x)$ in $\CF_N^+$ are supported on $p>0$ due to the analyticity of $\psi^{(N)}_n(x)$ in the lower $x$-plane when $n\ge N$. Therefore, the inner product defined via $\CK_N(x,y)$ is identically zero. Thanks to the  quotient space realization, i.e. $\CF^\pm_N=\CF^{\pm}_{1-N}/ P_N$, we can avoid using $\CK_N(x,y)$  by working in the space $\CF_{1-N}$. The idea is very simple and let's phrase it in a more general setup.  Given two representations $V_1$ and $V_2$ of some group $G$ and an intertwining map $\varphi: V_1\to V_2$ which can have a nontrivial kernel, then a positive semi-definite $G$-invariant inner product $(\,, \,)_1$ on $V_1$ induces a positive definite $G$-invariant inner product $(\,,\,)_2$ on $\Im\varphi$, if $\ker\varphi$ is orthogonal to the {\it whole} vector space $V_1$ with respect to $(\, ,\, )_1$ and $(\, ,\, )_1$ is positive for vectors not in $\ker\varphi$ . With these assumptions, the induced inner product $(\, , \,)_2$ is given by
$(v_2, v_2)_2\equiv (v_1, v_1)_1$
where $v_i\in V_i$ and $\varphi (v_1)=v_2$. The choice of $v_1$ is clearly not unique when $\varphi$ has a nontrivial kernel but the inner product is  independent of such a choice. We can fix $v_1$ by choosing a map $\mL:V_2\to V_1$ such that $\varphi\circ\mL$ is the identity operator on $V_2$. Then the inner product on $V_2$ becomes
\begin{align}
(v_2, v_2)_2=(\mL v_2, \mL v_2)_1
\end{align}
Now let's take $V_1=\CF^+_{1-N}, V_2=\CF^+_N$ and $\varphi=\partial_x^{2N-1}$. The relation (\ref{intertmap}) yields a positive semi-definite inner product on $V_1=\CF^+_{1-N}$  that satisfies the conditions discussed above
\begin{align}\label{in1}
(\psi_1, \psi_2)_1\equiv i (-)^{N+1}\, c_N\int\, dx\, \psi_1(x)^*\partial_x^{2N-1}\psi_2(x)
\end{align}
where $c_N$ is a positive constant. For example, the inner product of  $\psi^{(1-N)}_m$ and $\psi^{(1-N)}_n$ with respect to $(\,,\,)_1$ is
\begin{align}
(\psi^{(1-N)}_m, \psi^{(1-N)}_n)_1=c_N\frac{\Gamma(N+n)}{\Gamma(n-N+1)}\delta_{mn}
\end{align}
Next we choose a map $\mL_N: \CF_N^+\to \CF_{1-N}^+$ such that $\partial_x^{2N-1}\circ\mL_N$ is the identity operator on $\CF_N^+$. A very natural choice of $\mL_N$ is
\begin{align}\label{defmL}
(\mL_N \psi)(x)=\frac{1}{\Gamma(2N-1)}\int_{\mathbb R+i\epsilon}\frac{dy}{2\pi i }\,\mL_N(x-y)\psi(y) , \,\,\,\,\, \mL_N(z)=z^{2(N-1)}\log(z) 
\end{align}
where the limit $\epsilon\to 0^+$ is understood and the branch cut of $\log z$ is chosen to be the $z<0$ line. When acting $\partial_x^{2N-1}$ on $(\mL_N \psi)(x)$, the kernel $\mL_N(x-y)$ becomes $\frac{\Gamma(2N-1)}{x-y}$ and we can close the contour in the lower half plane, picking up the pole at $y=x$ since $\psi(y)$ is holomorphic in the lower half plane and decays fast enough at $\infty$. Alternatively, we can compute the action of $\mL_N$ on the basis $\psi^{(N)}_n$ directly, which is left to appendix \ref{mLapp}
\begin{align}
\left(\mL_N\psi^{(N)}_n\right)(x)=i(-)^{N+1}\frac{\Gamma(n-N+1)}{\Gamma(N+n)}\psi^{(1-N)}_n(x)+\text{Pol}_{N, n}(x)
\end{align}
where $\text{Pol}_{N, n}(x)$ is a polynomial annihilated by $\partial_x^{2N-1}$. Altogether, the inner product on the space $\CF_N^+$ induced by (\ref{in1}) and (\ref{defmL}) is 
\begin{align}\label{innerFN1}
(\psi, \psi)_{\CF^+_N}=(\mL_N\psi, \mL_N \psi)_1=\frac{(-)^N c_N}{2\pi \Gamma(2N-1)}\int_{\mathbb R}dx \int_{\mathbb R+i\epsilon}dy\,\psi(x)^*\mL_N(x\!-\!y)\psi(y)
\end{align}
In particular, choosing $c_N=1$, we get 
\begin{align}
(\psi^{(N)}_m, \psi^{(N)}_n)_{\CF^+_N}=\frac{\Gamma(n-N+1)}{\Gamma(N+n)}\delta_{mn}
\end{align}
consistent with the inner product (\ref{discketin}) for discrete series $\CD^+_N$.

Before moving to the next section, we want to make some remarks about the kernel $\mL_N(x,y)$.

 \begin{remark} 
 For complementary series of scaling dimension $\Delta$, the inner product is defined through the kernel $\CK_\Delta(x, y)\sim\frac{1}{|x-y|^{2\bar\Delta}}$. As a function of $\Delta$, $\CK_\Delta(x, y)$ is well-defined on the whole complex plane as long as $x\not=y$. Taylor expansion of $\CK_\Delta(x, y)$ around $\Delta=N$ yields
\begin{align}
\CK_\Delta(x, y)\stackrel{\Delta\to N}{\sim} |x-y|^{2(N-1)}+2(\Delta-N)|x-y|^{2(N-1)}\log(x-y)+\cdots
\end{align}
The first term on the R.H.S is  $\CK_N(x, y)$ and we have argued that it leads to a trivial inner product for $\CF^+_N$. The second term, with the numerical factor stripped off, is exactly the kernel $\mL_N(x-y)$ that defines the inner product for discrete series (up to a contour prescription).
\end{remark}

\begin{remark} Unlike $\CK_\Delta(x, y)$, $\mL_N(x-y)$ is not $\SO(1, 2)$-invariant. For example, under scaling transformation 
\begin{align}
e^{\lambda D}: \mL_N(x-y)\to e^{2\lambda(1-N)}\mL_N(e^\lambda x-e^\lambda y)=\mL_N(x-y)+\lambda (x-y)^{2(N-1)}
\end{align}
and under special conformal transformation 
\begin{align}
e^{b K}: \mL_N(x-y)&\to\mL_N\left(\frac{x}{1-bx}-\frac{y}{1-b y}\right) (1-b x)^{2(N-1)}(1-b y)^{2(N-1)}\nonumber\\
&\to \mL_N(x-y)+(x-y)^{2(N-1)}(\log(1-b x)+\log(1-b y))
\end{align}
However, these extra terms do {\it not} contribute to the inner product (\ref{innerFN1}).
\end{remark}

\subsubsection{Shadow transformations}\label{St12}

In the eq. (\ref{Shadow1}), we define the so-called shadow transformation as an intertwining map between the two representations with scaling dimension $\Delta$ and $\bar\Delta$ respectively. In the wavefunction picture, we want to realize $\CS_\Delta$ as a linear operator acting on any $\psi(x)\in\CF_\Delta$ by using a kernel function $S_\Delta(x, y)$
\begin{align}
\CS_\Delta: \,\, \psi(x)\longmapsto \int\, dy S_\Delta(x, y) \psi(y)
\end{align}
The intertwining condition imposes nontrivial constrains on $S_\Delta(x, y)$. These constrains turn out to be nothing but conformal Ward identities associated to the conformal group $\SO(1, 2)$. Thus $S_\Delta(x, y)$ is  the two-point function of a scalar primary operator with scaling dimension $\bar\Delta$ 
\begin{align}\label{shadownormalize}
S_\Delta(x,y)=\frac{2^{\bar\Delta}N_\Delta}{|x-y|^{2\bar\Delta}}, \,\,\,\,\, N_\Delta=\frac{1}{2^\Delta\sqrt{\pi}}\frac{\Gamma(\bar\Delta)}{\Gamma\left(\Delta-\frac{1}{2}\right)}
\end{align}
where the normalization constant $N_\Delta$ is chosen such that $\psi_n^{(\Delta)}$ is mapped to $\frac{\Gamma(\bar\Delta+n)}{\Gamma(\Delta+n)}\psi^{(\bar\Delta)}_n$ as in  (\ref{Shadow1}) with $s_\Delta\equiv 1$. For $\Delta\in \frac{1}{2}+i\mu$ or $0<\Delta<1$, the shadow transformation $\CS_\Delta$ is  an isomorphism between $\CF_\Delta$ and $\CF_{\bar\Delta}$ because it has a well-defined inverse given by $\CS_{\bar\Delta}$, i.e.
\begin{align}
\int_{-\infty}^\infty\, dz\, \frac{2^{\bar\Delta}\,N_\Delta}{|x-z|^{2\bar\Delta}}\frac{2^\Delta\,N_{\bar\Delta}}{|z-y|^{2\Delta}}=\delta(x-y)
\end{align}
This equation can be easily checked by Fourier transformation. Using eq. (\ref{Kinp}) and (\ref{shadownormalize}), it is clear that the shadow transformation $\CS_\Delta$ amounts to a rescaling in momentum space, namely $\CS_\Delta: \psi(p)\longmapsto p^{2\bar\Delta-1} \psi(p)$.

\begin{remark} For $\Delta\in\mathbb Z$, the shadow transformation is not an isomorphism between $\CF_\Delta$ and $\CF_{\bar\Delta}$. For example, when $\Delta=1-N$ is a nonpositive integer, $\psi^{(1-N)}_n$ with $n\le N-1$ are annihilated by $\CS_{1-N}$. Indeed, in this case, $\CS_{1-N}$ is equivalent to the differential operator $\partial^{2N-1}_x$ up to normalization.
\end{remark}
\,

\,

\subsubsection{Summary}
Given a complex constant $\Delta$ which we call {\it scaling dimension}, the $\SO(1, 2)$ generators act on the wavefunction space $\CF_\Delta$ spanned by $\left\{\psi^{(\Delta)}_n(x)=\frac{1}{\sqrt{2\pi}}(\frac{1-i x}{1+i x})^n\frac{2^\Delta}{(1+x^2)^\Delta}\right\}_{n\in\mathbb Z}$ as follows
\begin{align}
P\psi (x)=-\partial_x \psi(x), \,\,\,\,\, D\psi (x)=-(x\partial_x +\Delta)\psi(x), \,\,\,\,\, K\psi (x)=-(x^2\partial_x +2\Delta x)\psi(x)
\end{align}
The quadratic Casimir takes the value $\Delta(1-\Delta)$ acting on $\CF_\Delta$. All unitary irreducible representations of $\SO(1, 2)$ can be realized as either $\CF_\Delta$ or its invariant subspace for certain values of $\Delta$:

\begin{itemize}
\item \textbf{Case \rom{1}}: $\Delta=\frac{1}{2}+i\mu, \,\, \mu\in\mathbb R$ with $L^2(\mathbb R)$ inner product $(\psi, \psi)=\int dx \psi^*(x)\psi(x)$ for any $\psi(x)\in \CF_\Delta$. The representations of $\Delta=\frac{1}{2}\pm i\mu$ are equivalent. This is the {\it principal series}. 

\,

\,

\item \textbf{Case \rom{2}}: $\Delta=\frac{1}{2}+\nu, \,\, -\frac{1}{2}<\nu<\frac{1}{2}$ with  inner product $(\psi, \psi)=\int dx dy \frac{2^{\bar\Delta} N_\Delta}{|x-y|^{2\bar\Delta}}\psi(x)^*\psi(y)$ or equivalently  in momentum space $(\psi, \psi)=\int \frac{dp}{2\pi}\,  p^{1-2\Delta}|\psi(p)|^2$ for any $\psi(x)\in \CF_\Delta$, where $N_\Delta$ is defined in (\ref{shadownormalize}). The representations of $\Delta=\frac{1}{2}\pm \nu$ are equivalent. This is the {\it complementary series}. 

\,

\,
\item \textbf{Case \rom{3}}: $\Delta=N\in \mathbb Z_+$ with inner product
\begin{align}\label{innerFN}
(\psi, \psi)_{\CF^\pm_N}=\frac{(-)^N}{2\pi \Gamma(2N-1)}\int_{\mathbb R}dx \int_{\mathbb R\pm i\epsilon}dy\,\psi(x)^*(x-y)^{2(N-1)}\log(x-y)\psi(y)
\end{align}
where the contour $\mathbb R+i\epsilon$ works for $\psi(x)\in\CF^+_N$ which is spanned by $\{\psi^{(N)}_n(x)\}_{n\ge N}$ and the contour $\mathbb R-i\epsilon$ works for $\psi(x)\in\CF^-_N$ which is spanned by $\{\psi^{(N)}_n(x)\}_{n\le -N}$. This is the {\it discrete series}.
\end{itemize}

\subsection{Understand the two constructions from QFT}\label{QFTpicture}
In this subsection, we provide a physical picture to understand both constructions for the UIRs of $\SO(1,2)$ and their relation. 
For this purpose, let's consider a scalar field $\phi$ in $\text{dS}_2$ of a positive mass $m$. Parameterize the mass by $m^2=\Delta(1-\Delta)$,  with $\Delta\in(0, 1)$ for a light scalar, i.e. $0<m<\frac{1}{2}$,  and $\Delta\in\frac{1}{2}+i\mathbb R$ for a heavy scalar, i.e. $m\ge \frac{1}{2}$. In the following discussion, we will focus on {\it heavy} $\phi$  with $\Delta=\frac{1}{2}+i\mu$ and prove that its single-particle  Hilbert space $\CH_\Delta$ furnishes the principal series of scaling dimension  $\Delta=\frac{1}{2}+i\mu$. A similar computation can be found in \cite{Guijosa:2003ze,Joung:2006gj,Anous:2020nxu}. In addition, we will also illustrate how to realize the discrete basis $|n\rangle$, c.f. eq. (\ref{L-action}), and the continuous basis $|x\rangle$, c.f. eq. (\ref{onket}), on the Hilbert space $\CH_\Delta$. A similar story for light scalars is left to appendix \ref{light}. The QFT picture for $\SO(1,2)$ discrete series is still an open question.

\begin{figure}[h]
    \centering
    \includegraphics[width=0.6\textwidth]{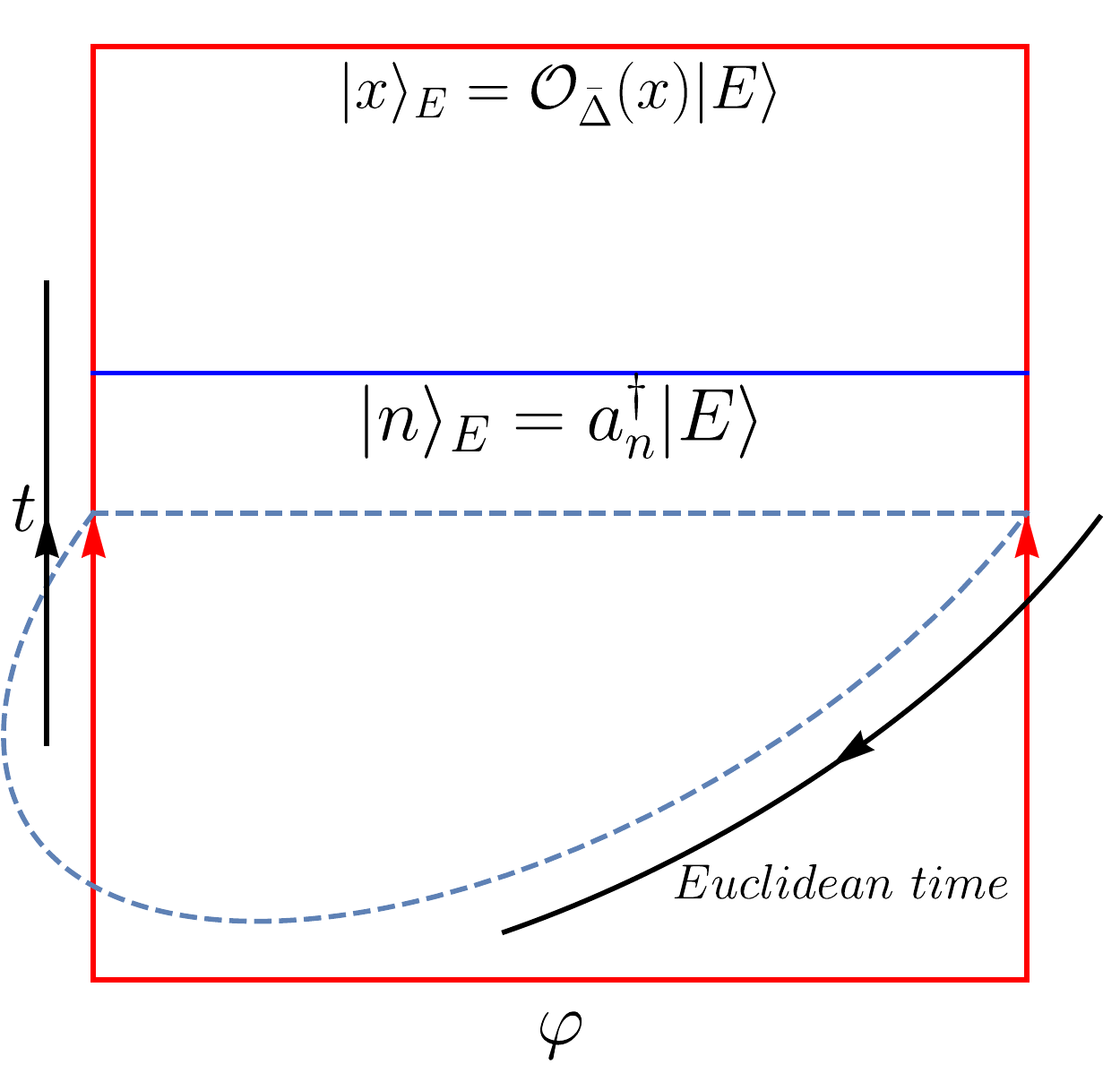}
    \caption{The Penrose diagram of $\text{dS}_2$. The two sides of the Penrose diagram are identified and hence every constant time slice is a circle. The Wick rotation of the $t\le 0$ part of $\text{dS}_2$ is represented by the region bounded by dashed lines. The Euclidean vacuum $|E\rangle$ can be computed by Euclidean path integral in this region.}
        \label{fig:Pen2D1}
\end{figure}

\subsubsection{Mode expansions}
We work in global conformal time coordinates $(t,\varphi)$ of $\text{dS}_2$, related to the embedding space coordinates by
\begin{align}\label{global}
X^0=\tan t, \,\,\,\,\, X^1=\frac{\cos\varphi}{\cos t}, \,\,\,\,\, X^2=\frac{\sin\varphi}{\cos t}, \,\,\,\,\,-\frac{\pi}{2}<t<\frac{\pi}{2}
\end{align}
in terms of which, the $\text{dS}_2$ metric is $ds^2=\frac{-dt^2+d\varphi^2}{\cos^2 t}$, conformally equivalent to a flat cylinder. The future boundary is at $t\to \frac{\pi}{2}$ and the past boundary is at $t\to -\frac{\pi}{2}$. The Wick rotation to  $S^2$ is realized by $t\to -i\tau$, with $\tau\to-\infty$ being the southern pole of $S^2$. The Penrose diagram of $\text{dS}_2$ is given by fig. (\ref{fig:Pen2D1}). Consider solutions to the wave equation $(\nabla^2-m^2)\phi=0$  of the form $\phi(t, \varphi)=f(t) e^{-in\varphi}, n\in\mathbb Z$, where $n$ labels angular momentum along the spatial circle. The equation for $f(t)$ is 
\begin{align}\label{ft}
 f''(t)+n^2 f(t)+ \frac{m^2}{\cos^2 t} f(t)=0
\end{align}
The general solution to this ODE is  $f(t)=c \,g_n(t)+c^* \,g_n(t)^*$, where 
\begin{align}\label{modeg}
g_n(t)=\frac{\Gamma(n+\bar\Delta)}{\sqrt{2}}\, e^{-i n \, t}\, \mathbf{F} \left(\Delta, \bar\Delta, n+1,\frac{1}{1+e^{2 i t}}\right), \,\,\,\,\, \bar\Delta=1-\Delta
\end{align}
Here we have used the regularized hypergeometric function $\mathbf{F}(a, b,c,z)\equiv \frac{1}{\Gamma(c)} F(a,b,c,z)$ because it is well-defined when $c$ is a nonpositive integer.

The Klein-Gordon(KG) inner product is defined as 
\begin{align}\label{KG}
(\phi, \phi')_{\text{KG}}=i\int_0^{2\pi}\,d\varphi\,\left( \phi^*\,\partial_t \phi'-\partial_t\phi^*\, \phi'\right)
\end{align}
For the modes $\phi_n=g_n(t)\frac{e^{-in\varphi}}{\sqrt{2\pi}}$, the KG inner product becomes $(\phi_n, \phi_m)_{\text{KG}}=\delta_{nm}$. A canonically normalized mode expansion of $\phi$ is thus
\begin{align}
\phi(t, \varphi)=\sum_{n\in\mathbb Z} \phi_n \, a_n +\phi_n^*\,  a_n^\dagger
\end{align}
The vacuum $|E\rangle$ of the theory is annihilated by all $a_n$. It is the so-called Euclidean vacuum because $\phi_n$ is regular on the lower hemisphere of the Wick rotated $\text{dS}_2$ \cite{Bousso:2001mw}. The states $|n\rangle_E=a_n^\dagger |E\rangle$, satisfying $_E\langle m|n\rangle_E=\delta_{mn}$, form a normalized orthonormal basis of global single-particle Hilbert space $\CH_\Delta$. Next, we will show that $\CH_\Delta$ is isomorphic to the principal series representation $\CP_{\Delta}$.

\subsubsection{$\so(1,2)$ action on $\CH_\Delta$}
Define the action of $L_{AB}$ on the quantum field $\phi(t,\varphi)$ to be 
\begin{align}\label{fieldact}
[L_{AB}, \phi(t,\varphi)]\equiv -\CL_{AB}\phi(t,\varphi), \,\,\,\,\, \CL_{AB}\equiv X_A\partial_B-X_B\partial_A
\end{align}
where $\CL_{AB}$ are Killing vectors of $\text{dS}_2$ in embedding space coordinates and an extra minus is introduced such that $L_{AB}$ satisfy the commutation relations (\ref{defiso}). In conformal global coordinates (\ref{global}), these Killing vectors are
\begin{align}
\CL_{01}=-\cos t\cos\varphi\partial_t+\sin t\,\sin\varphi\partial_{\varphi}, \,\,\,\,\, \CL_{02}=-\cos t\sin\varphi\partial_t-\sin t\,\cos\varphi\partial_{\varphi}, \,\,\,\,\, \CL_{12}=\partial_{\varphi}
\end{align}
and hence the differential operator realization of $\{L_0, L_\pm \}$ defined in subsection \ref{direct} becomes
\begin{align}
L_0\Leftrightarrow -\CL_0=-i\partial_{\varphi}, \,\,\,\,\, L_\pm\Leftrightarrow -\CL_\pm=-ie^{\mp i\varphi}(\cos\partial_t\mp i \sin t\,\partial_{\varphi})
\end{align}
Using the following recurrence relations of regularized hypergeometric functions
\begin{align}
&\frac{d}{dz}\left(z^{c-1}\mathbf{F}(a,b,c,z)\right)=z^{c-2}\mathbf{F}(a,b,c-1,z)\nonumber\\
&\frac{d}{dz}\left((1-z)^{a+b-c}\mathbf{F}(a,b,c,z)\right)=(c-a)(c-b)(1-z)^{a+b-c-1}\mathbf{F}(a,b,c+1,z)
\end{align}
one can check 
\begin{align}\label{CLpm}
\CL_\pm \phi_n=(n\pm\Delta)\phi_{n \pm 1}, \,\,\,\,\, \CL_\pm \phi_n^*=-(n\mp \bar\Delta)\phi^*_{n\mp 1}
\end{align}
Plug (\ref{CLpm}) into (\ref{fieldact}) and with some reshuffle of indices, we obtain
\begin{align}
[L_\pm, a_n]=-(n\mp \bar\Delta) a_{n \mp 1},\,\,\,\,\, [L_\pm, a^\dagger_n]=(n\pm \Delta) a^\dagger_{n \pm 1}
\end{align}
These commutation relations have some important implications. First, they are consistent with the expected reality condition $L^\dagger_\pm = L_\mp$. Second, together will the almost obvious commutation relation $[L_0, a_n^\dagger]=n a^\dagger_n$, they yield the $\so(1,2)$ action on $\CH_\Delta$
\begin{align}\label{actonH}
L_\pm |n\rangle_E=(n\pm \Delta) |n\pm 1\rangle_E, \,\,\,\,\, L_0|n\rangle_E=n|n\rangle_E
\end{align}
With the inner product $_E\langle m|n\rangle_E=\delta_{mn}$ and the $\so(1,2)$ action (\ref{actonH}), we are able to identify the single-particle Hilbert space $\CH_\Delta$ as the principal series representation $\CP_{\Delta}$ constructed in section \ref{direct}.

\subsubsection{The continuous basis}
In the CFT-type construction, we have used the basis $\{P, D,K\}$ of $\so(1,2)$. Let $\CP,\CD,\CK$ be their differential operator realizations respectively and they take the following form
\begin{align}\label{CPKD}
&\CP=-\cos t\cos\varphi\partial_t+(\sin t\,\sin\varphi-1)\partial_{\varphi}\nonumber\\
 &\CK=\cos t\cos\varphi\partial_t-(\sin t\sin\varphi+1)\partial_{\varphi}\nonumber\\
 & \CD=-\cos t\sin\varphi\partial_t-\sin t\cos\varphi\partial_\varphi
\end{align}
Next we push the scalar field $\phi$ to the future boundary and only keep track of the leading fall-offs. More precisely, put $t=\frac{\pi}{2}-\delta$ and expand $\phi$ in the limit $\delta\to 0$, up to $(1+\CO(\delta^2))$-corrections
\begin{align}\label{boundaryexp1}
\phi\left(t=\frac{\pi}{2}-\delta, \varphi\right)\overset{\delta\to 0}{\approx}C_{\Delta} \, \delta^\Delta\CO_\Delta(\varphi)+C_{\bar\Delta} \, \delta^{\bar\Delta}\CO_{\bar\Delta}(\varphi)
\end{align}
where 
\begin{align}
& C_\Delta=2^{i\mu} \frac{\Gamma(-2i\mu)}{\Gamma(\bar\Delta)}, \,\,\,\,\,\, C_{\bar\Delta}=C_\Delta^*=2^{-i\mu} \frac{\Gamma(2i\mu)}{\Gamma(\Delta)}\nonumber\\
\CO_\Delta(\varphi)= \sum_{n\in\mathbb Z} &(-i)^{n+\Delta} \frac{e^{-in\varphi}}{\sqrt{2\pi}}a_n+i^{n+\Delta} \frac{\Gamma(n+\Delta)}{\Gamma(n+\bar\Delta)}\frac{e^{in\varphi}}{\sqrt{2\pi}}a^\dagger_n, \,\,\,\,\, \CO_{\bar\Delta}(\varphi) =\CO_\Delta(\varphi)^*
\end{align}
Combining (\ref{fieldact}), (\ref{CPKD}) and (\ref{boundaryexp1}), we obtain how the conformal algebra $\{P,K,D\}$ acts on the two boundary operators $\CO_\Delta$ and $\CO_{\bar\Delta}$
\small
\begin{align}\label{manycomm}
&[P,\CO_\Delta(\varphi)]=\left((1-\sin\varphi)\partial_\varphi-\Delta\cos\varphi\right)\CO_\Delta(\varphi), \,\,\,\,\,\, [P,\CO_{\bar\Delta}(\varphi)]=\left((1-\sin\varphi)\partial_\varphi-\bar\Delta\cos\varphi\right)\CO_{\bar\Delta}(\varphi)\nonumber\\
&[K,\CO_\Delta(\varphi)]=\left((1+\sin\varphi)\partial_\varphi+\Delta\cos\varphi\right)\CO_\Delta(\varphi), \,\,\,\,\, [K,\CO_{\bar\Delta}(\varphi)]=\left((1+\sin\varphi)\partial_\varphi+\bar\Delta\cos\varphi\right)\CO_{\bar\Delta}(\varphi)\nonumber\\
&[D,\CO_\Delta(\varphi)]=\left(\cos\varphi\partial_\varphi-\Delta\sin\varphi\right)\CO_\Delta(\varphi), \,\,\,\,\,\,\,\,\,\,\,\,\,\,\,\,\,\, \,\,[D,\CO_{\bar\Delta}(\varphi)]=\left(\cos\varphi\partial_\varphi-\bar\Delta\sin\varphi\right)\CO_{\bar\Delta}(\varphi)
\end{align}
\normalsize
which implies that $\CO_\Delta$ and $\CO_{\bar\Delta}$ are two primary operators of scaling dimension $\Delta$ and $\bar\Delta$ respectively. In particular, their two-point functions with respect to the Euclidean vacuum are
\begin{align}\label{many2pt}
&\langle E|\CO_\Delta(\varphi_1)\CO_{\Delta}(\varphi_2)|E\rangle=\frac{N_{\bar\Delta}}{(1-\cos\varphi_{12})^\Delta}, \,\,\,\,\, \langle E|\CO_{\bar\Delta}(\varphi_1)\CO_{\bar\Delta}(\varphi_2)|E\rangle=\frac{N_{\Delta}}{(1-\cos\varphi_{12})^{\bar\Delta}} \nonumber\\
&\langle E|\CO_\Delta(\varphi_1)\CO_{\bar\Delta}(\varphi_2)|E\rangle=e^{\pi\mu}\delta(\varphi_{12}),\,\,\,\,\,\,\,\,\,\,\,\,\,\,\,\,\,\, \langle E|\CO_{\bar\Delta}(\varphi_1)\CO_{\Delta}(\varphi_2)|E\rangle=e^{-\pi\mu}\delta(\varphi_{12})
\end{align}
where the normalization constant $N_\Delta$ is defined in  eq. (\ref{shadownormalize}). Given these two different primary operators $\CO_\Delta$ and $\CO_{\bar\Delta}$, one would naively expect them to generate two UIRs of $\SO(1,2)$ at the boundary circle of $\text{dS}_{2}$.
However, this is {\it not} the case. The obvious reason is that they are both linear combinations of the annihilation operators $a_n$ and the creation operators $a_n^\dagger$.    Although different as operators, when acting with them on the Euclidean vacuum, the annihilation part drops out, so  the resulting single-particle excitations are nothing but the Hilbert space $\CH_\Delta$. To make this argument more precisely, consider the following two  families of states generated by $\CO_\Delta$ and $\CO_{\bar\Delta}$ respectively
\begin{align}
&|\Delta, \varphi\rangle\equiv \CO_{\Delta}(\varphi)|E\rangle=\sum_n i^{n+\Delta} \frac{\Gamma(n+\Delta)}{\Gamma(n+\bar\Delta)}\frac{e^{in\varphi}}{\sqrt{2\pi}}|n\rangle_E\nonumber\\
&|\bar\Delta, \varphi\rangle\equiv \CO_{\bar\Delta}(\varphi)|E\rangle=\sum_n i^{n+\bar\Delta}\frac{e^{in\varphi}}{\sqrt{2\pi}}|n\rangle_E
\end{align}
They are both complete basis of the single-particle Hilbert space $\CH_\Delta$ in the sense that 
\begin{align}\label{completeness}
\mathbb{1}_{\CH_\Delta}&\equiv \sum_{n}|n\rangle_{E\, E}\langle n|\nonumber\\
&=e^{\pi \mu}\int_0^{2\pi}d\varphi |\Delta, \varphi\rangle\langle \Delta, \varphi|=e^{-\pi \mu}\int_0^{2\pi}d\varphi |\bar\Delta, \varphi\rangle\langle \bar\Delta, \varphi|
\end{align}
Using the completeness condition (\ref{completeness}), we can relate the two basis as follows
\begin{align}
|\Delta,\varphi_1\rangle&=e^{-\pi \mu}\int_0^{2\pi} d\varphi_2  \langle \bar\Delta, \varphi_2 |\Delta,\varphi_1\rangle |\bar\Delta,\varphi_2\rangle\nonumber\\
&=e^{-\pi \mu} \int_0^{2\pi} d\varphi_2 \frac{N_{\bar\Delta}}{(1-\cos\varphi_{12})^\Delta}|\bar\Delta,\varphi_2\rangle
\end{align}
where the kernel function $ \frac{N_{\bar\Delta}}{(1-\cos\varphi_{12})^\Delta}$ is the $S^1$ counterpart of the kernel $S_{\bar\Delta}(x_{12})$.
Altogether, we have only one UIR at the boundary even though there exist two boundary primary operators, and we are free to choose either $|\Delta,\varphi\rangle$ or $|\bar\Delta,\varphi\rangle$ as a (continuous) basis of $\CH_\Delta$.

\begin{figure}
    \centering
    \includegraphics[width=0.6\textwidth]{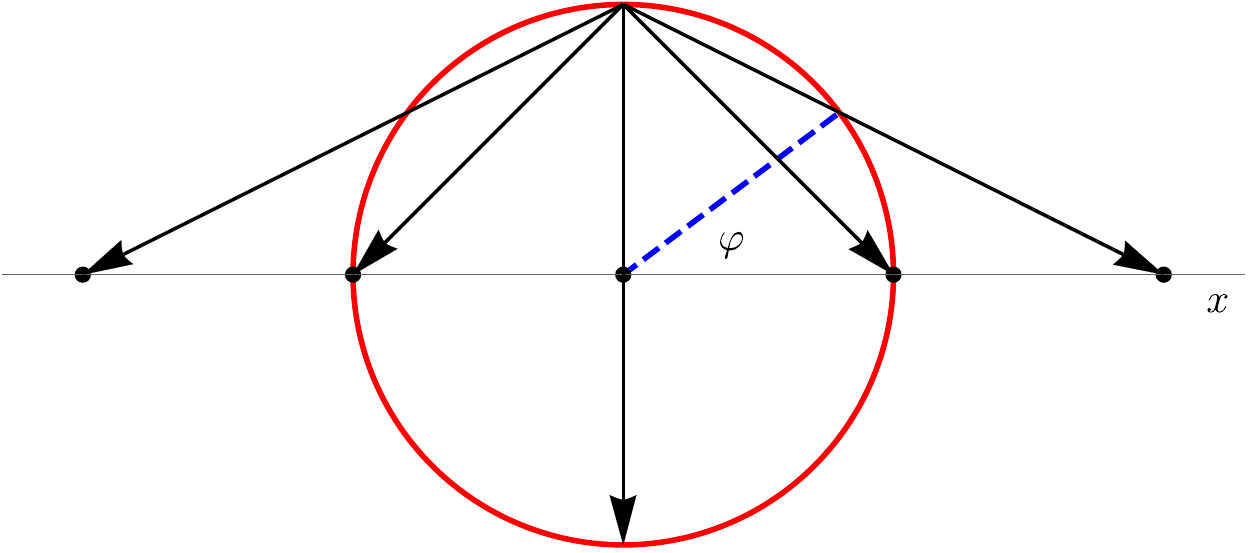}
    \caption{The stereographic map between $S^1$ and $\mathbb R$.}
        \label{fig:qiuji}
\end{figure}

 To make the connection with the $|x\rangle$ basis more explicit, let's switch to the stereographic coordinate of $S^1$, namely $\cos\varphi=\frac{2x}{1+x^2}, \sin\varphi=\frac{x^2-1}{x^2+1}$ (see fig. (\ref{fig:qiuji}) for an intuitive geometric picture of this coordinate transformation) and rescale the $|\bar\Delta, \varphi\rangle$ state by a Weyl factor associated to the map $x\to \varphi$:
 \begin{align}
 |\bar\Delta, x\rangle=\left(\frac{2}{1+x^2}\right)^{\bar\Delta} |\bar\Delta, \varphi(x)\rangle
 \end{align}
 Using the commutation relations in (\ref{manycomm}), it is straightforward to check that (dropping the label $\bar\Delta$ for $|\bar\Delta, x\rangle$)
 \begin{align}
 P|x\rangle=\partial_x |x\rangle, \,\,\,\,\, K|x\rangle=\left(x^2\partial_x+2\bar\Delta x\right)|x\rangle,\,\,\,\,\, D|x\rangle=(x\partial_x+\bar\Delta)|x\rangle
 \end{align}
 which is nothing but eq. (\ref{onket}) in the previous CFT-type construction. In addition, the inner product of the $|x\rangle$-basis can be derived from (\ref{many2pt})
  \begin{align}
 \langle x|y\rangle=\frac{2 \, e^{\pi\mu}}{(1+x^2)}\delta(\varphi(x)-\varphi(y))=e^{\pi\mu}\delta(x-y)
 \end{align}
 which is consistent with eq. (\ref{class12}) for $c=e^{\pi\mu}$.

\subsubsection{Summary}
The physical picture discussed above can be summarized as a dictionary between representation theory of $\SO(1,2)$ and local quantum field theory in $\text{dS}_2$:
\begin{align}
\text{Principal/Complementary Series}\,\, \CF_\Delta &\Longleftrightarrow \text{The single-Hilbert space}\,\,  \CH_\Delta\nonumber\\
\SO(1,2)\text{-invariant inner product of}\,\, \CF_\Delta &\Longleftrightarrow \text{The standard QFT inner product on} \,\,\CH_\Delta\nonumber\\
\text{The discrete}\,\, |n\rangle \,\, \text{basis}&\Longleftrightarrow \text{The global single-particle excitations}  \,\, a_n^\dagger|E\rangle \nonumber\\
\text{The continuous}\,\, |\bar\Delta, x\rangle \,\, \text{basis}&\Longleftrightarrow \text{The boundary  single-particle excitations}\,\, \CO_{\bar\Delta}(x)|E\rangle\nonumber\\
\text{Shadow transformation} &\Longleftrightarrow \text{A basis transformation of }\,\, \CH_\Delta
\end{align}

\section{UIRs of $\SO(1,d+1)$}\label{d+1}

Like in the $d=1$ case,  we start with a primary state $|\bar\Delta, 0\rangle_\alpha$ of  scaling dimension $\bar \Delta\equiv d-\Delta$. Apart from the scaling dimension, we also need to specify an $\SO(d)$ representation $\mY_{\bm{s}}$ carried by the index $\alpha$. Here $\mY_{\bm{s}}$ denotes a Young diagram associated to the $\SO(d)$ highest weight vector $\bm{s}=(s_1, s_2, \cdots, s_r), r\equiv \floor*{\frac{d}{2}}$ \footnote{Our convention for the highest weight vector $\bm s$ is $s_1\ge s_2\ge\cdots \ge s_{r-1}\ge |s_r|$. We also use the notation $\mY_{s_1}$ when $s_1\not=0, s_2=0$ and $\mY_{s_1, s_2}$ when $s_1\not=0, s_2\not=0, s_3=0$.}. In this work, we will focus on the single-row representation of $\SO(d)$, i.e. $\bm s=(s, 0,0,\cdots ), s\in\mathbb N$ which in bulk corresponds to scalar fields when $s=0$ and bosonic spin-$s$ fields when $s\ge 1$. Let  $|\bar\Delta, 0\rangle_{i_1\cdots i_s}$ be such a primary state where the indices $i_1\cdots i_s$ are symmetric and traceless. The action of conformal algebra is 
\begin{align}\label{defineprimary}
&D|\bar\Delta, 0\rangle_{i_1\cdots i_s}=\bar\Delta |\bar\Delta, 0\rangle_{i_1\cdots i_s}, \,\,\,\,\, K_i |\bar\Delta, 0\rangle_{i_1\cdots i_s}=0, \nonumber\\
& M_{k\ell}|\bar\Delta, 0\rangle_{i_1\cdots i_s}=\sum_{j=1}^s |\bar\Delta, 0\rangle_{i_1\cdots i_{j-1} k \,i_{j+1}\cdots i_s}\delta_{\ell i_j}-|\bar\Delta, 0\rangle_{i_1\cdots i_{j-1} \ell\, i_{j+1}\cdots i_s}\delta_{k i_j}
\end{align}
Acting by translations on this state produces a family of position-dependent states (dropping the label $\bar\Delta$ in the kets $|\bar\Delta, x\rangle_{i_1\cdots i_s}$again)
\begin{align}
|x\rangle_{i_1\cdots i_s}\equiv e^{x\cdot P} |\bar\Delta, 0\rangle_{i_1\cdots i_s}
\end{align}
The action of the $\so(1,d+1)$ algebra on these states is then easily computed to be:
\begin{align}\label{actonstate}
& P_i |x\rangle_{i_1\cdots i_s}=\partial_i |x\rangle_{i_1\cdots i_s}\nonumber\\
& D|x\rangle_{i_1\cdots i_s}=(x\cdot\partial_x +\bar\Delta)|x\rangle_{i_1\cdots i_s}\nonumber\\
& M_{k\ell}|x\rangle_{i_1\cdots i_s}=\left(x_\ell \partial_k-x_k\partial_\ell+\CM^{(s)}_{k\ell }\right)|x\rangle_{i_1\cdots i_s}\nonumber\\
& K_k |x\rangle_{i_1\cdots i_s}=\left(2x_k(x\cdot \partial_x+\bar\Delta)-x^2\partial_k-2x^\ell \CM_{k\ell}^{(s)}\right)|x\rangle_{i_1\cdots i_s}
\end{align}
where $\CM^{(s)}_{kl}$ is the spin-$s$ representation of $\so(d)$
\begin{align}
\CM^{(s)}_{kl}|x\rangle_{i_1\cdots i_s}=\sum_{j=1}^s |x\rangle_{i_1\cdots i_{j-1} k \,i_{j+1}\cdots i_s}\delta_{\ell i_j}-|x\rangle_{i_1\cdots i_{j-1} \ell\, i_{j+1}\cdots i_s}\delta_{k i_j}
\end{align} 
A general state in the vector space spanned by all $|x\rangle_{i_1\cdots i_s}$ is of the form 
\begin{align}
|\psi\rangle\equiv \int\, d^d x\, \psi_{i_1\cdots i_s}(x)|x\rangle_{i_1\cdots i_s}
\end{align}
where the wavefunction $\psi_{i_1\cdots i_s}(x)$ is smooth in $x^i$, and symmetric and traceless  with respect to the $i_\alpha$ indices. We package all components of $\psi_{i_1\cdots i_s}(x)$ into a $x$-dependent polynomial of degree $s$ by introducing a complex {\it null} vector $z^i \in \mathbb C^d$, i.e. 
\begin{align}
\psi(x, z)\equiv \frac{1}{s!}\psi_{i_1\cdots i_s} (x) \, z^{i_1}\cdots z^{i_s}
\end{align}
The null vector $z^i$ can be stripped off while respecting the nullness condition by the following {\it interior derivative}
\begin{align}\label{intder}
\CD_{z^i}\equiv \partial_{z^i}-\frac{1}{d+2(z\cdot\partial_z-1)}\, z_i\, \partial_z^2
\end{align}
The $\so(1, d+1)$ action on the wavefunctions $\psi(x, z)$ induced by eq. (\ref{actonstate}) is then given by 
\begin{align}\label{algebrarepd}
&P_i \psi(x, z)=-\partial_i \psi(x, z)\nonumber\\
&D \psi(x, z)=-(x\cdot \partial_x +\Delta)\psi(x, z)\nonumber\\
&M_{k\ell}\psi(x, z)=\left(x_k \partial_{\ell}-x_\ell \partial_k+z_k \partial_{z^\ell}-z_\ell \partial_{z^k}\right)\psi(x, z)\nonumber\\
&K_k\psi(x, z)=\left(x^2\partial_k-2x_k (x\cdot\partial_x+\Delta)-2x^\ell (z_k \partial_{z^\ell}-z_\ell \partial_{z^k})\right)\psi(x, z)
\end{align}
where the shorthand notation $\partial_i$ means {\it exclusively} the derivative with respect to $x^i$. To lift the $\so(1, d+1)$ action (\ref{algebrarepd}) to a group action, it suffices to exponentiate translations, dilatation, rotations and special conformal transformations separately because of the Bruhat decomposition,  i.e. $\SO(1, d+1)$ is the same as a product subgroups $\ttN\tN\tA\tM$, up to a lower dimensional submanifold \cite{Dobrev:1977qv}. The exponentiation of translations, dilatation and rotations is guaranteed by the smoothness condition imposed on the wavefunctions $\psi(x, z)$. However, the same does not hold for special conformation transformations.  Indeed, granting the exponentiation of $K_i$ to a group action on certain subspace $\CC_K$ of smooth functions, we obtain
\begin{align}\label{finiteK}
\left(e^{b\cdot K} \psi\right)(x, z)=\frac{1}{(1+2 b\cdot x +b^2 x^2)^\Delta}\,\psi\left(\frac{x^i+b^i x^2}{1+2 b\cdot x +b^2 x^2}, R(x_b)^i_{\,\,j} R(x)^j_{\,\,k} z^k\right)
\end{align}
where 
\begin{align}\label{xbR}
x^i_b\equiv \frac{x^i}{x^2}+b^i, \,\,\,\,\, R(x)^i_{\, \,j} \equiv \frac{2x^i x_j}{x^2}- \delta^{ij}
\end{align}
Replace $\psi$ by $\psi_1\equiv e^{-b\cdot K}\psi$ which also lives in the same subspace $\CC_K$ and rewrite eq. (\ref{finiteK}) in terms of $\tilde x^i \equiv \frac{x^i}{x^2}$ (except the combination $R(x)^i_{\,\, j} z^j$)
\begin{align}\label{finiteK1}
\psi (x, z)=\frac{1}{(x^2)^\Delta} \frac{1}{(\tilde x^2+2 b\cdot \tilde x +b^2)^\Delta}\,\psi_1\left(\frac{\tilde x^i+b^i }{\tilde x^2+2 b\cdot \tilde x +b^2}, R(b+\tilde x)^i_{\,\,j} R(x)^j_{\,\,k} z^k\right)
\end{align}
The R.H.S of (\ref{finiteK1}), apart from the factor $\frac{1}{(x^2)^\Delta}$, admits a  Taylor expansion in $\tilde x^i$ as $x\to \infty$ 
\begin{align}\label{asympd}
\psi(x, z)\stackrel{x\to\infty}{\approx}\frac{1}{(x^2)^{\Delta}}\sum_{n=0}^\infty C_{ns} \left( \tilde x^i,R(x)^i_{\,\, j} z^j\right)
\end{align}
where $C_{ns}( u, v)$ is a homogeneous polynomial of degree in $u^i$ and degree $s$ in $v^i$. This equation serves as the universal asymptotic boundary condition for all wavefunctions in $\CC_K$. Altogether, let $\CF_{\Delta, s}$ be the space of smooth functions on $\mathbb R^d$ which are also polynomials of degree $s$ in the null vector $z^i$ and satisfy the asymptotic condition (\ref{asympd}). It furnishes an $\SO(1, d+1)$ representation whose infinitesimal version is given by (\ref{algebrarepd}). 

We want to mention that it is straightforward to construct the more general representation $\CF_{\Delta, \bm s}$ for an arbitrary $\SO(d)$ highest weight vector $\bm s$. It suffices to choose the spin-$\bm s$ action of $M_{ij}$ on the primary state and then the wavefunction picture follows accordingly. In the mathematical literature, the representations $\CF_{\Delta, \bm s}$ are constructed using the induced representation method, which is reviewed in the appendix \ref{reviewinduce} where we also show explicitly the equivalence of the two constructions for $\CF_{\Delta, s}$. 

The  representations  $\CF_{\Delta, \bm s}$  are important for the following reasons  \cite{Dobrev:1977qv}: 
 \begin{itemize}
 \item Almost all $\CF_{\Delta, \bm s}$ are irreducible apart from some discrete values of $\Delta$. We give an elementary proof of this claim for $\CF_\Delta$ in the appendix \ref{irrCF}. A more general proof for $\CF_{\Delta, \bm s}$ can be found in \cite{10.3792/pja/1195523460}\footnote{The method in this paper is essentially equivalent to what we does in the appendix \ref{irrCF} except that the author used an abstract basis (instead of the spherical harmonics we use) for each $\SO(d+1)$ content of $\CF_{\Delta, \bm s}$, that works universally for any $\bm s$. In addition, this method heavily relies on the fact, which we will prove for $\CF_{\Delta, s}$ in the following section, that each $\SO(d+1)$ content contained in $\CF_{\Delta, \bm s}$ has multiplicity 1.} and a full list of irreducible representations  is given in a subsequent paper by the same author \cite{10.3792/pja/1195523378}. Indeed, there are only four types of reducible representations of this form when $\bm s=s$ is a single-row representation
 \begin{align}
 \CF_{1-t, s}, \,\,\,\,\ \CF_{1-s, t},\,\,\,\,\,  \CF_{d+t-1, s}, \,\,\,\,\ \CF_{d+s-1, t}
 \end{align}
 where $s=1, 2,3, \cdots $ and $t=0, 1,2,\cdots s-1$. We will see why these representations are reducible in  section \ref{Exc}.
Throughout this note, for a given spin $s$, a point in the $\Delta$-plane is called {\it generic} if $\CF_{\Delta, s}$ is irreducible and called {\it exceptional} if  $\CF_{\Delta, s}$ is reducible.
 \item Any irreducible representation of $\SO(1, d+1)$ is equivalent to some subrepresentations of $\CF_{\Delta,\bm s}$ (including $\CF_{\Delta,\bm s}$ itself when it is irreducible).
 \end{itemize}

\subsection{$\SO(d+1)$ content of $\CF_{\Delta, s}$}\label{content}
Before imposing the unitarity condition on $\CF_{\Delta, s}$, let's first focus on the $\SO(d+1)$ subgroup and see how $\CF_{\Delta, s}$ decomposes into irreducible representations of $\SO(d+1)$. In the $d=1$ case, it suffices to compute the eigenspectrum of $L_0$, the generator of $\SO(2)$.
For $d\ge 2$, a standard and elegant approach to this problem involves the Iwasawa decomposition, i.e. $\SO(1, d+1)=\tK\tN\tA$ \cite{Dobrev:1977qv},  and the induced representation construction which is reviewed in the appendix \ref{reviewinduce}. Using this approach, one can show that $\CF_{\Delta, s}$, when considered as an $\SO(d+1)$ representation, is equivalent to the induced representation $\text{ind}^{\SO(d+1)}_{\SO(d)}\mY_{s}$, whose $\SO(d+1)$ content follows from Frobenius reciprocity theorem \footnote{Roughly speaking, the Frobenius reciprocity theorem states that given a unitary irreducible representation $\sigma$ of $H$ and a unitary irreducible representation $\rho$ of $G$, then $\rho$ is contained in the induced representation $\text{ind}_H^G\sigma$ as many times as $\sigma$ contained in the restriction representation $\rho|_H$.} \cite{10.2307/j.ctt1bpm9sn}. Here we will present a more elementary argument based on our CFT-type construction of $\CF_{\Delta, s}$. The $\SO(d+1)$ generators are $L_{ij}=M_{ij}$ and $L_{d+1, i}=\frac{1}{2}(P_i+K_i)$, which act on $\psi(x, z)\in \CF_{\Delta, s}$ as 
\begin{align}
&L_{ij}\psi(x, z)=(x_i\partial_j-x_j\partial_i+z_i\partial_{z^j}-z_j\partial_{z^i})\psi(x, z)\nonumber\\
&L_{d+1, i}\psi(x, z)=\left(-\frac{1-x^2}{2} \partial_i-x_i(x\cdot\partial_x+\Delta)+(x\cdot z\, \partial_{z^i}-z_i \,x\cdot \partial_{z})\right)\psi(x, z)
\end{align}
One important observation is that the $\Delta$-dependence in the action of $L_{d+1, i}$ disappears if we perform a rescaling for the wavefunction $ \hat \psi(x, z)\equiv \left(\frac{1+x^2}{2}\right)^{\Delta-s} \psi(x, z)$. Another advantage of this scaling is that $\hat\psi(x)_{i_1\cdots i_s}$ has the same large $x$ asymptotic behavior as a spin-$s$ tensor field on $S^d$  in stereographic coordinates. The action of $\SO(d+1)$ on $\hat\psi$ is:
\begin{align}\label{SOd+1}
&L_{ij}\hat \psi(x, z)=(x_i\partial_j-x_j\partial_i+z_i\partial_{z^j}-z_j\partial_{z^i})\hat \psi(x, z)\nonumber\\
&L_{d+1, i}\hat\psi(x, z)=\left(-\frac{1-x^2}{2} \partial_i-x_i \, (x\cdot\partial_x+s)+(x\cdot z\, \partial_{z^i}-z_i \,x\cdot \partial_{z})\right)\hat\psi(x, z)
\end{align}
We claim that $\hat \psi(x, z)$ defines a spin-$s$ tensor on $S^d$. It suffices to show that any $L\in\so(d+1)$ acts on $\hat \psi(x, z)$ as a Lie derivative along the Killing vector generated by $L$ on $S^d$. In embedding space coordinates, $S^d$ is  a hypersurface
\begin{align}
Y_1^2+Y_2^2+\cdots + Y^2_{d+1}=1
\end{align}
with the space of Killing vectors spanned by $V_{ab}=Y_a\partial_{Y^b}-Y_b \partial_{Y^a}, \,\, 1\le a, b\le d+1$, where the vector field $V_{ab}$ is generated by the action of $L_{ab}$. The stereographic coordinates of $S^d$ correspond to choosing the following embedding
\begin{align}
Y^a=\left(\frac{2x^i}{x^2+1}, \frac{x^2-1}{x^2+1}\right)
\end{align}
which yields a conformally flat metric $g_{ij}=\left(\frac{2}{1+x^2}\right)^2\delta_{ij}$. Given an arbitrary vector field $V=V^{i}\partial_i$ and an arbitrary tensor $\varphi_{i_1\cdots i_s}$, the Lie derivative $\CL_V$ is defined as 
\begin{align}
\CL_V \varphi_{i_1\cdots i_s}=V^k\partial_k \varphi_{i_1\cdots i_s}+\sum_{j=1}^s \partial_{i_j}V^k \varphi_{i_1\cdots i_{j-1} k i_{j+1}\cdots i_s}
\end{align}
When $\varphi_{i_1\cdots i_s}$ is an symmetric and traceless tensor, we can use the index-free formalism and replace it by a polynomial $\varphi(x, z)\equiv \frac{1}{s!}\varphi_{i_1\cdots i_s} z^{i_1}\cdots z^{i_s}$, where $z^i$ is null. Then the Lie derivative $\CL_V$ acting on $\varphi(x, z)$ becomes
\begin{align}\label{Lieder}
\CL_V \varphi(x, z)=V^k\partial_k \varphi(x, z)+(z\cdot\partial_x V^k)\CD_{z^k} \varphi(x, z)
\end{align}
where $\CD_{z^i}$ is given by eq. (\ref{intder}). To compute the Lie derivatives $\CL_{V_{ab}}$, we need  to write out the Killing vectors $V_{ab}$ in terms of stereographic coordinates by using
\begin{align}
\left(V_{ab}\right)^k=g^{k\ell}(Y_a\partial_\ell Y_b-Y_b\partial_{\ell} Y_a)
\end{align} 
We have $\left(V_{ij}\right)^k=x_i \delta^k_j-x_j \delta^k_i$ and $\left(V_{d+1, i}\right)^k=-\frac{1-x^2}{2}\delta^k_i-x^k x_i$ respectively. Plugging these explicit forms of $V_{ab}$ into eq. (\ref{Lieder}) yields
\begin{align}\label{CLV}
&\CL_{V_{ij}}\varphi(x, z)=\left(x_i\partial_j-x_j\partial_i+z_i \partial_{z^j}-z_j \partial_{z^i}\right) \varphi(x, z)\nonumber\\
&\CL_{V_{d+1, i}}\varphi(x, z)=\left(-\frac{1-x^2}{2} \partial_i-x_i \, (x\cdot\partial_x+s)+(x\cdot z\, \partial_{z^i}-z_i \,x\cdot \partial_{z})\right)\varphi(x, z)
\end{align}
where the number $s$ in the second line comes from $z\cdot\partial_z$ acting on $\varphi(x, z)$. The agreement between eq. (\ref{SOd+1}) and eq. (\ref{CLV}) implies that the vector space $\CF_{\Delta, s}$ is isomorphic to the space of spin-$s$ tensors on $S^d$. From the latter, we can easily read off the $\SO(d+1)$ content of  $\CF_{\Delta, s}$. For example, when $s=1$, we can decompose a spin-1 tensor into a scalar function and a transverse spin-1 tensor, i.e. $\hat\psi_i(x)=\partial_i \chi(x)+\eta_i(x), \,\, \nabla_i \eta^i=0$, where $\chi$ admits an expansion in terms of scalar spherical harmonics except the constant one which is a zero mode of $\partial_i$ and $\eta_i$ admits an expansion in terms of transverse vector spherical harmonics. It is well known that the scalar harmonics on $S^d$ correspond to all single-row representations of $\SO(d+1)$, i.e. $\mY_n$,  while the transverse vector harmonics correspond to two-row representations with 1 box in the second row, i.e. $\mY_{n,1}$ \cite{doi:10.1063/1.526749}. Altogether, $\CF_{\Delta, 1}$ as an $\SO(d+1)$ representation contains the following irreducible components
\begin{align}
\left.\CF_{\Delta, 1}\right\vert_{\SO(d+1)}=\left(\bigoplus_{n\ge 1}\mY_n\right)\oplus\left(\bigoplus_{k\ge 1}\mY_{k, 1}\right)=\bigoplus_{n\ge 1}\bigoplus_{0\le m \le 1}\mY_{n,m}
\end{align}
For higher $s$, we decompose $\hat\psi_{i_1\cdots i_s}$ into a spin-$(s\!-\!1)$ tensor $\chi_{i_1\cdots i_{s-1}}$ and a transverse spin-s tensor $\eta_{i_1\cdots i_{s}}$, i.e.  
\begin{align}
\hat\psi_{i_1\cdots i_s}=\nabla_{(i_1}\chi_{i_2\cdots i_s)}-\text{trace}+\eta_{i_1\cdots i_s}, \,\,\,\,\, \nabla^{i_1}\eta_{i_1\cdots i_s}=0
\end{align}
The latter admits an expansion in terms of transverse spin-$s$ tensor harmonics which group theoretically correspond to all two-row representations of $\SO(d+1)$ with $s$ boxes in the second row. In $\chi_{i_1\cdots i_{s-1}}$, we should exclude the modes such that $\nabla_{(i_1}\chi_{i_2\cdots i_s)}$ is pure trace. These modes are spin-$(s\!-\!1)$ conformal Killing tensors on $S^d$ and they furnish the representation $\mY_{s-1, s-1}$ of $\SO(1, d+1)$ \cite{10.2307/3597366}, which becomes $\bigoplus_{0\le n\le s-1}\mY_{s-1,n}$ while restricted to the $\SO(d+1)$ subgroup. By  induction on $s$, we can immediately conclude that 
\begin{align}\label{decomposeCF}
\boxed{\left.\CF_{\Delta, s}\right\vert_{\SO(d+1)}=\bigoplus_{n\ge s}\bigoplus_{0\le m \le s}\mY_{n,m}}
\end{align}

\subsection{Shadow transformations}\label{shadownew}
In the $\SO(1, 2)$ case, we have shown that there exists  a linear intertwining operator $\CS_\Delta$, called shadow transformation, that maps $\CF_\Delta$ to $\CF_{\bar\Delta}$ and in particular, when $\Delta\notin\mathbb Z$ it is an isomorphism. In this subsection, we will show that a similar operator also exists for higher $d$. Assume that $\CS^{\Delta', s'}_{\Delta, s}: \CF_{\Delta, s}\to \CF_{\Delta', s'}$ is an intertwining operator defined by a kernel function $S^{\Delta', s'}_{\Delta,s}(x_1, x_2)_{i_1\cdots i_{s'}, j_1\cdots j_s}$
\begin{align}
\CS^{\Delta', s'}_{\Delta, s}: \psi_{i_1\cdots i_s}(x_1)\in\CF_{\Delta, s}\longmapsto \int\, d^dx_2 \,S^{\Delta', s'}_{\Delta,s}(x_1, x_2)_{i_1\cdots i_{s'}, j_1\cdots j_s} \psi_{j_1\cdots j_s}(x_2)
\end{align}
The requirement $\CS^{\Delta', s'}_{\Delta, s}[\psi_{i_1\cdots i_s}]\in\CF_{\Delta', s'}$ induces a set of differential equations for the kernel function (Suppress the spin indices of $S^{\Delta', s'}_{\Delta,s}(x, y)_{i_1\cdots i_{s'}, j_1\cdots j_s}$ for the simplicity of notation. It should be clear that $\CM^{(s')}_{k\ell}$ acts on the $i$ indices while $\CM^{(s)}_{k\ell}$ acts on the $j$ indices.):
\begin{gather}
(\partial_{x^i}+\partial_{y^i})S^{\Delta', s'}_{\Delta,s}(x, y)=0\\
(x\cdot\partial_{x}+y\cdot\partial_{y}+\Delta'+\bar\Delta)S^{\Delta', s'}_{\Delta,s}(x, y) =0\\
(x_\ell\partial_{x^k}-x_k\partial_{x^\ell}+y_\ell\partial_{y^k}-y_k\partial_{y^\ell}+\CM^{(s)}_{k\ell}+\CM^{(s')}_{k\ell})S^{\Delta', s'}_{\Delta,s}(x, y)=0\\
(x^2\partial_{x^k}-2x_k (x\cdot\partial_x+\Delta')+y^2\partial_{y^k}-2y_k (y\cdot\partial_y+\bar\Delta)+2x^\ell \CM^{(s')}_{k\ell}+2y^\ell \CM^{(s)}_{k\ell})S^{\Delta', s'}_{\Delta,s}(x, y) =0
\end{gather}
These differential equations are exactly the conformal Ward identities for a two-point function of two primary operators $\CO_{i_1\cdots i_s}$ and $\CO'_{i_1\cdots i_s'}$ with scaling dimension $\bar\Delta$ and $\Delta'$ respectively. It is well known that such a two-point function is vanishing unless the two operators have the same spin and scaling dimension. Therefore $\CS^{\Delta', s'}_{\Delta, s}$ exists only when $s=s', \Delta'=\bar\Delta$ and  in this case the corresponding kernel function, denoted by $S_{\Delta, s}(x_1-x_2)_{i_1\cdots i_s, j_1\cdots j_s}$, becomes the conformal two-point function $\langle \CO_{i_1\cdots i_s}(x_1)\CO_{j_1\cdots j_s}(x_2)\rangle$ which takes the following simple form in the index-free formalism
\begin{align}\label{conf2pt1}
S_{\Delta, s}(x_{12}; z, w)&\equiv S_{\Delta, s}(x_{12})_{i_1\cdots i_s, j_1\cdots j_s} \,z^{i_1}\cdots z^{i_s} w^{j_1}\cdots w^{j_s}\nonumber\\
&=N_{\Delta, s}\frac{(-z\cdot R(x_{12})\cdot w)^s}{\left(\frac{1}{2}x^2_{12}\right)^{\bar\Delta}}, \,\,\,\,\, x^i_{12}\equiv x^i_1-x_2^i
\end{align}
where $N_{\Delta, s}$ is a normalization constant, $z^i$ and $w^i$ are null vectors and $R(x)_{ij}=\frac{2x_i x_j}{x^2}-\delta_{ij}$ is defined in the eq. (\ref{xbR}). 

Given shadow transformations $\CS_{\Delta, s}$ and $\CS_{\bar\Delta, s}$, the composition $\CS_{\bar\Delta, s}\circ\CS_{\Delta, s}$ is an intertwining map that maps $\CF_{\Delta, s}$ to itself. When $\CF_{\Delta, s}$ is irreducible\footnote{At those exceptional points where $\CF_{\Delta, s}$ is reducible, the corresponding shadow transformations are not invertible. We will comment more on these cases later.}, which is true for almost all $\Delta$, the composition should be proportional to identity map due to Schur's lemma. We want to (partially) fix the normalization by requiring $\CS_{\bar\Delta, s}\circ\CS_{\Delta, s}=1_{\CF_{\Delta, s}}$, i.e. 
\begin{align}
\int\, d^d x_2 S_{\bar\Delta, s}(x_{12})_{i_1\cdots i_s, j_1\cdots j_s}S_{\Delta, s}(x_{23})_{ j_1\cdots j_s,k_1\cdots k_s}=\delta^d(x_{13})\delta_{i_1\cdots i_s,k_1\cdots k_s}
\end{align}
or equivalent in momentum space 
\begin{align}\label{SSbar}
S_{\bar\Delta, s}(p)_{i_1\cdots i_s, j_1\cdots j_s}S_{\Delta, s}(p)_{ j_1\cdots j_s,k_1\cdots k_s}=\delta_{i_1\cdots i_s,k_1\cdots k_s}
\end{align}
The Fourier transformation  $S_{\Delta, s}(p; z,  w)\equiv \int d^d x \, e^{ip\cdot x}\,S_{\Delta, s}(x; z,  w)$ can be performed by using the binomial expansion for the numerator of $S_{\Delta, s}(x; z,  w)$ and then applying the following formula to each term
\begin{align}\label{Fourier22}
\int\, d^d x\, \frac{1}{|x|^{2\bar\Delta}} e^{ip\cdot x}=2^{d-2\bar\Delta}\pi^{\frac{d}{2}}\, \frac{\Gamma(\frac{d}{2}-\bar\Delta)}{\Gamma(\bar\Delta)} p^{2\bar\Delta-d}
\end{align}
The final expression of $S_{\Delta, s}(p;z,w)$admits a harmonic expansion with respect to $\SO(d-1)$, which is the little group of a fixed  momentum $p$. For a detailed derivation of such an expansion, we refer to the book \cite{Dobrev:1977qv}.
Here we present the result 
\small
\begin{align}\label{spinskernel}
S_{\Delta, s}(p; z, w)= \pi^{\frac{d}{2}}\, N_{\Delta, s}\, \frac{2^\Delta\Gamma(\frac{d}{2}-\bar\Delta)\,p^{2\bar\Delta-d}}{(\bar\Delta+s-1)\Gamma(\bar\Delta-1)}\sum_{\ell =0}^s\kappa_{s\ell}(\Delta)\Pi^{s \ell}(\hat p; z,w), \,\,\,\,\, \kappa_{s\ell}(\Delta)=\frac{(\Delta+\ell-1)_{s-\ell}}{(\bar\Delta+\ell -1)_{s-\ell }}
\end{align}
\normalsize
where $\Pi^{s\ell}(\hat p; z, w)\equiv \Pi^{s\ell}(\hat p)_{i_1\cdots i_s, j_1\cdots j_s}z^{i_1}\cdots z^{i_s}w^{j_1}\cdots w^{j_s}, 0\le \ell\le s$,  are $(s+1)$ {\it projection operators} that only depend on the unit vector $\hat p$ in the direction of $p^i$ and satisfy the following properties
\begin{align}\label{Piproperty}
&\textbf{Orthogonality}: \Pi^{s\ell}(\hat p)_{i_1\cdots i_s, j_1\cdots j_s}\Pi^{s\ell'}(\hat p)_{j_1\cdots j_s, k_1\cdots k_s}=\delta_{\ell \ell'}\Pi^{s\ell}(\hat p)_{i_1\cdots i_s, k_1\cdots k_s}\nonumber\\
&\textbf{Completeness}: \sum_{\ell=0}^s  \Pi^{s\ell}(\hat p; z, w)= (z\cdot w)^s\nonumber\\
&\textbf{Helicity}\,\, \ell: p^{i_\ell} p^{i_{\ell+1}}\cdots p^{i_s}\Pi^{s\ell}(\hat p)_{i_1\cdots i_s, j_1\cdots j_s}=p^{j_\ell} p^{j_{\ell+1}}\cdots p^{j_s}\Pi^{s\ell}(\hat p)_{i_1\cdots i_s, j_1\cdots j_s}=0, \,\, \ell\ge 1
\end{align}
For example, when $s=1$, we have
\begin{align}
\Pi^{10}(\hat p)_{i, j}=\hat p_i \hat p_j, \,\,\,\,\, \Pi^{11}(\hat p)_{i, j}=\delta_{ij}-\hat p_i \hat p_j
\end{align}
\begin{remark} The set of $\Pi^{s\ell}$ can be thought as the manifestation of the branching rule from $\SO(d)$ (the full rotation group of $p$) to $\SO(d-1)$ (the little group of $p$). More explicitly, the spin-$s$ representation of $\SO(d)$ can be decomposed into the direct sum of the spin-$\ell$ representations of $\SO(d-1)$ with $\ell$ ranging from $0$ to $s$. Each $\Pi^{s\ell}$ projects the spin-$s$ $\SO(d)$ representation to its  $\SO(d-1)$ spin-$\ell$ summand. For example, in the $s=1$ case, $\Pi^{10}$ projects a vector to its component along the direction of $p$ which is clearly invariant under the little group and $\Pi^{11}$ yields the transverse part which carries the spin-1 representation of the little group.
\end{remark}

Using the orthogonality and completeness of $\{\Pi^{s\ell}\}$ and noticing $\kappa_{s\ell}(\Delta)\kappa_{s\ell}(\bar\Delta)=1$, we find that the normalization condition (\ref{SSbar}) is equivalent to 
\begin{align}
\frac{(2\pi)^d\, \Gamma(\frac{d}{2}-\bar\Delta)\,\Gamma(\frac{d}{2}-\Delta)\,N_{\Delta}\, N_{\bar\Delta}}{(\bar\Delta+s-1)(\Delta+s-1)\Gamma(\bar\Delta-1)\Gamma(\Delta-1)}=1
\end{align}
Apparently, this equation cannot fix the normalization constant $N_{\Delta, s}$ completely. Two convenient solutions which we will use are 
\begin{align}
N^+_{\Delta, s}=\frac{(\bar\Delta+s-1)\Gamma(\bar\Delta-1)}{2^\Delta\pi^{\frac{d}{2}}\, \Gamma(\frac{d}{2}-\bar\Delta)}, \,\,\,\, N^-_{\Delta, s}=\frac{(\Delta+s-1)\Gamma(\Delta-1)}{2^\Delta\pi^{\frac{d}{2}}\, \Gamma(\frac{d}{2}-\bar\Delta)}
\end{align}
and the corresponding shadow transformations will be denoted by $\CS^\pm_{\Delta, s}$. When $d=1$ and $s=0$, $N^+_{\Delta,s}$ is reduced to $N_\Delta=\frac{\Gamma(\bar\Delta)}{2^\Delta\sqrt{\pi}\Gamma(\Delta-\frac{1}{2})}$, the normalization we have chosen in eq. (\ref{shadownormalize}) for shadow transformations of $\SO(1,2)$.

Let's write out the  kernel for $\CS^\pm_{\Delta, s}$ in momentum space explicitly:
\begin{align}\label{Spm}
&S^+_{\Delta, s}(p; z, w)= p^{2\bar\Delta-d}\sum_{\ell =0}^s\frac{(\Delta+\ell-1)_{s-\ell}}{(\bar\Delta+\ell -1)_{s-\ell }}\Pi^{s \ell}(\hat p; z,w)\nonumber\\
&S^-_{\Delta, s}(p; z, w)= p^{2\bar\Delta-d}\frac{\Gamma(\Delta-1)}{\Gamma(\bar\Delta-1)}\sum_{\ell =0}^s\frac{(\Delta+\ell-1)_{s-\ell+1}}{(\bar\Delta+\ell -1)_{s-\ell+1 }}\Pi^{s \ell}(\hat p; z,w)
\end{align}
For a generic $\Delta$, the two choices are equivalent since they only differ by a normalization factor. So we will stick to $S^+_{\Delta, s}$ in this case. However, at exceptional points, $S^\pm_{\Delta, s}$ are completely different and the corresponding properties are summarized in the table (\ref{table:1}), where $s=1,2,\cdots$ and $t=0,1,\cdots, s-1$.

\begin{table}[h!]
\centering
\begin{tabular}{ |c|c|c|} 
\hline
 & $S_{\Delta', s'}^+$ & $S_{\Delta', s'}^-$ \\
\hline
$(\Delta', s')=(1-t, s)$ & only contains $\Pi^{s\ell}$ with $t\!+\!1\le\ell\le s$ & ill-defined \\
\hline
$(\bar\Delta', s')=(1-t, s)$ &  ill-defined & only contains $\Pi^{s\ell}$ with $0\le\ell\le t$ \\
\hline
$(\Delta', s')=(1-s, t)$ &  contains all $\Pi^{t \ell}, 0\le\ell\le t$ & ill-defined \\
\hline
$(\bar\Delta', s')=(1-s, t)$ &   contains all $\Pi^{t\ell}, 0\le\ell\le t$ & $\delta$-function in momentum space \\
\hline
\end{tabular}
\caption{Properties of $S^\pm_{\Delta, s}$ at exceptional points.}
\label{table:1}
\end{table}

\noindent{}Notice that $S^-_{d+s-1, t}$ is special because it is a polynomial in $x^i$ of degree $2s-2$
\begin{align}
S^-_{d+s-1, t}(x;z,w)=\frac{(d+s+t-2)\Gamma(d+s-2)}{2^{d+2(s-1)}\pi^{\frac{d}{2}}\Gamma(\frac{d}{2}+s-1)}\left(x^2\right)^{s-t-1}\left(x^2\,  z\cdot w-2\, x\cdot z\, x\cdot w\right)^{t}
\end{align}
Therefore, its Fourier transformation is a $\delta$-function with $(2s-2)$ derivatives in momentum space. 

\subsection{(Nonexceptional) unitary scalar representations}
Recall that a wavefunction  $\psi\in\CF_\Delta$ defines a ket $|\psi\rangle=\int d^d x\, \psi(x)|x\rangle$. So an inner product on $\CF_\Delta$ is fixed by defining a pairing $\langle x|y\rangle$:
\begin{align}\label{psi1psi2}
(\psi_1, \psi_2)\equiv \int d^d x\int d^d y\, \psi_1(x)^* \CK_\Delta(x, y) \psi_2(y), \,\,\,\,\, \CK_\Delta(x, y)=\langle x|y\rangle
\end{align} 
Imposing the dS reality condition $L_{AB}^\dagger=-L_{AB}$ as in section \ref{cou}, we find that the function $\CK_\Delta(x, y)$ exists only in the following two cases
\begin{align}
&\text{\rom{1}}:\Delta=\frac{d}{2}+i\mu, \,\, \mu\in\mathbb R,  \,\,\,\,\, \CK_\Delta(x, y)=\, \delta^d(x-y) \nonumber\\
&\text{\rom{2}}: \Delta\in\mathbb R,  \,\,\,\,\, \,\,\,\,\, \,\,\,\,\, \,\,\,\,\, \,\,\,\,\, \,\,\,\,\, \,\,\,\,\,\,\CK_\Delta(x,y)=\frac{ 2^{\bar\Delta}c_\Delta}{|x-y|^{2\bar\Delta}}
\end{align}
where $c_\Delta$ is a normalization constant. 
In case (\rom{1}),  (\ref{psi1psi2}) is the standard $L^2$ inner product on $\mathbb R^d$. It is  positive definite and normalizable since $\psi(x)\in\CF_{\Delta}$ falls off as $\frac{1}{|x|^{2\Delta}}$ for large $x$. Therefore, the function space $\CF_{\frac{d}{2}+i\mu}$ equipped with the standard $L^2$ inner product furnishes a unitary irreducible representation of $\SO(1, d+1)$, which is known as the (unitary) {\it scalar principal series representation} of scaling dimension $\Delta=\frac{d}{2}+i\mu$. In the case (\rom{2}), we choose $c_\Delta=N^+_\Delta=\frac{\Gamma(\bar\Delta)}{2^\Delta\pi^{\frac{d}{2}}\Gamma(\frac{d}{2}-\bar\Delta)}$, so that $\boxed{\CK_\Delta(x,y)=S^+_\Delta(x,y)}$, i.e. the kernel of shadow transformation $\CS^+_\Delta: \CF_\Delta\to \CF_{\bar\Delta}$. Using the Fourier transformation of $S^+_\Delta(x)$, c.f. eq.(\ref{Spm}), we rewrite the inner product (\ref{psi1psi2}) in momentum space as 
\begin{align}\label{1212}
(\psi_1, \psi_2)=\frac{1}{(2\pi)^d}\int\, d^d p\, p^{2\bar\Delta-d}\psi_1(p)^*\psi_2(p)
\end{align}
where $\psi(p)\equiv \int\, d^d x \, e^{ip\cdot x} \psi(x)$ is the Fourier transformation of $\psi(x)$. This inner product is positive definite as long as it is convergent. Due to the smoothness of $\psi(x)$, its Fourier transformation $\psi(p)$ decays exponentially for large $p$. So the $p$-integral in (\ref{1212}) is convergent around $p\to \infty$. For small $p$, as we have argued in the $\SO(1, 2)$ case, $\psi(p)$ has two types of leading fall-offs: $p^0$ and $p^{2\Delta-d}$. Then the requirement of convergence near $p\to 0$ yields $\boxed{0<\Delta<d}$. Therefore, the representations $\CF_\Delta$ with $0<\Delta<d$ are unitary, known as the (unitary) {\it scalar complementary series}.

Before moving to spinning representations, we want to present a different way to expand the kernel $\CK_\Delta$ based on our discussion in the section \ref{content}. It yields the same constraint on the scaling dimension.  Using the Weyl transformation $\psi(x)=\left(\frac{2}{1+x^2}\right)^\Delta\hat\psi(x)$, where $\hat\psi(x)$ is a function on $S^d$, we rewrite the inner product (\ref{psi1psi2}) as 
\begin{align}\label{12122}
(\psi_1,\psi_2)=N^+_\Delta\int\, \frac{d^dx}{\left(\frac{1+x^2}{2}\right)^d}\frac{d^dy}{\left(\frac{1+y^2}{2}\right)^d}\left(\frac{(1+x^2)(1+y^2)}{2(x-y)^2}\right)^{\bar\Delta}\hat\psi(x)^*\hat\psi(y)
\end{align}
Switch to spherical coordinates $\Omega^i=(\sin\theta\omega^i, \cos\theta), \omega^i\in S^{d-1}$ which is related the stereographic coordinates $x^i$ by $x^i=\cot\frac{\theta}{2}\omega^i$ and eq. (\ref{12122}) becomes an integral on $S^d\times S^d$
\begin{align}\label{12123}
(\psi_1, \psi_2)= \int d^d\Omega_1 \, d^d \Omega_2 \, \hat\CK(\Omega_1, \Omega_2)\hat\psi(\Omega_1)^*\hat\psi_2(\Omega_2), \,\,\,\,\, \hat \CK_\Delta(\Omega_1, \Omega_2)=\frac{N^+_\Delta}{(1-\Omega_1\cdot\Omega_2)^{\bar\Delta}}
\end{align}
Now we need to perform a harmonic expansion for the new kernel $\hat\CK_\Delta$, along the lines of (\ref{harmexp2}). First, write it as an integral using Schwinger's trick
\begin{align}
\hat\CK_\Delta(\Omega_1, \Omega_2)=\frac{N^+_\Delta}{\Gamma(\bar\Delta)}\int_0^\infty \frac{ds}{s}\, s^{\bar\Delta} e^{-s+s\Omega_1\cdot \Omega_2}
\end{align}
Then plug it in the expansion of plane waves in Gegenbauer polynomials \cite{doi:10.2991/jnmp.2004.11.s1.22}
\begin{align}
e^{s\Omega_1\cdot \Omega_2}=\Gamma(\alpha)\left(\frac{s}{2}\right)^{-\alpha}\sum_{\ell\ge 0}(\alpha+\ell) I_{\alpha+\ell}(s) C^\alpha_{\ell}(\Omega_1\cdot\Omega_2), \,\,\,\,\, \alpha=\frac{d-1}{2}
\end{align}
Using the addition theorem, Gegenbauer polynomials can  thus be expanded in spherical harmonics on $S^d$ \cite{doi:10.1063/1.526621}
\begin{align}
C^\alpha_{\ell}(\Omega_1\cdot\Omega_2)=\frac{2\pi^{\frac{d+1}{2}}}{\Gamma(\alpha)(\ell+\alpha)}\sum_{\bm m} Y_{\ell \bm m}(\Omega_1)Y_{\ell\bm m}(\Omega_2)^* 
\end{align}
Performing the $s$-integral, we obtain the desired expansion of the kernel $\hat\CK_\Delta$ in spherical harmonics
\begin{align}\label{hatK1}
\hat\CK_\Delta(\Omega_1, \Omega_2)=\sum_{\ell\ge 0}\sum_{\bm m}\frac{\Gamma(\ell+\bar\Delta)}{\Gamma(\ell+\Delta)}Y_{\ell \bm m}(\Omega_1)Y_{\ell\bm m}(\Omega_2)^* 
\end{align}
where we have plugged in the explicit form of $N^+_\Delta$.
For $\hat\CK_\Delta$ to be positive definite, the sign of $\frac{\Gamma(\ell+\bar\Delta)}{\Gamma(\ell+\Delta)}$ should {\it not} oscillate with $\ell$, which requires $\frac{\ell+\bar\Delta}{\ell+\Delta}>0$ for any $\ell\ge 0$. The most stringent constraint comes from $\ell=0$ and it is $\boxed{0<\Delta<d}$, in agreement with what we have found in momentum space. With this constraint satisfied, $\frac{\Gamma(\ell+\bar\Delta)}{\Gamma(\ell+\Delta)}$ stays positive for all $\ell$ and hence $\hat\CK_\Delta$ is positive definite.

\begin{remark}
When $\Delta$ is a nonpositive integer, $\frac{\Gamma(\ell+\bar\Delta)}{\Gamma(\ell+\Delta)}$ vanishes for $0\le\ell\le -\Delta$. Thus the spherical harmonics $Y_{\ell \bm m}$ with $0\le\ell\le -\Delta$ are null states. They furnish an irreducible representation of $\SO(1, d+1)$, which is checked explicitly in  appendix \ref{irrCF}. When $\bar\Delta$ is a nonpositive integer, we choose $c_\Delta=N^-_\Delta$. It amounts to replacing $\frac{\Gamma(\ell+\bar\Delta)}{\Gamma(\ell+\Delta)}$ by $\frac{(\bar\Delta)_\ell}{(\Delta)_\ell}$ in the expansion (\ref{hatK1}).
In this case, all $Y_{\ell \bm m}$ with $\ell\ge 1-\bar\Delta$ are null and carry an irreducible representation, which is again checked in the appendix \ref{irrCF}.
\end{remark}

\subsection{(Nonexceptional) unitary spinning representations}
For a spinning representation $\CF_{\Delta, s}$, inner product is defined by the kernel
\begin{align}
\CK_{\Delta,s}(x_1, x_2)_{i_1\cdots i_s, j_1\cdots j_s}\equiv\,  _{i_1\cdots i_s}\langle x_1|x_2\rangle_{j_1\cdots j_s}
\end{align}
In the index free formalism, we contract the $i$-indices with a null vector $z$ and contract the $j$-indices with a different null vector $w$
\begin{align}
\CK_{\Delta, s}(x_1, z; x_2, w)\equiv \,\CK_{\Delta,s}(x_1, x_2)_{i_1\cdots i_s, j_1\cdots j_s}\,z^{i_1}\cdots z^{i_s} w^{j_1}\cdots w^{j_s}
\end{align}With the de Sitter reality condition imposed, $\CK_{\Delta, s}(x_1, z; x_2, w)$ exists only for two cases (\rom{1}): $\Delta=\frac{d}{2}+i\mu$ and (\rom{2}): $\Delta\in\mathbb R$. In the first case, $\CK_{\Delta, s}$ is simply a $\delta$-function in both spacetime coordinate and spin indices, i.e.
\begin{align}
\CK_{\Delta, s}(x_1, z; x_2, w)=\delta^d(x_1, x_2)(z\cdot w)^s
\end{align}
and hence the inner product defined by this $\CK_{\Delta, s}$ becomes the $L^2$-inner product for spin-$s$ tensors on $\mathbb R^d$
\begin{align}
\boxed{(\psi, \varphi)_\CP\equiv \int d^d x\, \psi_{i_1\cdots i_s}^*(x)\, \varphi_{i_1\cdots i_s}(x)}
\end{align}
The normalizability of this inner product on $\CF_{\frac{d}{2}+i\mu, s}$ is guaranteed by the asymptotic behavior  (\ref{asympd}). Therefore, $\CF_{\frac{d}{2}+i\mu, s}$  is a unitary irreducible representation,  belonging to the so-called (unitary) {\it spinning principal series}. In the second case, the reality condition is actually equivalent to  conformal Ward identities for two-point functions. Therefore $\CK_{\Delta, s}$ is the same as $S_{\Delta, s}$ up to normalization, i.e. 
\begin{align}\label{conf2pt}
\CK_{\Delta, s}(x_1, z; x_2, w)=c_{\Delta, s}\frac{(-z\cdot R(x_{12})\cdot w)^s}{\left(\frac{1}{2}x^2_{12}\right)^{\bar\Delta}}
\end{align}
For a generic $\Delta$, we choose $c_{\Delta, s}=N^+_{\Delta,s}$ and hence  $\CK_{\Delta, s}=S^+_{\Delta, s}$, which in momentum space is (c.f. eq. (\ref{Spm}))
\begin{align}
S^+_{\Delta, s}(p; z, w)= p^{2\bar\Delta-d}\sum_{\ell =0}^s\kappa_{s\ell}(\Delta)\Pi^{s \ell}(\hat p; z,w), \,\,\,\,\, \kappa_{s\ell}(\Delta)=\frac{(\Delta+\ell-1)_{s-\ell}}{(\bar\Delta+\ell -1)_{s-\ell }}
\end{align} 
As in the scalar case, the inner product defined by $S^+_{\Delta, s}$ is normalizable when $0<\Delta<d$.  Additionally, for $S^+_{\Delta, s}$ to be positive, we need all $\kappa_{s\ell}(\Delta)$ to have a fixed sign (insert an overall minus sign if negative) which yields 
\begin{align}\label{kappasign}
\frac{\kappa_{s,\ell+1}(\Delta)}{\kappa_{s,\ell}(\Delta)}=\frac{\bar\Delta+\ell-1}{\Delta+\ell-1}>0
\end{align}
The eq. (\ref{kappasign}) holds for all $\ell$ if and only if $\boxed{1<\Delta<d-1}$. It is also straightforward to check that all $\kappa_{s\ell}(\Delta)$ are indeed positive for $\Delta$ in this range. Altogether, $\CF_{\Delta, s}$ with the following inner product
\begin{align}
\boxed{(\psi, \varphi)_\CC=\int\, d^d x_1\, d^d x_2\, \psi^*(x_1)_{i_1\cdots i_s} \, S^+_{\Delta, s}(x_{12})_{i_1\cdots i_s,j_1\cdots j_s} \varphi(x_2)_{j_1\cdots j_s}}
\end{align}
 is a unitary irreducible representation of $\SO(1, d+1)$ when $1<\Delta<d-1$. It belongs to the so-called (unitary) {\it spinning complementary series}.

\subsection{Exceptional series}\label{Exc}
We have studied the constraint of unitarity on $\CF_{\Delta, s}$ for generic $\Delta$ and managed to identify the (unitary) principal and complementary series. In this subsection, we will look into the four types of $\CF_{\Delta, s}$ at exceptional points, mainly following  \cite{Dobrev:1977qv}:
\begin{align}
\CF_{1-t, s}, \,\,\,\,\, \CF_{d+t-1, s}, \,\,\,\,\,\CF_{1-s, t}, \,\,\,\,\,\CF_{d+s-1, t}
\end{align}
The first important observation is that the four representations share the same $\SO(1, d+1)$ quadratic Casimir
\begin{align}\label{cashw}
\CC_2=-(s-1)(s+d-1)-t(t+d-2)
\end{align}
which is also the Casimir associated to the  finite dimensional representation $\mY_{s-1, t}$ of $\SO(1, d+1)$. This  observation suggests that these representations are related by a chain of intertwining maps. For representations in the two pairs $(\CF_{1-t, s},\CF_{d+t-1, s})$ and $(\CF_{1-s, t},\CF_{d+s-1, t})$, this is certainly true due to shadow transformations, which have been explored in the subsection \ref{shadownew}. To find intertwining maps relating the two pairs, let's revisit the shadow transformations $\CS^+_{1-t, s}$. At the end of  section \ref{shadownew} (see  table (\ref{table:1}) in particular), we find that $S^+_{1-t, s}(p)$ only contains the projection operators $\Pi^{s\ell}$ with $\ell\ge t+1$. Due to the third property in  eq. (\ref{Piproperty}), it means
\begin{align}\label{S^++}
p^{i_1}\cdots p^{i_{s-t}}S^+_{1-t, s}(p)_{i_1\cdots i_s, j_1\cdots j_s}=0
\end{align}
So the shadow transformation $\CS^+_{1-t, s}$  has a nontrivial kernel which consists functions of the form 
\begin{align}
p_{(i_1}\cdots p_{i_{s-t}} f_{i_{s-t+1}\cdots i_s)}-\text{trace}
\end{align}
 in momentum space. Switch back to position space and the kernel can be alternatively expressed as $\{(z\cdot\partial_x)^{s-t} f(x, z)\}$ in the index-free formalism. Requiring that $(z\cdot\partial_x)^{s-t} f(x, z)$ transforms under the $\CF_{1-t,s}$, we find $f(x, z)$ is actually an element in $\CF_{1-s,t}$. Therefore, $(z\cdot\partial_x)^{s-t}$ is an intertwining operator mapping $\CF_{1-s, t}$ to $\CF_{1-t, s}$. 
 Similarly, the eq. (\ref{S^++}) also implies that the image of $\CS^+_{1-t, s}$ is annihilated by $(p\cdot\CD_z)^{s-t}$ or equivalently in position space $(\partial_x\cdot\CD_z)^{s-t}$. Again, it is straightforward to check that $\boxed{\CD_{s,t}\equiv (\partial_x\cdot\CD_z)^{s-t}}$ is an intertwining operator mapping $\CF_{d+t-1, s}$ to $\CF_{d+s-1, t}$.
 Altogether, the diagram (\ref{fig:exact sequence}) shows the six intertwining maps that relate the four types of exceptional representations
\begin{figure}[H]
\centering
\begin{tikzcd}[scale cd=1.5, column sep=huge, row sep=huge]
\CF_{d+s-1, t}  \arrow[d, "\CS^-_{d+s-1,t}"']
& \CF_{d+t-1,s}  \arrow[l, "\CD_{s,t}"' ]  \arrow[d, shift left=1.5 ex, "\CS^-_{d+t-1,s}"]\\
\CF_{1-s,t} \arrow[r,  "(z\cdot\partial_x)^{s-t}"' ] \arrow[u, shift right=1.5 ex, "\CS^+_{1-s,t}"']
& \CF_{1-t,s} \arrow[u, "\CS^+_{1-t,s}"]
\end{tikzcd}
\caption{A sequence of intertwining maps for the exceptional representations}
\label{fig:exact sequence}
\end{figure}
Apart from commuting with group actions, these intertwining maps are important for the following reasons 
\begin{itemize}
\item Each directed sequence of homomorphisms in the diagram \ref{fig:exact sequence} is {\it exact}. For example, we have just shown the two sequences  
$\CF_{1-t,s}\xrightarrow{\CS^+_{1-t,s}}\CF_{d+t-1,s}\xrightarrow{\CD_{s,t}} \CF_{d+s-1,t}$ and $\CF_{1-s,t}\xrightarrow{(z\cdot\partial_x)^{s-t}}\CF_{1-t,s}\xrightarrow{\CS^+_{1-t,s}}\CF_{d+t-1,s}$ are exact. A detailed proof for the rest sequences can be found in \cite{Dobrev:1977qv}.
\item The kernel and image of each map are irreducible representations of $\SO(1, d+1)$. This claim can be checked by comparing with the list of irreducible representations given in \cite{10.3792/pja/1195523378}. 
\end{itemize}
The exactness of these intertwining maps implies that there are only three inequivalent irreducible subrepresentations contained in the four exceptional representations 
\begin{align}\label{3irreps}
&\ker (z\cdot\partial_x)^{s-t}, \,\,\,\,\,\  \CU_{s, t}\equiv \ker \CD_{s, t}\cong \CF_{1-t, s}/\Im (z\cdot\partial_x)^{s-t}\nonumber\\
&\CV_{s, t}\equiv\Im(z\cdot\partial_x)^{s-t}\cong \Im \CS^+_{1-s, t}\cong \CF_{1-s, t}/\ker \CS^+_{1-s, t}
\end{align}
where $\ker (z\cdot\partial_x)^{s-t}$ carries the finite dimensional representation  $\mY_{s-1, t}$ of $\SO(1, d+1)$\footnote{When $d=2$ and $t\ge 1$, the kernel of $(z\cdot\partial_{x})^{s-t}$ carries a reducible representation $\mY_{s-1,t}\oplus\mY_{s-1,-t}$ of $\SO(1,3)$.\label{d=2special}} which explains the Casimir (\ref{cashw}). For example, when $s=2, t=1$, $\ker (z\cdot\partial_x)^{s-t}$ is spanned by $z_i, x\cdot z, x_i z_j-x_j z_i, 2x_i\, x\cdot z-x^2 z_i$. Since $\ker (z\cdot\partial_x)^{s-t}$ is finite dimensional, it cannot be unitary unless it is a trivial representation which corresponds to $s=1, t=0$. For the rest two irreducible representations $\CU_{s, t}$ and $\CV_{s,t}$, by analyzing their $\SO(d+1)$ content, i.e. 
\begin{align}
\left.\CU_{s, t}\right|_{\SO(d+1)}=\bigoplus_{n\ge s}\bigoplus_{t+1\le m\le s}\mY_{n,m}, \,\,\,\,\, \left.\CV_{s, t}\right|_{\SO(d+1)}=\bigoplus_{n\ge s}\bigoplus_{0\le m\le t}\mY_{n,m}
\end{align}
we manage to identify them as the irreducible representation $\text{D}^{\lceil\frac{d}{2}\rceil-2}_{(\mY_{s,t+1};0)}$ and $\text{D}^{\lceil\frac{d}{2}\rceil-1}_{(\mY_{s};t)}$ on Hirai's list respectively \cite{10.3792/pja/1195523378}. 

To define inner product on $\CU_{s, t}$ and $\CV_{s,t}$, it is more convenient to use their quotient space realization given in the eq. (\ref{3irreps}).
For example, we can use $\CS^+_{1-s, t}$ to define a pairing on $\CF_{1-s, t}$. Since $\ker\CS^+_{1-s, t}$ drops out by construction, it naturally induces a pairing on the quotient space $\CV_{s, t}$. The explicit form of $\CS^+_{1-s, t}$ in momentum space is given by the eq. (\ref{Spm})
\begin{align}
S^+_{1-s, t}(p;z,w)=p^{d+2s-2}\sum_{\ell=0}^t (-)^{t-\ell}\frac{(s+1-t)_{t-\ell}}{(d+s+\ell-2)_{t-\ell}}\Pi^{t\ell}(\hat p; z,w)
\end{align}
Due to the alternating sign $(-)^{t-\ell}$, the inner product induced by $\CS^+_{1-s, t}$ is not positive definite unless $t=0$. Therefore, among the irreducible representations $\CV_{s, t}$, only $\CV_{s, 0}$ is unitary. In this case, the inner product becomes more transparent if we write it in spherical coordinate using the harmonic expansion (\ref{hatK1})
\begin{align}
(\psi_1,\psi_2)&\equiv \int d^d x_1 \, d^d x_2\, \psi_1^*(x_1) S^+_{1-s,0}(x_{12})\psi_2(x_2)\nonumber\\
&=\sum_{\ell\ge s}\sum_{\bm m}\frac{\Gamma(d+s+\ell-1)}{\Gamma(\ell+1-s)}\left(\int_{S^d}\, d^d \Omega_1\hat\psi_1(\Omega_1) Y_{\ell\bm m}(\Omega_1)^*\right)^*\left(\int_{S^d}\, d^d \Omega_2\hat\psi_2(\Omega_2) Y_{\ell\bm m}(\Omega_2)^*\right)
\end{align}
where $\hat\psi(\Omega)=\left(\frac{1+x^2}{2}\right)^{1-s}\psi(x)$.
Since the sum starts from $\ell=s$, the spherical harmonics on $S^d$ with $\ell\le s-1$ are projected out. These spherical harmonics carry the $\mY_{s-1}$ representation of $\SO(1, d+1)$ and span the kernel of $\CS^+_{1-s, 0}$.

Similarly for $\CU_{s, t}$, we can define a paring on $\CF_{1-t, s}$ by using $\CS^+_{1-t, s}$ and it naturally induces a paring on $\CU_{s, t}$. Since 
\begin{align}
S^+_{1-t, s}(p; z, w)=p^{d+2t-2}\sum_{\ell =t+1}^s\frac{(\ell-t)_{s-\ell}}{(d+t+\ell -2)_{s-\ell }}\Pi^{s \ell}(\hat p; z,w)
\end{align}
is manifestly positive definite with its kernel being factored out, the paring on $\CU_{s, t}$ defined in this way is  positive,  assuming its normalizability. However, the normalizability is not obvious in this case. For example, when $s=2, t=1$, the asymptotic behavior of any wavefunction $\psi(x, z)$ in $\CF_{1-t, s}$ is 
\begin{align}\label{st21}
\psi(x, z)\stackrel{x\to\infty}{\approx}\sum_{k=0}^\infty C_{k} \left( \tilde x^i,R(x)^i_{\,\, j} z^j\right)
\end{align}
where  $C_{k}( u, v)$ is a homogeneous polynomial of degree $k$ in $u^i$ and degree 2 in $v^i$. The $k=0$ term is potentially problematic  because its Fourier transform has a $p^{-d}$ behavior around $p=0$ by a simple power counting and the higher $k$ terms are normalizable. Fortunately any second order tensor in $v^i \equiv R(x)^i_{\,\,j} z^j$ is pure gauge, i.e. 
\begin{align}
v^i v^j =z\cdot \partial_x\left(-\frac{x^i}{x^2}\,(2\, x\cdot z \, x^j-x^2 z^j)\right)
\end{align}
So the $k=0$ term in (\ref{st21}) is killed by $\CS^+_{0,2}$. A general proof about the normalizability for any $s$ and $t$ is given in appendix \ref{normaliz}.
The book \cite{Dobrev:1977qv} contains a different proof regrading the normalizability. 
Altogether, $\CU_{s, t}$ is a unitary irreducible representation for any $s=1, 2,3,\cdots$ and $t=0,1,\cdots, s-1$. Both $\CU_{s,t}$ and $\CV_{s,0}$ belong to the (unitary) exceptional series.

\,

\,

\subsection{$\SO(1,4)$}\label{SO14}
The group $\SO(1,4)$ deserves a separate discussion for two reasons. First, it is the isometry group of the 4-dimensional de Sitter spacetime and 4 is the physically most relevant dimension. Second, some  results derived above are degenerate when $d\le 3$ and need some further clarifications.

At the beginning of this section, we have learned that any irreducible representation of $\SO(1,d+1)$ is equivalent to some subrepresentation of $\CF_{\Delta, \bm s}$. When $d=3$, the highest weight vector $\bm s$ of $\SO(d)$ is reduced to a nonnegative integer $s$ and hence it suffices to consider all $\CF_{\Delta,s}$. The $\SO(d+1)$ content of $\CF_{\Delta,s}$ is in general given by eq. (\ref{decomposeCF}). However, this decomposition needs a minor modification at $d=3$:
\begin{align}
\left.\CF_{\Delta, s}\right\vert_{\SO(4)}=\bigoplus_{n\ge s}\bigoplus_{0\le |m| \le s}\mY_{n,m}
\end{align}
because transverse  spin-$s$ tensor harmonics on $S^3$ correspond to all $\mY_{n, \pm s}, n\ge s$ of $\SO(4)$. Similarly, the $\SO(4)$ content of $\CU_{s,t}$ becomes 
\begin{align}
\left.\CU_{s, t}\right|_{\SO(4)}=\bigoplus_{n\ge s}\bigoplus_{t+1\le |m|\le s}\mY_{n,m}
\end{align}
Unlike in higher dimensions,  $\CU_{s,t}$ is actually reducible with respect to $\SO(1, 4)$. It can be decomposed into two irreducible components $\CU_{s,t}^\pm$ with the following $\SO(4)$ content respectively
\begin{align}
\left.\CU_{s, t}^\pm\right|_{\SO(4)}=\bigoplus_{n\ge s}\bigoplus_{t+1\le m\le s}\mY_{n,\pm m}
\end{align}
$\CU^{\pm}_{s, t}$ are related by the spatial reflection because one of the $\SO(4)$ Cartan generators $L_{34}=-\frac{1}{2}(P_3+K_3)$ is mapped to $-L_{34}$ under the spatial reflection while the other Cartan generator $L_{12}=M_{12}$ stays invariant, which means that the $\SO(4)$ highest weight vector $(n,m)$ is mapped to $(n,-m)$. Therefore $\CU_{s,t}$ is still irreducible with respect to the bigger group $\text{O}(1, 4)$. In addition, $\CU^\pm_{s,t}$ belong to the {\it discrete series} of $\SO(1,4)$\cite{Dobrev:1977qv}. For higher $d$, the exceptional series is completely different from the discrete series (which only exists when $d$ is odd), e.g. the highest weight vector $\bm s$ for a discrete series representation cannot have any vanishing entry \cite{Basile:2016aen}.

\subsection{Summary and bulk QFT correspondence}\label{summ}
We have identified all unitary irreducible representations contained in $\CF_{\Delta,s}$ for the conformal group $\SO(1, d+1)$ with $d\ge 3$. In this section, we summarize these representations and briefly comment on their bulk QFT realizations. For a spin-$s$ field of mass $m$ in $\text{dS}_{d+1}$, its scaling dimension $\Delta$ satisfies the following equations (see e.g. \cite{Hinterbichler:2016fgl})
\begin{align}\label{mD}
&s=0: \,\,\,\,\, m^2=\Delta (d-\Delta)\nonumber\\
&s\ge 1: \,\,\,\,\, m^2=(\Delta+s-2)(d+s-2-\Delta)
\end{align}
With this convention, photon, graviton and their higher spin generalizations have $m=0$.

\begin{itemize}
\item \textbf{Trivial representation}.
\item \textbf{Scalar principal series}: $\CF_{\frac{d}{2}+i\mu}$ ($\mu\in\mathbb R$) equipped with the $L^2(\mathbb R^d)$ inner product. It describes a massive scalar field in $\text{dS}_{d+1}$ with mass $m\ge\frac{d}{2}$. 
\item \textbf{Scalar complementary series}: $\CF_{\Delta}$ ($0<\Delta<d$) equipped with the inner product 
\begin{align}
(\psi_1,\psi_2)=\int\, d^d x_1\, d^d x_2\, \psi_1(x_1)^* S^+_\Delta(x_{12})\,\psi_{2}(x_2)
\end{align}
It describes a massive scalar field in $\text{dS}_{d+1}$ with mass $0<m<\frac{d}{2}$.
\item \textbf{Spinning principal series}: $\CF_{\frac{d}{2}+i\mu, s}$ ($\mu\in\mathbb R$ and $s\in\mathbb Z_+$) equipped with the $L^2$ inner product for tensors on $\mathbb R^d$. It describes a massive spin-$s$ in $\text{dS}_{d+1}$ with mass $m\ge s+\frac{d}{2}-2$. 
\item \textbf{Spinning complementary series}: $\CF_{\Delta, s}$ ($1<\Delta<d-1$ and $s\in\mathbb Z_+$) equipped with the inner product 
\begin{align}
(\psi_1,\psi_2)=\int\, d^d x_1\, d^d x_2\, \psi_1(x_1)_{i_1\cdots i_s}^* S^+_{\Delta,s}(x_{12})_{i_1\cdots i_s,j_1\cdots j_s}\,\psi_{2}(x_2)_{j_1\cdots j_s}
\end{align}
It describes a massive spin-$s$ field in $\text{dS}_{d+1}$ with mass $\sqrt{(s-1)(s+d-3)}<m< s+\frac{d}{2}-2 $. The lower bound $\sqrt{(s-1)(s+d-3)}$ is the so-called Higuchi bound \cite{Higuchi:1986py} which is equivalent to the requirement of no negative norm states in bulk canonical quantization.
\item \textbf{Exceptional series} \rom{1}: $\CV_{s, 0}=\CF_{1-s}/\ker \CS^+_{1-s}$ ($s\in\mathbb Z_+$) equipped with the inner product
\begin{align}
(\psi_1,\psi_2)=\int\, d^d x_1\, d^d x_2\, \psi_1(x_1)^* S^+_{1-s}(x_{12})\,\psi_{2}(x_2)
\end{align}
A local QFT realization of these representations are not known. 
They might be described by  shift symmetric scalars of mass square $m^2=-(s-1)(s+d-1)$ \cite{Bonifacio:2018zex} \footnote{In  \cite{Bonifacio:2018zex}, the authors also argue that the shift symmetric tensor fields, which carry the $\CV_{s,t}$ representations for nonzero $t$, are nonunitary by considering the decoupling limit of massive higher spin fields. }, with the shift symmetry being gauged.  We will leave this problem to future investigation.
\item \textbf{Exceptional series} \rom{2}: $\CU_{s, t}=\CF_{1-t, s}/\Im(z\cdot\partial_x)^{s-t}$ ($s\in\mathbb Z_+$ and $t=0,1,\cdots, s-1$) equipped with the inner product
\begin{align}
(\psi_1,\psi_2)=\int\, d^d x_1\, d^d x_2\, \psi_1(x_1)^*_{i_1\cdots i_s} S^+_{1-t, s}(x_{12})_{i_1\cdots i_s, j_1\cdots j_s}\,\psi_{2}(x_2)_{ j_1\cdots j_s}
\end{align}
It describes a partially massless spin-$s$ gauge field with mass square $m^2_{s, t}=(s-1-t)(d+s+t-3)$ \cite{Deser:1983mm,Brink:2000ag,Deser:2001pe,Deser:2001us,Deser:2001wx,Deser:2001xr,Zinoviev:2001dt,Dolan:2001ih,Hinterbichler:2016fgl}. The parameter $t$ is called {\it depth}, which is also the spin of the corresponding ghost field. In particular, when $t=s-1$, $\CU_{s,s-1}$ describes a massless spin-$s$ gauge field, e.g. photon, graviton, etc.
\end{itemize}
\begin{remark}
For principal and complementary series, $\CF_{\Delta, s}$ and $\CF_{\bar\Delta,s}$ are isomorphic due to the shadow transformations. So it suffices to consider $\mu\ge 0$  in principal series and $\Delta>\frac{d}{2}$ in complementary series, as shown in fig. (\ref{fig:unitarityregion}).

\begin{figure}
     \centering
     \begin{subfigure}[b]{0.48\textwidth}
         \centering
         \includegraphics[width=\textwidth]{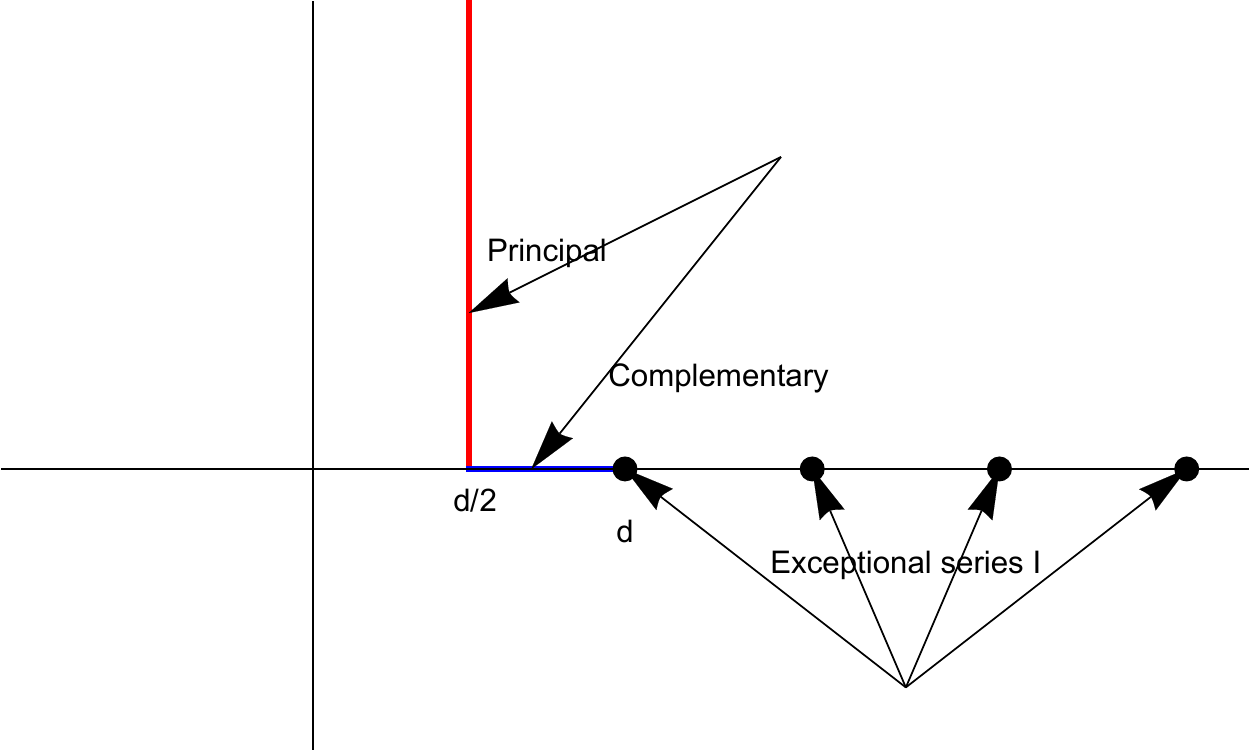}
         \caption{$\Delta$ in scalar UIRs}
         \label{fig:s1}
     \end{subfigure}
     \hfill
     \begin{subfigure}[b]{0.48\textwidth}
         \centering
         \includegraphics[width=\textwidth]{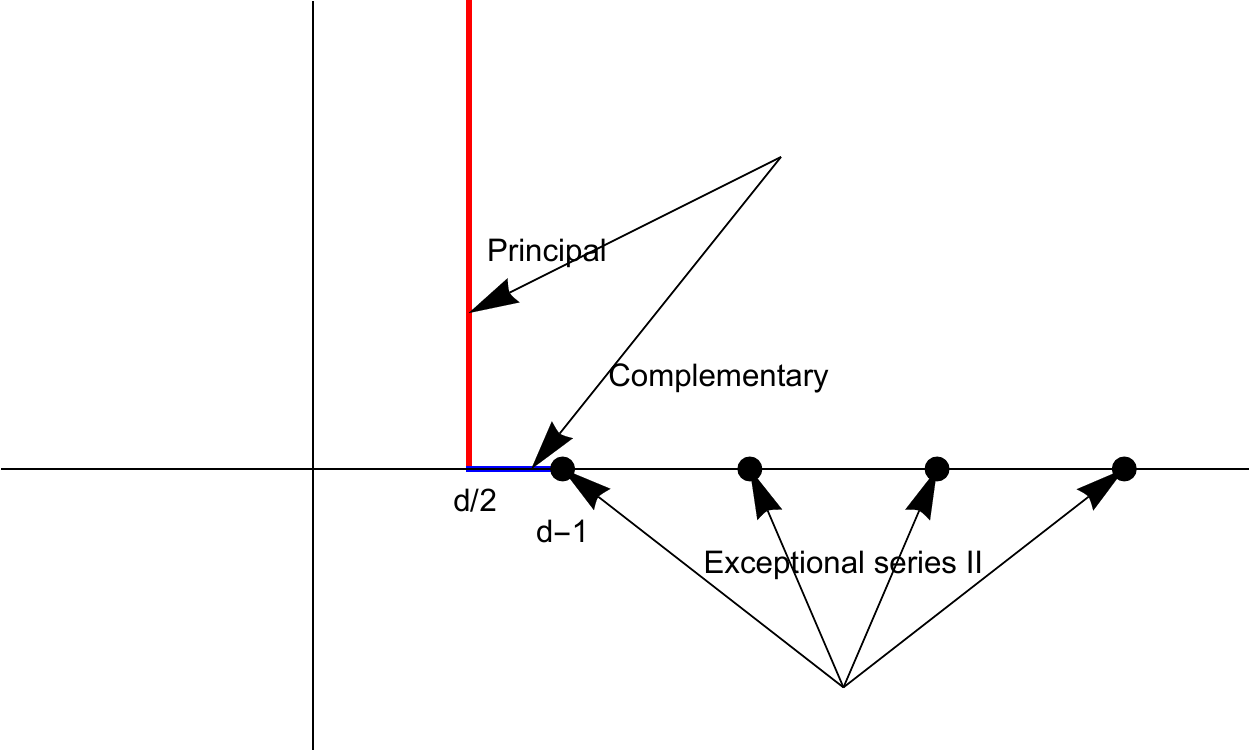}
         \caption{$\Delta$ in spinning UIRs}
         \label{fig:s2}
     \end{subfigure}
    \caption{The domain of unitarity in complex $\Delta$-plane for both scalar (i.e. $s=0$) and spinning (i.e. $s\ge 1$) representations.}
        \label{fig:unitarityregion}
\end{figure}

\end{remark}
\begin{remark}\label{d=2vanishing}
When $d=2$, the exceptional series \rom{2} only contains $\CU_{s,0}$. The reason is as follows. $\CS^+_{1-t,s}$ only involves projection operators $\Pi^{s\ell}$ with $\ell\ge t+1$. However,  $\Pi^{s\ell}$  vanishes identically for $\ell\ge 2$ when $d=2$ \cite{Dobrev:1977qv}, implying that $\CS^+_{1-t,s}$ is actually a zero map for $t\ge 1$. Therefore $\ker\CS^+_{1-t,s}=\Im (z\cdot \partial_x)^{s-t}=\CF_{1-t,s}$. This is  consistent with the fact that graviton in $\text{dS}_3$ has no propagating degrees of freedom. 
\end{remark}

\subsection{Comments on the physical picture}
The dictionary between UIRs of $\SO(1, d+1)$ and QFT in $\text{dS}_{d+1}$ is provided above. In this subsection, we'd like to give a more elaborate and detailed discussion about the realization of the representations in a physical context. We will use  massive and massless spin-1 fields to illustrate the basis idea of the realization,  and the discussion can be straightforwardly generalized to higher spin fields.

First, consider a massive spin-1 field $A_\mu$ of mass $m>0$ in the future planar patch of $\text{dS}_{d+1}$, which is parameterized by
\begin{align}
X^0=-\frac{1+x^2-\eta^2}{2\eta}, \,\,\,\,\, X^i=-\frac{x^i}{\eta}, \,\,\,\,\, X^{d+1}=\frac{1-x^2+\eta^2}{2\eta}
\end{align}
where $-\infty<\eta<0$ and $\eta\to 0^-$ corresponds to the future boundary of $\text{dS}_{d+1}$. The planar patch covers  half of the de Sitter space, i.e. the region $X^0>X^{d+1}$, and its metric is given by
\begin{align}
ds^2=\frac{-d\eta^2 + dx^2}{\eta^2}
\end{align} 
In the planar patch, Killing vector fields associated to the conformal generators $\{P_i, K_i, D, M_{ij}\}$ are
\begin{align}\label{vecfield}
&\CP_i=-\partial_i, \,\,\,\,\, \CK_i=(x^2-\eta^2)\partial_i-2x_i (x\cdot\partial_x+\eta\partial_\eta)\nonumber\\
&\CD=-(x\cdot\partial_x+\eta\partial_\eta),\,\,\,\,\, \CM_{ij}=x_i\partial_j -x_j\partial_i
\end{align}
The  mass $m$ of $A_\mu$ and its scaling dimension $\Delta$  is related by eq. (\ref{mD}), i.e. $m^2=(\Delta-1)(\bar\Delta-1)$.  The leading asymptotic behavior of $A_i$ near the future is: 
\begin{align}\label{Aia}
A_i(\eta, x)\stackrel{\eta\to 0^-}{\approx} (-\eta)^{\Delta-1} \alpha_i(x)+(-\eta)^{\bar\Delta-1} \beta_i(x)
\end{align}
where the operators $\alpha_i$ and $\beta_i$ are different linear combinations of creation and annihilation operators.  Eq. (\ref{Aia}) completely determines the leading behavior of $A_\eta$ via the constraint $\nabla_\mu A^\mu=0$:
\begin{align}
\nabla_\mu A^\mu=0\Leftrightarrow \partial_i A_i = \frac{1}{\eta}(\eta\partial_\eta+1-d)A_\eta
\end{align}
When $\Delta\not=1,d-1$, which holds for $m>0$, we are allowed to invert the operator $\frac{1}{\eta}(\eta\partial_\eta+1-d)$ in the following sense 
\begin{align}\label{Aeta}
A_\eta(\eta, x)&= (\eta\partial_\eta+1-d)^{-1}\left(\eta\,  \partial_i A_i(\eta, x)\right)\nonumber\\
&\approx \frac{1}{\bar\Delta-1}(-\eta)^{\Delta} \partial_i \alpha_i(x)+\frac{1}{\Delta-1}(-\eta)^{\bar\Delta} \partial_i\beta_i(x), \,\,\,\,\, \eta\to 0^-
\end{align}
Using (\ref{vecfield}), (\ref{Aia}) and (\ref{Aeta}), one can show that $\alpha_i$ and $\beta_i$ are primary operators of scaling dimension $\Delta$ and $\bar\Delta$ respectively. For example, the action of $K_k$ on $\alpha_i$ can be computed as follows:
\begin{align}\label{KA}
[K_k, A_i]&\equiv -\CL_{\CK_k} A_i=-\CK_k A_i-\partial_i \CK_k^\mu A_\mu\nonumber\\
&=\left[(\eta^2-x^2)\partial_k+2x_k (x\cdot\partial_x+\eta\partial_\eta+1)\right]A_i+2\delta_{ik}(x\cdot A+\eta A_\eta)-2 x_i A_k
\end{align}
Taking the $\eta\to 0^-$ limit and keeping the leading asymptotics, (\ref{KA}) is reduced to 
\begin{align}
&[K_k, \alpha_i(x)]=\left[2x_k (x\cdot\partial_x+\Delta)-x^2\partial_k\right]\alpha_i+2 x^\ell (\delta_{ik} \alpha_\ell-\delta_{i\ell} \alpha_k)\nonumber\\
&[K_k, \beta_i(x)]=\left[2x_k (x\cdot\partial_x+\bar\Delta)-x^2\partial_k\right]\beta_i+2 x^\ell (\delta_{ik} \beta_\ell-\delta_{i\ell} \beta_k)
\end{align}
Define $|\Delta, x\rangle_i =\alpha_i(x)|E\rangle$ and $|\bar\Delta, x\rangle_i =\beta_i(x)|E\rangle$.  $|\bar\Delta, x\rangle_i$ transforms under $\so(1,d+1)$ according to (\ref{actonstate}) and $|\bar\Delta, x\rangle_i$ transforms similarly except $\bar\Delta$ is replaced by $\Delta$. They both span the single-particle Hilbert space $\CH_{\Delta,1}$ of $A_\mu$ (for the same reason explained in subsection \ref{QFTpicture}) and hence we can expand an arbitrary state in $\CH_{\Delta,1}$ by using either basis
\begin{align}
|\Psi\rangle=\int\,d^d x\, \psi^{(1)}_i (x)|\Delta, x\rangle_i=\int\,d^d x\, \psi^{(2)}_i (x)|\bar\Delta, x\rangle_i
\end{align}
By construction, $\psi^{(1)}_i (x)$ belongs to $\CF_{\bar\Delta,1}$ and $\psi^{(2)}_i (x)$ belongs to $\CF_{\Delta,1}$. In this formalism, the isomorphism between $\CF_{\Delta,1}$ and $\CF_{\bar\Delta,1}$ is almost obvious because $\psi^{(1)}_i$ and $\psi^{(2)}_i$ are related by a basis transformation, which is actually a shadow transformation.

Next, we consider a massless vector field. In this case, the Maxwell action has a gauge symmetry $A_\mu\to A_\mu+\partial_\mu \phi$. Imposing Lorentz gauge $\nabla_\mu A^\mu=0$,  the leading asymptotic behavior of $A_i$ near the future is: 
\begin{align}
A_i(\eta, x)\stackrel{\eta\to 0^-}{\approx} (-\eta)^{d-2}\alpha_i(x)+ \beta_i(x)
\end{align}
Compared to the massive case, there are two main differences. First, the Lorentz gauge leads to $\partial_i\alpha_i(x)=0$ because $\frac{1}{\eta}(\eta\partial_\eta+1-d)$ is not invertible while acting on $(-\eta)^{d-2}$. Second, there is  a residual on-shell gauge redundancy $A_\mu\to A_\mu+\partial_\mu\phi$, where $\phi$ satisfies $\nabla^2\phi=0$. Since the leading fall-offs of $\phi$ at the future boundary are $(-\eta)^0\lambda$ and $(-\eta)^d\xi$, $\beta_i$ behaves like a boundary gauge field, i.e. $\beta_i \to \beta_i+\partial_i\lambda$. The constraint $\partial_i \alpha_i=0$ implies that we should exclude wavefunctions in $\CF_{0,1}$ that are total derivatives, i.e. $\psi^{(1)}_i(x)=\partial_i \psi^{(1)}$.
 On the other hand, in order to kill the boundary gauge symmetry of $\beta_i$  we should restrict ourselves to wavefunctions in $\CF_{d,1}$ that satisfy $\partial_i \psi^{(2)}_i=0$. Altogether, the single-particle Hilbert space $\CH_{\text{Max}}$ of a Maxwell field is isomorphic to $\CF_{0,1}$ with total derivatives modded out, or the divergence free subspace  of $\CF_{d,1}$. Identifying $\CH_{\text{Max}}$ as the exceptional series representation $\CU_{1,0}$, the first realization of $\CH_{\text{Max}}$  corresponds to $\CU_{1,0}\cong \CF_{0,1}/\Im(z\cdot\partial_x)$ and the second realization corresponds to $\CU_{1,0}\cong\ker (\partial_x\cdot\CD_z)$, c.f. eq. (\ref{3irreps}).

\section{Harish-Chandra characters}\label{HCchar}
\subsection{General theory}
With the UIRs constructed, the next step is to compute their group characters which collect the information about the representations in a simple function. The characters associated to finite dimensional representations can be defined unambiguously. For example, given a finite dimensional representation $\rho$ of a group $G$, the corresponding group character associated to certain element $g\in G$ is defined as a trace of $\rho(g)$ over the representation space $V_\rho$, i.e. $\Theta_\rho(g)\equiv \Tr_{V_\rho} \rho(g)$.  However, such a trace does not necessarily make sense for an infinite dimensional representation like $\CF_{\Delta, s}$. To tell when a ``trace'' can be defined in the infinite dimensional case, we  need some deep notions and theorems in representation theory \cite{10.2307/j.ctt1bpm9sn}:

\begin{definition}
Let $G$ be a connected reductive Lie group and let $\tK$ be a maximal compact subgroup. A representation $\pi$ of $G$ on a Hilbert space $V$ is called \textbf{admissible} if $\pi|_{\tK}$ is unitary and if each unitary irreducible representation $\tau$ of $K$ occurs with only finite multiplicity in $\pi|_{\tK}$.
\end{definition}
\noindent{}In particular, $\SO(1, d+1)$  is a connected reductive Lie group  and  all $\CF_{\Delta, s}$ (including their unitary irreducible subrepresentations)  are admissible.

\begin{definition}
We say an admissible representation $\pi$ of a linear connected reductive group $G$ has a \textbf{Harish-Chandra character} (or \textbf{global character}) $\Theta_\pi$ if $\pi (\varphi)$ is of trace class for any compact supported function $\varphi$ on $G$ and if $\varphi\to \Tr \pi(\varphi)\equiv \Theta_\pi(\varphi)$ is a distribution. In this case, the character $\Theta_\pi$ is clearly conjugation invariant.
\end{definition}
\noindent{}Then the following theorem tells us when an admissible representation $\pi$ has a Harish-Chandra character
\begin{theorem}
Every admissible representation $\pi$ of a linear connected reductive group $G$ whose decomposition $\pi |_{\tK}=\bigoplus_{\tau} n_\tau \, \tau$ satisfies $n_\tau\le C \dim \tau$ has a Harish-Chandra character.
\end{theorem}
\noindent{}Since each $\SO(d+1)$ content of  $\CF_{\Delta, s}$  has multiplicity 1, $\CF_{\Delta, s}$ and its  irreducible components admit a Harish-Chandra character. We shall also use the heuristic notations
\begin{align}
\Theta_{\CF_{\Delta,s}}(g)\equiv \Tr_{\CF_{\Delta,s}}g
\end{align}
which only exist in the distribution sense. In this section, we will mainly focus on $g=q^D$, namely $\Theta_{\CF_{\Delta,s}}(q)\equiv \Theta_{\Delta, s}(q^D)$. The more general case, with $q^D$ being decorated by angular momenta, i.e. $\Theta_{\CF_{\Delta,s}}(q, \bm u)\equiv \Theta_{\Delta, s}(q^D e^{u_1 M_{12}}e^{u_2 M_{34}}\cdots)$, will be discussed in appendix \ref{fullchar}.

\subsection{Compute $\Theta_{\CF_{\Delta,s}}(q)$}\label{comchar}
In this subsection, we show how to compute the character $\Theta_{\CF_{\Delta,s}}(q)$. Let's start from the $s=0$ case. By definition, one would find an orthonormal basis of $\CF_\Delta$ and compute the matrix element of $D$ with respect to this basis (which is actually  done in the appendix \ref{irrCF} with the basis being spherical harmonics). Then exponentiate the infinite dimensional matrix $D$ and compute the trace. However, we can make life much easier by directly using the ket basis $| x \rangle$. The action of $q^D$ on $|x\rangle$ is 
\begin{align}
q^D |x\rangle=q^{\bar\Delta}\left|q x\right\rangle=\int d^d y\, q^{\bar\Delta}\delta^d(qx-y)|y \rangle
\end{align}
from which we  can directly read off the matrix elements of $q^D$ with respect to the basis $|x\rangle$. The integral of the diagonal entries of $q^D$ is localized on the fixed points of $q^D$, i.e. $0$ and $\infty$ or equivalently the southern and northern poles when conformally mapped to $S^d$. But the $x^i$ coordinate system only covers the $x=0$ point and hence we would not be able to correctly capture the contribution from $\infty$ if we naively computed $\int\, d^d x\, q^\Delta\delta^d(qx-x)$. We can bypass this problem by conjugating $q^D$ with special conformal transformations, i.e. $q^D\to e^{-b\cdot K} q^De^{b\cdot K}$. This conjugation maps the two fixed points $0$ and $\infty$ to  $0$ and $\frac{b^i}{b^2}$ respectively. More explicitly, the action of $e^{-b\cdot K} q^D e^{b\cdot K}$ on $|x\rangle$ is given by
\begin{align}
e^{-b\cdot K} q^D e^{b\cdot K} |x\rangle=\Omega(q, x,b)^{\bar\Delta} |\tilde x(q,x, b)\rangle, \,\,\,\,\, \tilde x^i=q\frac{x^i+\left(q-1\right) x^2 b^i}{1+2(q-1)\, x\cdot b+\left(q-1\right)^2 b^2 x^2}
\end{align}
where 
\begin{align}
\Omega(q, x, b)=\frac{q}{1+2(q-1)\, x\cdot b+\left(q-1\right)^2 b^2 x^2}
\end{align}
Therefore the character $\Theta_{\CF_{\Delta}}(q)$ should be 
\begin{align}\label{chartheta}
\Theta_{\CF_{\Delta}}(q)=\int d^d x\, \Omega(q, x,b)^{\bar\Delta}\delta^d(\tilde x-x)
\end{align}
This integral is localized at $x^i=0$ and $x^i=\frac{b^i}{b^2}$.
At these points, the scaling factor $\Omega(q, x, b)$ becomes 
\begin{align}
\Omega(q, 0, b)=q, \,\,\,\,\, \Omega(q, b^i/b^2, b)= q^{-1}
\end{align}
and the Jacobian associated to the map $x^i\to \tilde x^i-x^i$ becomes $|q-1|^d$ and $|1-q^{-1}|^d$ respectively. Altogether, the integral in eq. (\ref{chartheta}) yields
\begin{align}\label{chartheta1}
\Theta_{\CF_{\Delta}}(q)=\frac{q^{\bar\Delta}}{|q-1|^d}+\frac{q^{-\bar\Delta}}{|1-q^{-1}|^d}=\frac{q^\Delta+q^{\bar\Delta}}{|1-q|^d}
\end{align}
where the $b$ dependence drops out explicitly as expected from the conjugation invariance.

\begin{remark} The character $\Theta_{\CF_{\Delta}}(q)$ given by  eq. (\ref{chartheta1})  is symmetric under $q\to q^{-1}$. This property, which holds for all $\CF_{\Delta, \bm s}$,  is a result of the conjugation invariance of $\Theta$ since $D=L_{0, d+1}$ is mapped to $-D$ by the conjugation of $e^{i\pi L_{d+1, i}}$. In view of this, we assume $\boxed{0<q<1}$ in the remaining part of this section, without loss of generality.
\end{remark}
The generalization of our computation to the spinning case is almost straightforward. Let's write the ket basis of $\CF_{\Delta, s}$ as $|x\rangle_\alpha$ where the index $\alpha$ carries the spin-$s$ representation of $\SO(d)$. The action of $e^{-b\cdot K} e^{t D} e^{b\cdot K}$ on this basis can be schematically expressed as 
\begin{align}
e^{-b\cdot K} q^D e^{b\cdot K} |x\rangle_\alpha = \CO_{\alpha\beta}(q ,x, b)\Omega(q, x,b)^{\bar\Delta} |\tilde x(q,x, b)\rangle_\beta
\end{align}
where $\CO_{\alpha\beta}(q, x, b)$ is an $\SO(d)$ rotation matrix in the spin-$s$ representation. In general, $\CO_{\alpha\beta}(q, x, b)$ takes a very sophisticated form but while evaluated at the two fixed points, i.e. $x^i=0$ and $x^i=\frac{b^i}{b^2}$, it becomes an identity matrix. Altogether, we have 
\begin{align}
\Theta_{\CF_{\Delta,s}}(q)&=\sum_{\alpha}\int d^d x\, \CO_{\alpha\alpha}(q, x, b) \Omega(q, x,b)^{\bar\Delta}\delta^d(\tilde x-x)\nonumber\\
&=\Theta_{\CF_{\Delta}}(q)\times \left(\sum_\alpha\, 1\right)=D^d_s\, \frac{q^\Delta+q^{\bar\Delta}}{(1-q)^d}
\end{align}
where $D^d_s$ is the dimension of the spin-$s$ representation of $\SO(d)$. Thus $D^d_s\, \frac{q^\Delta+q^{\bar\Delta}}{(1-q)^d}$ is the Harish-Chandra character for both principal and complementary series.

\subsection{Harish-Chandra characters of exceptional series}
Now let's proceed to compute the Harish-Chandra characters of exceptional series $\CV_{s, 0}$ and $\CU_{s, t}$. For $\CV_{s, 0}$, recall that $\CV_{s, t}\equiv \Im(z\cdot \partial_x)^{s-t}\cong \CF_{1-s, t}/\ker(z\cdot\partial_x)^{s-t}$, which yields
\begin{align}
\Theta_{\CV_{s, 0}}(q)=\Theta_{\CF_{1-s}}(q)-\Theta_{\ker(z\cdot\partial_x)^{s}}(q)
\end{align}
Since $\ker(z\cdot\partial_x)^{s}$ carries the $\mY_{s-1}$ of $\SO(1, d+1)$, $\Theta_{\ker(z\cdot\partial_x)^{s}}(q)$ is nothing but the usual $\SO(d+2)$ character corresponding to the highest weight representation $\mY_{s-1}$, denoted by $\Theta^{\SO(d+2)}_{\mY_{s-1}}(q)$. Thus we obtain 
\begin{align}\label{CVchar}
\Theta_{\CV_{s, 0}}(q)=\frac{q^{1-s}+q^{d+s-1}}{(1-q)^d}-\Theta^{\SO(d+2)}_{\mY_{s-1}}(q)
\end{align}
Similarly for $\CU_{s,t}$, using the isomorphism $\CU_{s, t}\cong\CF_{1-t, s}/\CV_{s, t}$, we obtain
\begin{align}\label{CUchar}
\Theta_{\CU_{s, t}}(q)=D^d_s\frac{q^{1-t}+q^{d+t-1}}{(1-q)^d}-D^d_t\frac{q^{1-s}+q^{d+s-1}}{(1-q)^d}+\Theta^{\SO(d+2)}_{\mY_{s-1, t}}(q)
\end{align}
In general, one can write out the characters like $\Theta^{\SO(d+2)}_{\mY_{s-1, t}}(q)$ explicitly by using Weyl character formula \cite{10.2307/j.ctt1bpm9sn}. However, in order to compare with the paper \cite{Anninos:2020hfj}, we derive a slightly different expression (c.f. eq. (\ref{newcharcomp})) for $\Theta^{\SO(d+2)}_{\mY_{s-1, t}}(q)$ in  appendix \ref{SOd2}. Plugging this new expression into eq. (\ref{CVchar}) and eq. (\ref{CUchar}) yields
\begin{align}\label{flippingformula}
\Theta_{\CV_{s, 0}}(q)=\left[ \frac{q^{1-s}+q^{d+s-1}}{(1-q)^d} \right]_+, \,\,\,\,\, \Theta_{\CU_{s, t}}(q)=\left[D^d_s\frac{q^{1-t}+q^{d+t-1}}{(1-q)^d}-D^d_t\frac{q^{1-s}+q^{d+s-1}}{(1-q)^d}\right]_+
\end{align}
where the ``flipping'' operator $[\,\,\,]_+$ acts on an arbitrary Laurent series $\sum_k c_k q^k$ as
\begin{align}
\left[\sum_k c_k q^k\right]_+=\sum_{k<0} (-c_k) q^{-k}+\sum_{k>0} c_k q^k
\end{align}
We have checked that Harish-Chandra characters $\Theta_{\CF_{\Delta, s}}$, $ \Theta_{\CV_{s,0}}$ and $\Theta_{\CU_{s,t}}$ agree with the results in literature \cite{10.3792/pja/1195522333,Basile:2016aen, Dobrev:1977qv}. Some lower $d$ examples are
\begin{align}
&d=3: \,\, \Theta_{\CU^\pm_{s, t}}(q)= \frac{(2s+1) q^{t+2}-(2t+1)q^{s+2}}{(1-q)^3}\nonumber\\
&d=4:\,\,   \Theta_{\CU_{s, t}}(q)=2 (s-t) (s+t+2) \frac{q^2}{(1-q)^4}=\left(D^d_{s,t+1}+D^d_{s,-(t+1)}\right)\frac{q^2}{(1-q)^4}
\end{align}
where $D^d_{s, t+1}=D^d_{s,-(t+1)}=(s-t) (s+t+2)$. In contrast to the $d=3$ case, the small $q$ expansion of  $\Theta_{\CU_{s, t}}(q)$ for $\SO(1,5) $ starts with a quadratic term, independent of the  spin or depth. Actually, by using eq. (\ref{flippingformula}), one can check that this property persists in any higher dimensions
\begin{align}
d\ge 5:\,\,  \Theta_{\CU_{s, t}}(q)=D^d_{s, t+1} q^2+\CO(q^3)
\end{align}
In \cite{Sun:2020sgn}, we give a physical explanation for this difference between $d=3$ and $d\ge 4$, by counting quasinormal modes of spin-$s$ gauge fields in the static path of $\text{dS}_{d+1}$ \footnote{The  static patch metric is given by $ds^2=-(1-r^2) dt^2+\frac{dr^2}{1-r^2}+r^2 d\Omega^2_{d-1}$, where $0\le r\le 1$. The dS horizon is at $r=1$. A quasinormal mode in the static patch is defined as an on-shell mode with the in-falling boundary condition $e^{-i\omega (t-\rho)}$ near horizon, where $\rho=\tanh^{-1}(r)$.}. More generally, we show in  \cite{Sun:2020sgn} that  Harish-Chandra characters encode precisely  de Sitter quasinormal modes as follows. Let $\varphi$ be a unitary field in $\text{dS}_{d+1}$ and let $R$ be the corresponding $\SO(1, d+1)$ UIR. Let $\Sigma=\{\omega_{QN}\}$ be the set of quasinormal frequencies of $\varphi$ and $d_{\omega_{QN}}$ be the degeneracy of quasinormal modes with quasinormal frequency $\omega_{QN}$. Then the Harish-Chandra character $\Theta_R(q)$ can be recovered as 
\begin{align}
\Theta_R(q)=\sum_{\omega_{QN}\in\Sigma}\, d_{\omega_{QN}} \, q^{i\omega_{QN}},\,\,\,\,\, 0<q<1
\end{align}
where the condition $0<q<1$ is necessary for the infinite sum over $\omega_{QN}$ to be convergent because $\Im(\omega_{QN})>0$. These quasinormal modes are eigenmodes of the hermitian operator $-iD$ but with complex eigenvalues  $\omega_{QN}$. They do not belong to the single-particle Hilbert space of $\varphi$. Instead, they appear as resonances \cite{Anninos:2020hfj}.

\begin{remark}
For the discrete series of $\SO(1, 2)$, we use the quotient space realization $\CD^+_N\bigoplus\CD^-_N=\CF_{1-N}/P_N$ where $P_N$ carries the spin-($N-1$) representation of $\SO(1, 2)$
\begin{align}
\Theta_{\CD^+_N\bigoplus\CD^-_N}(q)=\Theta_{\CF_{1-N}}(q)-\sum_{k=1-N}^{N-1} q^k=\frac{2 \,q^N}{1-q}
\end{align}
The character for each summand $\CD^\pm_N$ is $\frac{q^N}{1-q}$ \cite{10.2307/j.ctt1bpm9sn}.
\end{remark}

\begin{remark}
In the $d=2$ case, according to  footnote \ref{d=2special}, we should include $\Theta^{\SO(4)}_{\mY_{s-1,-t}}(q)$ on the R.H.S of eq. (\ref{CUchar}) when $t\ge 1$:
\begin{align}
&t\ge 1: \,\,\,\,\,\Theta_{\CU_{s,t}}=\frac{2q}{(1-q^2)}\left(q^t+q^{-t}-q^s-q^{-s}\right)+\Theta^{\SO(4)}_{\mY_{s-1,t}}(q)+\Theta^{\SO(4)}_{\mY_{s-1,-t}}(q)=0\\
&t=0:\,\,\,\,\,\Theta_{\CU_{s,0}}=\frac{4\, q-q^{1+s}-q^{1-s}}{(1-q)^2}+\Theta^{\SO(4)}_{\mY_{s-1}}=\frac{2\,q}{(1-q)^2}
\end{align}
where we have used $\Theta^{\SO(4)}_{\mY_{s-1,t}}(q)=\frac{q}{(1-q^2)}\left(q^s+q^{-s}-q^t-q^{-t}\right)$ and $\Theta^{\SO(4)}_{\mY_{s-1}}(q)=\frac{q}{(1-q)^2}\left(q^s+q^{-s}-2\right)$. The vanishing of $\Theta_{\CU_{s,t}}$ for $t\ge 1$ is expected due to  remark \ref{d=2vanishing}.
\end{remark}

\section{$\SO(1,d+1)$ vs. $\SO(2,d)$}\label{versus}
In this section, we will comment on distinctions between UIRs of $\SO(2, d)$ and $\SO(1, d+1)$.  $\SO(2, d)$ is the isometry group of $(d+1)$ dimensional AdS and the conformal group of $\mathbb{R}^{1, d-1}$. It shares the same Lie algebra as $\SO(1,d+1)$ but with 
different reality conditions (see, for example \cite{Simmons-Duffin:2016gjk})
\begin{align}
D^\dagger=D, \,\,\,\,\, P_i^\dagger=K_i, \,\,\,\,\, M_{ij}^\dagger=-M_{ij}
\end{align}
$D$ generates time translation in the global coordinates of AdS, playing the role of a Hamiltonian. So the physically relevant representations of $\SO(2, d)$ are the lowest-weight ones, where the Hamiltonian  $D$ is bounded from  below.  These representations are fixed by a primary state and the corresponding representation spaces are spanned by  primary states together with their descendants. In contrast,  primary states in the dS sense (e.g. $|\bar\Delta,0\rangle_{i_1\cdots i_s}$ in e.q. (\ref{defineprimary})) are not normalizable and do not belong to the representation spaces $\CF_{\Delta, s}$. The $\SO(1, d+1)$ UIRs given by the list in section \ref{summ} are not lowest-weight representations.
We also want to mention that $\SO(2, d)$ does have UIRs that are not lowest-weight representations, for example principal series representations, which describe continuous-spin fields \cite{Kravchuk:2018htv,Metsaev:2019opn}.

Since $D$ is bounded from below, the most natural  way to define an $\so(2, d)$ character is \cite{Dolan:2005wy,Basile:2016aen}\footnote{One can also turn on the $\SO(d)$ generators, i.e. $\chi_R(q, \bm x)=\Tr_R q^D x_1^{L_{12}} x_2^{L_{34}}\cdots $}
\begin{align}\label{AdSchar}
\chi_R(q)\equiv \Tr_R q^D, \,\,\,\,\, 0<|q|<1
\end{align}
where $R$ is a lowest-weight representation of $\SO(2, d)$. Compared to the Harish-Chandra character of $\SO(1, d+1)$, the trace in (\ref{AdSchar}) makes perfect sense given $0<|q|<1$ but $\chi_R(q)$ fails to be a {\it group} character because $q^D$ is {\it not} an element of $\SO(2, d)$ for $0<|q|<1$. Some examples of the $\so(2,d)$ characters are \cite{Basile:2016aen}:
\begin{itemize}
\item The scalar representation of scaling dimension $\Delta>\max\{0,\frac{d-2}{2}\}$, which describes a scalar field in $\text{AdS}_{d+1}$ of mass $m^2=\Delta(\Delta-d)$:
\begin{align}
\chi_\Delta(q)=\frac{q^\Delta}{(1-q)^d}
\end{align}
It takes a similar form as the Harish-Chandra character $\Theta_{\CF_\Delta}(q)$ of $\SO(1, d+1)$, except the absence of the  $q^{\bar\Delta}$ term. The difference arises from the fact that  in AdS we should choose either standard quantization or alternate quantization.
\item The spin-$s$ representation of scaling dimension $\Delta=d+s-2$, which describes a massless spin-$s$ gauge field in $\text{AdS}_{d+1}$ or a spin-$s$ conversed current in boundary $\text{CFT}_d$:
\begin{align}
\chi_s(q)=\frac{D^d_s q^{d+s-2}-D^d_{s-1}q^{d+s-1}}{(1-q)^d}
\end{align}
It is very different from its $\SO(1, d+1)$ counterpart
\begin{align}
\Theta_{\CF_{s,s-1}}(q)=\chi_s(q)+\frac{D_s^d q^{2-s}-D^d_{s-1}q^{1-s}}{(1-q)^d}+\Theta^{\SO(d+2)}_{\mY_{s-1,s-1}}(q)
\end{align}
In \cite{Sun:2020sgn}, we attribute the discrepancy between $\Theta_{\CF_{s,s-1}}(q)$ and $\chi_s(q)$ to gauge-invariant quasinormal modes sourced by a boundary spin-$s$ gauge field.
\end{itemize}

\section*{Acknowledgments} 
I am grateful to Frederik Denef, Austin Joyce and Bob Penna for discussions and explanations. I especially thank Dionysios Anninos, Frederik Denef and Albert Law for a thorough reading and comments on the draft. 
Z.S. is supported by NSF grant PHY-1914860 and the  Gravity Initiative at Princeton University.

\appendix

\section{UIRs of $\SL(2,\mathbb R)$}\label{SLrev}
In this appendix, we review the UIRs of $\SL(2,\mathbb R)$, following the constructions in section \ref{direct}.  The group $\SL(2,\mathbb R)$ consists of real $2\times 2$ matrices with determinant 1. Its Lie algebra $\mathfrak{sl}(2,\mathbb R)$  is isomorphic to $\so(1,2)$. The latter is spanned by three generators $\{P,K,D\}$, satisfying the commutation relation (\ref{PDKcom}). We will also use them as a basis of $\mathfrak{sl}(2,\mathbb R)$. In the defining representation of $\SL(2,\mathbb R)$, the generators $\{P,K,D\}$ can be realized as the following $2\times 2$ matrices
\begin{align}
P=\begin{pmatrix}0&-1\\0&0\end{pmatrix}, \,\,\,\,\, K=\begin{pmatrix}0&0\\1&0\end{pmatrix},\,\,\,\,\, D=\frac{1}{2}\begin{pmatrix}1&0\\0&-1\end{pmatrix}
\end{align}
With this choice, $L_0=-\frac{i}{2}(P+K)=-\frac{\sigma_2}{2}$, where $\sigma_2$ is the second Pauli matrix. 

For $\SO(1,2)$, $e^{2\pi i L_0}$ is the identity element  and hence the spectrum of $L_0$ must be integers in any $\SO(1,2)$ representation. In contrast, $e^{2\pi i L_0}=-\small\begin{pmatrix}1&0\\0&1\end{pmatrix}$ in  $\SL(2,\mathbb R)$. As an involutory central element,  $e^{2\pi i L_0}$ is either 1 (even ``parity'') or $-1$ (odd ``parity'') in an irreducible representation of $\SL(2,\mathbb R)$.  The UIRs of $\SL(2,\mathbb R)$ with an even parity are equivalent to  UIRs of $\SO(1,2)$. Since the latter have been analyzed extensively in section \ref{SO12}, we will focus on parity odd UIRs in this appendix. In these representations, the eigenvalues of $L_0$ are half-integers. Denote the eigenstates of $L_0$ by $|r\rangle$, i.e. $L_0|r\rangle=r|r\rangle$, where $r\in\mathbb Z+\frac{1}{2}$. The action of $L_\pm$ on these states are given by  (\ref{L-action}) and (\ref{L+action}), with $n$ replaced by $r$:
\begin{align}
L_\pm |r\rangle=(r\pm \Delta)|r\pm 1\rangle
\end{align}
The reality condition $L^\dagger_\pm =L_\mp$ implies 
\begin{align}
\frac{\langle r+1|r+1\rangle}{\langle r|r\rangle}=\frac{r+\bar\Delta}{r+\Delta^*}
\end{align}
For $\lambda_r\equiv \frac{r+\bar\Delta}{r+\Delta^*}$ to be real, $\Delta$ lies on either  the real line or $\frac{1}{2}+i\mathbb R$, i.e.
\begin{align}
(i) \, \Delta=\frac{1}{2}+i\mu, \,\,\mu\in\mathbb R, \,\,\,\,\, (ii)\, \Delta\in\mathbb R
\end{align}
The positivity of $\lambda_r$ requires separate discussions for different values of $\Delta$:
\begin{itemize}
\item $\lambda_r$ is identically equal to 1 when $\Delta=\frac{1}{2}+i\mu, \mu\not=0$. We can consistently choose $\langle r|r\rangle=1$ for all $r\in\mathbb Z+\frac{1}{2}$ and hence the resulting representation, denoted by $\CP^-_{\Delta}$, is unitary. $\CP^-_{\Delta}$ is a parity odd  principal series representation. The irreducibility of $\CP^-_{\Delta}$ is shown pictorially in fig. (\ref{CPodd}). $\CP^-_{\Delta}$ and $\CP^-_{\bar\Delta}$ are  equivalent since we can easily generalize the shadow transformation, c.f. (\ref{Shadow1}), to the parity odd case.
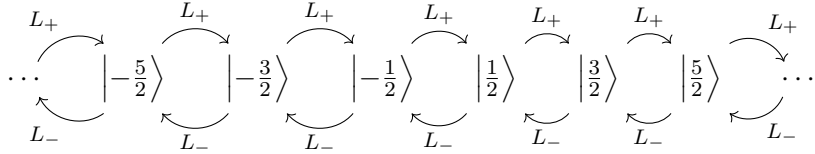
\begin{figure}[H]
\centering
\begin{tikzcd}[column sep=small]
\cdots \arrow[bend left=50]{r}{L_+}& \left|-\frac{5}{2}\right\rangle  \arrow[bend right=-50]{l}{L_-} \arrow[bend left=50]{r}{L_+} & \left|-\frac{3}{2}\right\rangle  \arrow[bend right=-50]{l}{L_-} \arrow[bend left=50]{r}{L_+} & \left|-\frac{1}{2}\right\rangle \arrow[bend right=-50]{l}{L_-} \arrow[bend left=50]{r}{L_+} & \left|\frac{1}{2}\right\rangle  \arrow[bend right=-50]{l}{L_-} \arrow[bend left=50]{r}{L_+} & \left|\frac{3}{2}\right\rangle \arrow[bend right=-50]{l}{L_-} \arrow[bend left=50]{r}{L_+} & \left|\frac{5}{2}\right\rangle  \arrow[bend right=-50]{l}{L_-} \arrow[bend left=50]{r}{L_+} &\cdots \arrow[bend right=-50]{l}{L_-}
\end{tikzcd}
\caption{The action of $L_\pm$ in principal series $\CP^-_{\Delta}$. }
\label{CPodd}
\end{figure}

\item When $\Delta\in\mathbb R$ but $\Delta\notin\mathbb Z+\frac{1}{2}$, $r$ takes value in all half-integers and we need 
\begin{align}
\lambda_r=\frac{(r+\frac{1}{2})^2-(\Delta-\frac{1}{2})^2}{(r+\Delta)^2}>0
\end{align}
to hold for all $r\in\mathbb Z+\frac{1}{2}$. However, $\lambda_{-\frac{1}{2}}<0$. So there does {\it not} exist unitary complementary series with odd parity. The physical reason for the absence of these representations is that the principal series representation $\CP^{-}_{\frac{1}{2}+i\mu}$ describes a Dirac fermion of mass $\mu$ in $\text{dS}_2$. The  Wick rotation  $\mu\to i\mu$ yields a fermion with an imaginary mass, spoiling the hermiticity of the Dirac action.

\item When $\Delta\in\mathbb N+\frac{1}{2}$, the vector space spanned by all $|r\rangle$ is reducible, c.f. diagram (\ref{Delta=1odd}), and contains two  {\it irreducible} $\mathfrak{sl}(2,\mathbb R)$-invariant subspaces:
\begin{align}
&\{|r\rangle\}_{r\ge \Delta}: \text{carries a lowest-weight representation}\,\, \CD^{-+}_\Delta\nonumber\\
&\{|r\rangle\}_{r\le -\Delta}: \text{carries a highest-weight representation}\,\, \CD^{--}_\Delta
\end{align}
where the first ``-'' in the superscripts  of $\CD^{-\pm}_\Delta$ indicates that these representations are parity odd.

\begin{figure}[H]
\centering
\captionsetup{width=.85\linewidth}
\begin{tikzcd}[column sep=small]
\cdots \arrow[bend left=50]{r}{L_+}& \left|-\frac{7}{2}\right\rangle  \arrow[bend right=-50]{l}{L_-} \arrow[bend left=50]{r}{L_+} &\left|-\frac{5}{2}\right\rangle  \arrow[bend right=-50]{l}{L_-} \arrow[bend left=50]{r}{L_+} & \left|-\frac{3}{2}\right\rangle  \arrow[bend right=-50]{l}{L_-} & \left|-\frac{1}{2}\right\rangle \arrow[bend right=-50]{l}{L_-} \arrow[bend left=50]{r}{L_+}  & \left|\frac{1}{2}\right\rangle  \arrow[bend right=-50]{l}{L_-} \arrow[bend left=50]{r}{L_+} & \left|\frac{3}{2}\right\rangle  \arrow[bend left=50]{r}{L_+} & \left|\frac{5}{2}\right\rangle  \arrow[bend right=-50]{l}{L_-} \arrow[bend left=50]{r}{L_+} & \left|\frac{7}{2}\right\rangle \arrow[bend right=-50]{l}{L_-} \arrow[bend left=50]{r}{L_+} &  \cdots  \arrow[bend right=-50]{l}{L_-} 
\end{tikzcd}
\caption{The action of $L_\pm$ for $\Delta=\frac{3}{2}$. $ \{|\frac{3}{2}\rangle, |\frac{5}{2}\rangle, \cdots\}$ furnishes an irreducible lowest-weight representation and $\{|-\frac{3}{2}\rangle, |-\frac{5}{2}\rangle, \cdots \}$ furnishes an irreducible highest-weight representation.}
\label{Delta=1odd}
\end{figure}
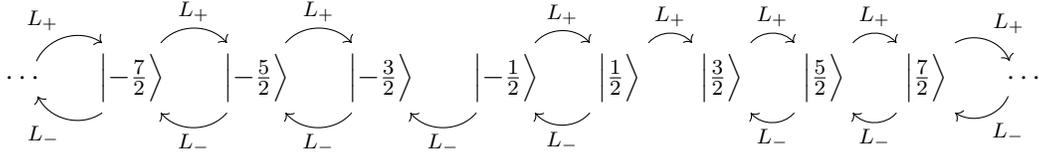
It is clear that all $\lambda_r$ are positive when restricted to these two subspaces. Thus $\CD^{-\pm}_\Delta$ are also UIRs, belonging to discrete series.
In each $\CD^{-\pm}_\Delta$, we choose the normalization to be 
\begin{align}\label{discketin}
\langle r|r\rangle_{\CD^{-\pm}_\Delta}=\frac{\Gamma(\pm r+\bar\Delta)}{\Gamma(\pm r+\Delta)}
\end{align}
\item When $\Delta\in\mathbb Z_{<0}+\frac{1}{2}$, the vector space spanned by all $|r\rangle$ contains only one {\it irreducible} $\mathfrak{sl}(2,\mathbb R)$-invariant subspace which furnishes a finite dimensional nonunitary representation. Fig. (\ref{Delta=-1odd}) is an example for $\Delta=-\frac{1}{2}$.
\begin{figure}[H]
\centering
\captionsetup{width=.85\linewidth}
\begin{tikzcd}[column sep=small]
\cdots \arrow[bend left=50]{r}{L_+}&  \left|-\frac{5}{2}\right\rangle  \arrow[bend right=-50]{l}{L_-} \arrow[bend left=50]{r}{L_+} & \left|-\frac{3}{2}\right\rangle \arrow[bend right=-50]{l}{L_-}  \arrow[bend left=50]{r}{L_+} & \left|-\frac{1}{2}\right\rangle   \arrow[bend left=50]{r}{L_+} & \left|\frac{1}{2}\right\rangle \arrow[bend right=-50]{l}{L_-}  & \left|\frac{3}{2}\right\rangle  \arrow[bend right=-50]{l}{L_-} \arrow[bend left=50]{r}{L_+} & \left|\frac{5}{2}\right\rangle \arrow[bend right=-50]{l}{L_-} \arrow[bend left=50]{r}{L_+} &  \cdots  \arrow[bend right=-50]{l}{L_-} 
\end{tikzcd}
\caption{The action of $L_\pm$ when $\Delta=-\frac{1}{2}$. In this case, $|\pm\frac{1}{2}\rangle$  span the spin-$\frac{1}{2}$ representation of $\SL(2,\mathbb{R})$.}
\label{Delta=-1odd}
\end{figure}
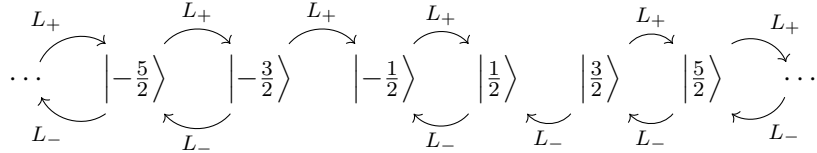
\end{itemize}
Altogether, the UIRs of $\SL(2,\mathbb R)$ (up to unitary equivalence) can be summarized as follows:
\begin{itemize}
\item The trivial representation.
\item For each $\Delta\in\frac{1}{2}+i\mathbb R_{ >0}$, there exist {\it two} inequivalent principal series representations $\CP_\Delta$ and $\CP^-_\Delta$,  with opposite parities. For $\Delta=\frac{1}{2}$, only the parity even one, i.e. $\CP_{\frac{1}{2}}$, exists. 
\item For each $\frac{1}{2}<\Delta<1$, there exists {\it one} complementary series representation $\CC_\Delta$, with an even parity.
\item For each $\Delta\in\frac{1}{2}\mathbb Z_{>0}$, there exist {\it two} inequivalent discrete series representations, which are complex conjugate of each other. When $\Delta\in\mathbb Z_{>0}$, the two representations $\CD^\pm_\Delta$ are parity even, and when $\Delta\in\mathbb N+\frac{1}{2}$, the two representations $\CD^{-\pm}_\Delta$ are parity odd.
\end{itemize}
In this list, the $\Delta=\frac{1}{2}$ point is very special because it represents three inequivalent UIRs, one in the principal series with an even parity, which describes a scalar field of mass $\frac{1}{2}$ in $\text{dS}_2$,  and two in the discrete series with an odd parity, which describe the left-handed and right-handed Majorana-Weyl fermions in $\text{dS}_2$.

\section{The explicit action of $\mathfrak{L}_N$ on $\psi^{(N)}_n(x)$}\label{mLapp}
In the section \ref{disd=1}, we have defined an operator $\mL_N: \CF^+_N\to \CF^+_{1-N}$ such that the composition $\partial_x^{2N-1}\circ\mL_N$ gives the identity operator on $ \CF^+_N$. In this appendix, we aim to compute the action of $\mL_N$ on the basis $\psi^{(N)}_n(x)$ of $\CF^+_{N}$, c.f. eq.(\ref{defmL})
\begin{align}\label{repeatn}
\left(\mL_N\psi^{(N)}_n\right)(x)=\frac{1}{\Gamma(2N-1)}\int_{\mathbb R+i\epsilon}\frac{dy}{2\pi i }\,(x-y)^{2(N-1)}\log(x-y)\psi^{(N)}_n(y)
\end{align}
\begin{figure}
\centering
\begin{subfigure}{.4\textwidth}
  \centering
  \includegraphics[width=1.0\linewidth]{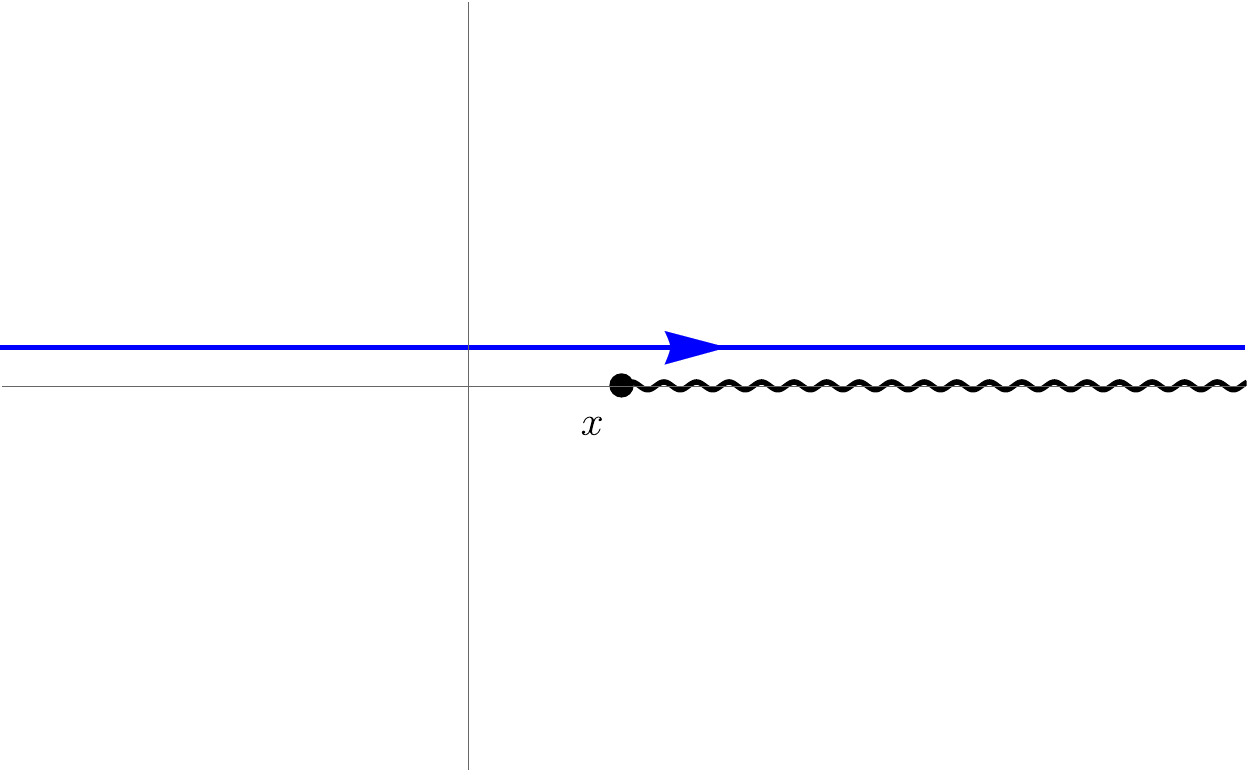}
  \caption{Original contour}
  \label{fig:original}
\end{subfigure}\,\,\,\,\,
\begin{subfigure}{.4\textwidth}
  \centering
  \includegraphics[width=1.0\linewidth]{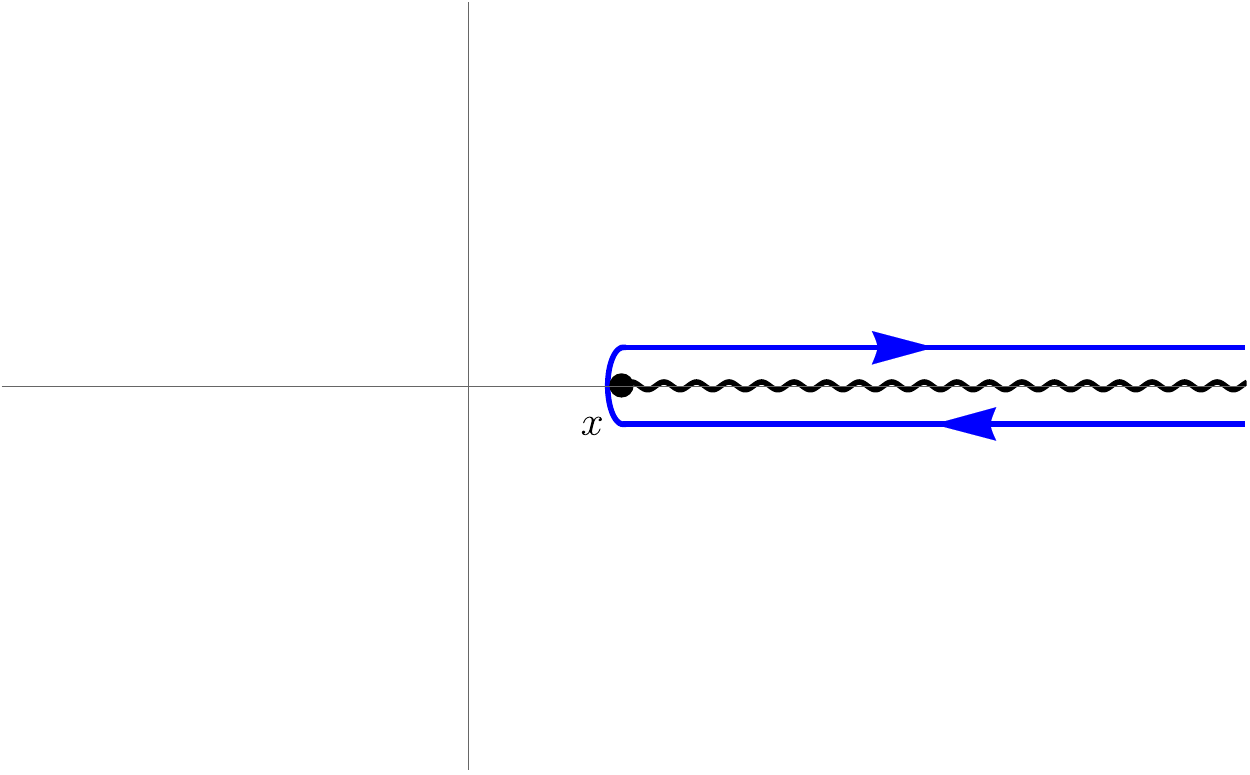}
  \caption{Deformed contour}
  \label{fig:deformed}
\end{subfigure}
\caption{The contour deformation for the $y$-integral in eq. (\ref{repeatn}). The black wavy line denotes the branch cut of $\log(x-y)$.}
\label{fig:contourdeform}
\end{figure}
As shown in the fig. (\ref{fig:contourdeform}), we can deform the contour to go around the branch cut and then $\log(x-y)$ is replaced by its discontinuity along the branch cut
\begin{align}
\left(\mL_N\psi^{(N)}_n\right)(x)=-\frac{2^N}{\sqrt{2\pi}\, \Gamma(2N-1)}\int_x^\infty dy\, (y-x)^{2(N-1)}\frac{(1-i y)^{n-N}}{(1+i y)^{n+N}}
\end{align} 
Next, we rewrite $(1-i y)^{n-N}$ as a polynomial of $1+iy$ and evaluate the  $y$-integral for each monomial by using integration by part repeatedly
\begin{align}
\left(\mL_N\psi^{(N)}_n\right)(x)&=\frac{(-)^{n-N+1}2^N}{\sqrt{2\pi}\Gamma(2N-1)}\sum_{k=0}^{n-N}(-2)^k\binom{n-N}{k}\int_x^\infty dy\, \frac{(y-x)^{2(N-1)}}{(1+iy)^{2N+k}}\nonumber\\
&=\frac{2^N i}{\sqrt{2\pi}}\frac{(-)^{n+1}}{1+i x}\sum_{k=0}^{n-N}\binom{n-N}{k}\frac{k!}{(2N+k-1)!}\left(\frac{-2}{1+i x}\right)^k
\end{align}
To compute the remaining series of $k$, we can extend the sum to $k=1-2N$ and subtract these extra terms by hand. The sum from $k=1-2N$ to $k=n-N$ is nothing but a binomial expansion.  Altogether, we obtain
\begin{align}\label{mLexplicit}
\left(\mL_N\psi^{(N)}_n\right)(x)=i(-)^{N+1}\frac{\Gamma(n-N+1)}{\Gamma(N+n)}\psi^{(1-N)}_n(x)+\text{Pol}_{N, n}(x)
\end{align}
where $\text{Pol}_{N, n}(x)$ is a polynomial in $x$ annihilated by $\partial_x^{2N-1}$
\begin{align}\label{extrapol}
\text{Pol}_{N, n}(x)=\frac{2^{N-1}(-)^{n+1}i}{\sqrt{2\pi}}\frac{\Gamma(n-N+1)}{\Gamma(N+n)}\sum_{k=0}^{2(N-1)}\binom{n+N-1}{2(N-1)-k}\left(\frac{1+i x}{-2}\right)^k
\end{align}
As a quick consistency check, combining eq. (\ref{intertmap}) and (\ref{mLexplicit}), we obtain the expected property
\begin{align}
\partial_x^{2N-1}\left(\mL_N\psi^{(N)}_n\right)(x)=\psi^{(N)}_n(x)
\end{align}

\section{Light scalars in $\text{dS}_2$}\label{light}
In this appendix, we  quantize a light scalar field $\phi$ of scaling dimension $\Delta=\frac{1}{2}+\nu, \nu\in(-\frac{1}{2},\frac{1}{2})$ in the conformal global coordinates of $\text{dS}_2$, along the lines of subsection \ref{QFTpicture}. Expand $\phi$ as follows 
\begin{align}
\phi=\sum_{n} \phi_n a_n+\phi_n^* a_n^\dagger
\end{align}
where the mode functions are chosen to be
\begin{align}
\phi_n=\frac{\Gamma(n+\Delta)}{\sqrt{2}}\, e^{-i n \, t}\, \mathbf{F} \left(\Delta, \bar\Delta, n+1,\frac{1}{1+e^{2 i t}}\right)\frac{e^{-int}}{\sqrt{2\pi}}
\end{align}
These modes are {\it not} unit normalized with respect to the Klein-Gordon inner product (\ref{KG})
\begin{align}
(\phi_n,\phi_m)_{\text{KG}}=\frac{\Gamma(n+\Delta)}{\Gamma(n+\bar\Delta)}\delta_{nm}
\end{align}
and hence the norm of single-particle states $|n\rangle_E=a_n^\dagger |E\rangle$ becomes
\begin{align}
_E\langle n|m\rangle_E=\frac{\Gamma(n+\bar\Delta)}{\Gamma(n+\Delta)}\delta_{nm}
\end{align}
Since we have changed the $\Gamma$-function factor in the definition of $\phi_n$ compared to the heavy scalar case, i.e. $\Gamma(n+\bar\Delta)\to \Gamma(n+\Delta)$, the action of $\CL_\pm$ on $\phi_n$ should change accordingly (but the action on $\phi^*_n$ remains the same)
\begin{align}
\CL_\pm \phi_n=(n\pm\bar\Delta)\phi_{n \pm 1}, \,\,\,\,\, \CL_\pm \phi_n^*=-(n\mp \bar\Delta)\phi^*_{n\mp 1}
\end{align}
which yields
\begin{align}\label{La}
[L_\pm, a_n]=-(n\mp \Delta) a_{n \mp 1},\,\,\,\,\, [L_\pm, a^\dagger_n]=(n\pm \Delta) a^\dagger_{n \pm 1}
\end{align}
$L_\pm$ satisfying (\ref{La}) are hermitian conjugate of each other. These equations also complete the $\so(1,2)$ action on the single-particle Hilbert space $\CH_\Delta=\text{span}\{|n\rangle_E\}_n$
\begin{align}\label{actonH1}
L_\pm |n\rangle_E=(n\pm \Delta) |n\pm 1\rangle_E, \,\,\,\,\, L_0|n\rangle_E=|n\rangle_E
\end{align}
With the inner product $_E\langle m|n\rangle_E=\frac{\Gamma(n+\bar\Delta)}{\Gamma(n+\bar\Delta)}\delta_{nm}$ and the $\so(1,2)$ action (\ref{actonH1}), we are able to identify the single-particle Hilbert space $\CH_\Delta$ as the complementary series representation $\CC_{\Delta}$ constructed in section \ref{direct}.

Next we read off the leading behaviors of $\phi$ near the future boundary 
\begin{align}\label{boundaryexp}
\phi\left(t=\frac{\pi}{2}-\delta, \varphi\right)\overset{\delta\to 0}{\approx}C_{\Delta} \, \delta^\Delta\CO_\Delta(\varphi)+C_{\bar\Delta} \, \delta^{\bar\Delta}\CO_{\bar\Delta}(\varphi)
\end{align}
where $ C_\Delta=2^{\nu} \, \frac{\Gamma(-2\nu)}{\Gamma(\bar\Delta)}, C_{\bar\Delta}=2^{-\nu} \, \frac{\Gamma(2\nu)}{\Gamma(\Delta)}$ and
\small
\begin{align}
\CO_\Delta(\varphi)= \sum_{n\in\mathbb Z} \frac{\Gamma(n+\Delta)}{\Gamma(n+\bar\Delta)}\left((-i)^{n+\Delta}\frac{e^{-in\varphi}}{\sqrt{2\pi}}a_n+\text{h.c.}\right), \,\,\,\,\, \CO_{\bar\Delta}(\varphi)= \sum_{n\in\mathbb Z}\left((-i)^{n+\bar\Delta}\frac{e^{-in\varphi}}{\sqrt{2\pi}}a_n+\text{h.c.}\right)
\end{align}
\normalsize
$\CO_\Delta$ and $\CO_{\bar\Delta}$ defined this way are primary operators of scaling dimension $\Delta$ and $\bar\Delta$ respectively, satisfying eq.
(\ref{manycomm}). Their two-point functions with respect to the Euclidean vacuum are
\begin{align}\label{many2pt1}
&\langle E|\CO_\Delta(\varphi_1)\CO_{\Delta}(\varphi_2)|E\rangle=\frac{N_{\bar\Delta}}{(1-\cos\varphi_{12})^\Delta}, \,\,\,\,\, \langle E|\CO_{\bar\Delta}(\varphi_1)\CO_{\bar\Delta}(\varphi_2)|E\rangle=\frac{N_{\Delta}}{(1-\cos\varphi_{12})^{\bar\Delta}} \nonumber\\
&\langle E|\CO_\Delta(\varphi_1)\CO_{\bar\Delta}(\varphi_2)|E\rangle=e^{-i\pi\nu}\delta(\varphi_{12}),\,\,\,\,\,\,\,\,\,\,\,\,\,\,\,\,\,\, \langle E|\CO_{\bar\Delta}(\varphi_1)\CO_{\Delta}(\varphi_2)|E\rangle=e^{i\pi\nu}\delta(\varphi_{12})
\end{align}
where the normalization constant $N_\Delta$ is defined in eq. (\ref{shadownormalize}). Define the single-particle excitation of $\CO_\Delta$ and $\CO_{\bar\Delta}$ as
\begin{align}
&|\Delta, \varphi\rangle\equiv \CO_{\Delta}(\varphi)|E\rangle=\sum_n i^{n+\Delta} \frac{\Gamma(n+\Delta)}{\Gamma(n+\bar\Delta)}\frac{e^{in\varphi}}{\sqrt{2\pi}}|n\rangle_E\nonumber\\
&|\bar\Delta, \varphi\rangle\equiv \CO_{\bar\Delta}(\varphi)|E\rangle=\sum_n i^{n+\bar\Delta}\frac{e^{in\varphi}}{\sqrt{2\pi}}|n\rangle_E
\end{align}
They can be used to decompose the identity operator on $\CH_\Delta$
\begin{align}
\mathbb{1}_{\CH_\Delta}=e^{-i\pi\nu}\int\, d\varphi|\Delta,\varphi\rangle\langle \bar\Delta, \varphi|=e^{i\pi\nu}\int\, d\varphi|\bar\Delta,\varphi\rangle\langle \Delta, \varphi|
\end{align}
which leads to the completeness of both $|\Delta, \varphi\rangle$ and $|\bar\Delta, \varphi\rangle$ in $\CH_\Delta$
\begin{align}
\mathbb{1}_{\CH_\Delta}&=\int d\varphi_1 d\varphi_2 |\Delta,\varphi_1 \rangle\langle \bar\Delta, \varphi_1|\bar\Delta,\varphi_2\rangle\langle \Delta, \varphi_2|\nonumber\\
&= N_\Delta\int d\varphi_1 d\varphi_2 \frac{1}{(1-\cos\varphi_{12})^{\bar\Delta}}|\Delta,\varphi_1 \rangle\langle \Delta, \varphi_2|\nonumber\\
&= N_{\bar\Delta}\int d\varphi_1 d\varphi_2 \frac{1}{(1-\cos\varphi_{12})^{\Delta}}|\bar\Delta,\varphi_1 \rangle\langle \bar\Delta, \varphi_2|
\end{align}
Therefore $|\Delta, \varphi\rangle$ and $|\bar\Delta, \varphi\rangle$ are simply two different basis of the single-particle Hilbert space.

Switching to the stereographic coordinate of $S^1$, which is shown in fig. (\ref{fig:qiuji}), we define the $|x\rangle$ basis as
 \begin{align}
 |x\rangle=\left(\frac{2}{1+x^2}\right)^{\bar\Delta} |\bar\Delta, \varphi(x)\rangle
 \end{align}
The $\so(1,2)$ action on this basis is given by  eq. (\ref{onket}). In addition, the inner product of the $|x\rangle$-basis can be derived from (\ref{many2pt1})
  \begin{align}
 \langle x|y\rangle=\frac{2^{\bar\Delta}}{(1+x^2)^{\bar\Delta}}\frac{2^{\bar\Delta}}{(1+y^2)^{\bar\Delta}}\frac{N_\Delta}{(1-\cos(\varphi(x)-\varphi(y)))^{\bar\Delta}}=\frac{2^{\bar\Delta}N_\Delta}{(x-y)^{2\bar\Delta}}
 \end{align}
 which is consistent with eq. (\ref{class12}) for $c=N_\Delta$ and is also the same as the shadow transformation $S_\Delta(x,y)$, c.f. (\ref{shadownormalize}).

\section{Induced representations}\label{reviewinduce}
In this appendix, we review  the method of induced representations and its applications  in the $\SO(1, d+1)$ case. Let $H$ be a subgroup of $G$ and let $\rho$ be a representation of $H$ on some vector space $V$. The  \textbf{induced representation} $\text{ind}^G_H \rho$ of $G$ operates on the following space of equivariant maps from $G$ to $V$
\begin{align}
\text{Map}_H(G, V)\equiv \left\{ \Psi: G\to V \vert \Psi(g h)=\rho(h)^{-1} \Psi(g), \,\, g\in G, h\in H\right\}
\end{align}
by
\begin{align}
(g \circ \Psi)(g')\equiv \Psi \left(g^{-1}g'\right)
\end{align}
In the case of $G=\SO(1, d+1)$, we take $H$ to be the minimal parabolic subgroup $\tS\equiv \tN\tA\tM$ and $V$ to be the space $\text{P}_s[z^i]$ of homogeneous polynomials in a null vector $z^i$ of degree $s$, which is also equivalent to the space of symmetric and traceless tensors of rank $s$. The representation $\rho=\rho_{\Delta, s}$ of $\tS$ is chosen to be 
\begin{align}
\rho_{\Delta, s}\left(n e^{\lambda D} m \right) f(z)= e^{-\lambda\Delta} f \left( \rho_1(m)^{-1} z\right)
\end{align}
where $n\in \tN$, $m\in\tM$ and $\rho_1(m)$ is the fundamental representation of $\tM=\SO(d)$, i.e. 
\begin{align}
\rho_1\left(e^{\frac{1}{2}\theta^{ij}M_{ij}}\right): z^k\to z^k+\theta^{kj} z_j+\CO(\theta^2)
\end{align}
The induced representation $\text{ind}^{G}_{\tS} \rho_{\Delta, s}$ then acts on the space of polynomial-valued functions $\Psi(g, z)$ that satisfy
\begin{align}\label{covcon1}
\Psi(gs, z)=\rho_{\Delta, s}(s)^{-1} \Psi(g, z), \,\,\,\,\, s\in \tS
\end{align}
Due to the Bruhat decomposition of $\SO(1, d+1)$ \cite{Dobrev:1977qv} and the equivariant condition (\ref{covcon1}), such a function $\Psi(g, z)$ is completely fixed by its value on $\ttN$,  which is geometrically a flat space $\mathbb R^d$.\footnote{The Bruhat decomposition says that $\SO(1, d+1)$, up to a lower dimensional submanifold, can be written as a product $\ttN\tS$ . This lower dimensional manifold corresponds to the {\it infinity} point of $\mathbb R^d$ when we quotient out the $\tS$ dependence.} Define $\psi(x, z)\equiv \Psi(e^{x\cdot P}, z)$ and we will show that the (infinitesimal version of) induced representation $\text{ind}^{G}_{\tS} \rho_{\Delta, s}$ agrees with the CFT-type construction, c.f. eq.(\ref{algebrarepd}).
\begin{itemize}
\item Translations:
\begin{align}
\left(e^{a\cdot P}\circ\psi\right)(x, z)=\left(e^{a\cdot P}\circ\Psi\right)(e^{x\cdot P}, z)=\Psi(e^{(x-a)\cdot P}, z)=\psi(x-a, z)
\end{align}
Derivative with respect to $a^i$ yields $\boxed{P_i \psi(x, z)=-\partial_i \psi(x, z)}$
\item Dilatations:
\begin{align}
\left(e^{\lambda D}\circ\psi\right)(x, z)=\Psi(e^{-\lambda D} e^{x\cdot P}, z)
\end{align}
Using $e^{-\lambda D} P_i e^{\lambda D}=e^{-\lambda} P_i$, we obtain
\begin{align}
\left(e^{\lambda D}\circ\psi\right)(x, z)=\rho_{\Delta, s}(e^{\lambda D})\psi(e^{-\lambda} x, z)=e^{-\lambda \Delta}\psi(e^{-\lambda} x, z)
\end{align}
which leads to $\boxed{D\psi(x, z)=-(x\cdot \partial_x+\Delta)\psi(x, z)}$.
\item Rotations $m=e^{\frac{1}{2}\theta^{ij}M_{ij}}$:
\begin{align}
m\circ \psi(x, z)= \rho_{\Delta, s}(m) \Psi(m^{-1} e^{x\cdot P} m, z)=\Psi(e^{x_m\cdot P},\rho_1(m)^{-1} z)
\end{align}
where $x_m$ is defined via $m^{-1} e^{x\cdot P} m= e^{x_m \cdot P}$. Infinitesimally, we have 
\begin{align}
x_m^i=x^i-\theta^{ij}x_j+\CO(\theta^2), \,\,\,\,\, \rho_1(m)^{-1} z^i=z^i-\theta^{ij}z_j+\CO(\theta^2)
\end{align}
and hence $\boxed{M_{ij}\psi(x, z)=(x_i\partial_j-x_j\partial_i+z_i\partial_{z^j}-z_j\partial_{z^i})\psi(x, z)}$.
\item Special conformal transformations:
\begin{align}
\left(e^{b\cdot K}\circ\psi\right)(x, z)= \Psi(e^{-b\cdot K} e^{x\cdot P}, z)
\end{align}
For our purpose, we can replace $e^{-b \cdot K}$ by $1-b\cdot K$ and then use 
\begin{align}
-e^{-x\cdot P}\,b\cdot K \,e^{x\cdot P}=-b\cdot K-2 x\cdot b \, D+2b^i x^j M_{ij}+(x^2 b^i-2 x\cdot b\, x^i)P_i
\end{align}
which yields 
\begin{align}
\left(e^{b\cdot K}\circ\psi\right)(x, z)&=\Psi\left(e^{((1-2x\cdot b)x+x^2b)\cdot P} e^{-b\cdot K} e^{-2x\cdot b D} e^{2b_i x_j M_{ij}}, z\right)+\CO(b^2)\nonumber\\
&=e^{-2\Delta x\cdot b}\psi((1-2 x\cdot b)x+x^2 b, 2 \, x\cdot z \, b-2\, b\cdot z\, x)+\CO(b^2)
\end{align}
Take derivative with respect to $b^i$ and we obtain
\begin{align}
\boxed{K_i \psi(x, z)=\left(x^2 \partial_i-2x_i(x\cdot\partial_x+\Delta)+2 (x\cdot z\, \partial_{z^i}-z_i \,x\cdot \partial_{z})\right)\psi(x, z)}
\end{align}
\end{itemize}

\section{Irreducibility of $\CF_\Delta$}\label{irrCF}
In this appendix we give an elementary proof about the irreducibility of $\CF_\Delta$ for a generic $\Delta$.  The argument heavily relies on the result of subsection \ref{content} where we have learned that the $\SO(d+1)$ content of $\CF_\Delta$ consists of all single-row representations $\SO(d+1)$. Assuming the existence of a nontrivial $\SO(1, d+1)$ subspace $\CW_\Delta$ of $\CF_{\Delta}$, it admits an $\SO(d+1)$ decomposition of the following form
\begin{align}
\CW_\Delta=\bigoplus_{n\in\sigma} \mY_n
\end{align}
where $\sigma$ is a nontrivial subset of $\mathbb N$ and $\mY_n$ is the spin-$n$ representation of $\SO(d+1)$. We will show that this assumption is invalid because given an arbitrary wavefunction $\psi(x)$ in certain $Y_n$, by acting  dilatation operator on $\psi(x)$ repeatedly, it can get components in any $\SO(d+1)$ content of $\CF_\Delta$.

Pick any $\psi(x)\in\CF_\Delta$. According to the analysis of  subsection \ref{content}, it is a function $\hat\psi$ on $S^d$ (in stereographic coordinate) multiplied by a Weyl factor, i.e. $\psi(x)=\left(\frac{2}{1+x^2}\right)^\Delta\hat\psi(x)$.
Switching to spherical coordinates $x^i =\omega^i\cot\frac{\theta}{2}, \omega^i\in S^{d-1}$, the action of  dilatation operator on $\hat\psi(\theta,\omega)$ becomes 
\begin{align}
D\,\hat\psi(\theta, \omega)=\left(\sin\theta\partial_\theta+\Delta\cos\theta\right)\hat\psi(\theta, \omega)
\end{align}
The function $\hat\psi$ can be expanded in terms of spherical harmonics on $S^d$
\begin{align}
Y_{n\ell\bm m}(\theta, \omega)=\CN_{n\ell}\,  \sin^\ell\theta\, C^{\ell+\frac{d-1}{2}}_{n-\ell}(\cos\theta) Y_{\ell\bm m}(\omega), \,\,\,\, n\ge \ell
\end{align}
where $Y_{\ell\bm m}(\omega)$ denote the normalized spherical harmonics on $S^{d-1}$ and $\CN_{n\ell}$ is a constant such that $Y_{n\ell \bm m}$ is normalized
\begin{align}
\CN_{n\ell}=2^{\ell-1+\frac{d}{2}}\,\Gamma\left(\ell+\frac{d-1}{2}\right)\sqrt{\frac{(n-\ell)!(n+\frac{d-1}{2})}{\pi\, \Gamma(n+\ell+d-1)}}
\end{align}
For a fixed $n$, all $\left(\frac{2}{1+x^2}\right)^\Delta Y_{n\ell\bm m}(\theta, \omega)$ span the $\mY_n$ part of $\CF_\Delta$. Next, we need to figure out the action of $D$ on all the spherical harmonics. This computation can be done by using the following recurrence relations of Gegenbauer polynomials
\begin{align}
&\partial_x C^\lambda_n (x)=2\lambda C^{\lambda+1}_{n-1}(x), \,\,\,\,\, x \, C^\lambda_n(x)=\frac{n+1}{2(n+\lambda)}C^\lambda_{n+1}(x)+\frac{n-1+2\lambda}{2(n+\lambda)}C^\lambda_{n-1}(x)\nonumber\\
&(1-x)^2C^{\lambda+1}_{n-1}(x)=\frac{n+2\lambda}{2\lambda}x C^\lambda_n(x)-\frac{(n+1)}{2\lambda}C^\lambda_{n+1}(x)
\end{align}
which yield
\small
\begin{align}
\boxed{D Y_{n\ell\bm m}=(\Delta+n)\, \sqrt{\frac{(n+1-\ell)(n+\ell+d-1)}{(2n+d-1)(2n+d+1)}} Y_{n+1, \ell\bm m}+(1-\bar\Delta-n)\, \sqrt{\frac{(n-\ell)(n+\ell+d-2)}{(2n+d-3)(2n+d-1)}} Y_{n-1, \ell\bm m}}
\end{align}
The coefficients of $Y_{n\pm 1,\ell\bm m}$ are nonvanishing except $\Delta\in\{d, d+1, d+2,\cdots\}\cup\{0,-1,-2,\cdots\}$. Due to the reasoning given at the beginning of this appendix, the representation $\CF_\Delta$ is irreducible when $\Delta$ is away form these integers. For $\Delta\in\{d, d+1, d+2,\cdots\}$, the subspace with $\SO(d+1)$ content $\bigoplus_{n\ge 1-\bar\Delta} \mY_n$ is mapped to itself by $D$ and this claim also holds for any  $L_{0, i}$, because $L_{0,i}$ can be related to $D=L_{0,d+1}$ by $\SO(d+1)$ conjugations. Therefore, this subspace is irreducible with respect to $\SO(1, d+1)$. Using the same argument, we can claim that when $\Delta\in\{0,-1,-2,\cdots\}$, the finite dimensional subspace $\bigoplus_{0\le n\le -\Delta}\mY_n$ is irreducible, carrying the spin $(-\Delta)$ representation of $\SO(1,d+1)$. These finite dimensional representations are nonunitary except the $\Delta=0$ one which is the trivial representation.

\section{Normalizability of the inner product on $\CU_{s,t}$}\label{normaliz}
In the subsection \ref{Exc}, we introduced an inner product for the exceptional series representation $\CU_{s,t}=\CF_{1-t,s}/\Im(z\cdot\partial_x)^{s-t}$:
\begin{align}\label{innerUst}
(\psi,\varphi)_{\CU{s,t}}=\int\, d^d x_1 d^d x_2\psi(x_1)^*_{i_1\cdots i_s}S^+_{1-t, s}(x_{12})_{i_1\cdots i_2, j_1\cdots j_2}\varphi(x_2)_{j_1\cdots j_s}
\end{align}
where $\psi_{i_1\cdots i_s}, \varphi_{j_1\cdots j_s}\in\CF_{1-t,s}$ and $S^+_{1-t, s}$ defines the shadow transformation $\CS^+_{1-t,s}$ from $\CF_{1-t,s}$ to $\CF_{d+t-1,s}$, whose Fourier transformation is given by eq. (\ref{Spm}) in the index-free formalism:
\begin{align}
S^+_{1-t, s}(p; z, w)=p^{d+2t-2}\sum_{\ell =t+1}^s\frac{(\ell-t)_{s-\ell}}{(d+t+\ell -2)_{s-\ell }}\Pi^{s \ell}(\hat p; z,w)
\end{align}
However, we did not justify the normalizability of this inner product. In this appendix, we will prove that the inner product (\ref{innerUst}) is normalizable, by a detailed study of the asymptotic behaviors of wavefunctions in $\CF_{1-t,s}$. 

By definition, the large $x$ behavior of $\psi(x,z)\in\CF_{1-t,s}$ is 
\begin{align}
\psi(x, z)\stackrel{x\to\infty}{\approx}\sum_{k=0}^\infty x^{2(t-1)} C_{k} \left( \tilde x^i,R(x)^i_{\,\, j} z^j\right)
\end{align}
where $C_k(u,v)$ is a homogeneous polynomial of degree $k$ in the first argument and degree $s$ in the second. With some simple power counting, one can learn that the first $t$ polynomials, i.e. $\{C_0, C_1, \cdots C_{t-1}\}$, might lead to divergence at small $p$ in (\ref{innerUst}). We will show this does not happen because  all $x^{2(t-1)}C_k$ with $0\le k\le t-1$ lie in the image of $(z\cdot \partial_x)^{s-t}$, which is annihilated by $\CS^+_{1-t,s}$. In particular, it suffices to show the following proposition:
\begin{proposition}\label{Qpure}
Let $P_k(x^i)$ be an arbitrary degree-$k$ homogeneous polynomial in $x^i$ and define $\hat P_k(x,v)=x^{2(t-1-k)}P_k(x^i) v^{i_1}\cdots v^{i_s}$, where $t\in\{0,1,\cdots s-1\}$ and $v^i=R(x)^i_{\,\,j} z^j=\frac{2 x\cdot z}{x^2}x^i -z^i$. When $k=0,1,\cdots t-1$, $\hat P_k(x,v)$ can be expressed as $\hat P_k(x,v)=(z\cdot\partial_x)^{s-t} Q_k(x,v)$ for some polynomial $Q_k(x,v)$.
\end{proposition} 
\noindent{}In order to prove the proposition \ref{Qpure}, we need some lemmas:
\begin{lemma}\label{xtov}
\begin{align}\label{xtov1}
(z\cdot \partial_x)^n\frac{x^{i_1}\cdots x^{i_n}}{x^2}=\frac{(-)^n \, n!}{x^2}v^{i_1}\cdots v^{i_n}
\end{align}
\end{lemma}
\noindent{}\textbf{Proof}: Since the L.H.S is a totally symmetric tensor, it suffices to compute $(z\cdot \partial_x)^n\frac{(x\cdot \xi)^n}{x^2}$, where $\xi_i$ is an auxiliary vector:
\begin{align}
(z\cdot \partial_x)^n\frac{(x\cdot \xi)^n}{x^2}&=\sum_{\ell=0}^n\binom{n}{\ell}(z\cdot \partial_x)^{n-\ell} (x\cdot\xi)^n\,(z\cdot \partial_x)^{\ell} \frac{1}{x^2}\nonumber\\
&=\frac{n!}{x^2}\sum_{\ell=0}^n\binom{n}{\ell} (z\cdot\xi)^{n-\ell}\left(\frac{-2x\cdot z\, x\cdot\xi}{x^2}\right)^\ell=\frac{(-)^n \,n!}{x^2}(v\cdot \xi)^n
\end{align}
Stripping off $\xi_i$, we recover eq. (\ref{xtov1}).

\begin{lemma}\label{zxmove}
For a degree-$k$ polynomial $P_k(x)$, we have  
\begin{align}\label{Pkg}
P_k(x) (z\cdot\partial_x)^{n+k}g(x)=(z\cdot\partial_x)^n\left\{\sum_{\ell=0}^k (-)^\ell\binom{n+\ell-1}{\ell}(z\cdot\partial_x)^\ell P_k(x)\,  (z\cdot\partial_x)^{k-\ell} g(x)\right\}
\end{align}
\end{lemma}
\noindent{}\textbf{Proof}: Define 
\begin{align}
A(\bm a)=\sum_{\ell=0}^k a_\ell (z\cdot\partial_x)^\ell P_k(x)\,  (z\cdot\partial_x)^{k-\ell} g(x)
\end{align}
We want to find $\bm a=(a_0, a_1,\cdots, a_k)$ such that $A(\bm a)$ is mapped to $P_k(x) (z\cdot\partial_x)^{n+k}g(x)$ under $(z\cdot\partial_x)^n$. Let's first act $z\cdot\partial_x$ on $A(\bm a)$ and expand the result as follows
\begin{align}
z\cdot\partial_x A(\bm a)=\sum_{\ell=0}^k a^{(1)}_\ell (z\cdot\partial_x)^\ell P_k(x)\,  (z\cdot\partial_x)^{k-\ell+1} g(x)
\end{align}
where $a^{(1)}_\ell=a_{\ell}+a_{\ell-1}$ for $1\le\ell\le k $ and $a_0^{(1)}=a_0$. The relation between $\bm a$ and $\bm a^{(1)}$ can be captured by a $(k+1)\times (k+1)$ matrix $\Lambda$, i.e. $\bm a^{(1)}=(\mathbb 1+\Lambda)\,  \bm a$, where 
\begin{align}
\Lambda=
\begin{pmatrix} 
0&0& 0&\cdots &0& 0\\ 1&0&0&\cdots & 0&0\\ 0&1&0&\cdots &0&0\\ \cdots&\cdots&\cdots&\cdots&\cdots&\cdots\\ 0&0&0&\cdots&1&0
\end{pmatrix}
\end{align}
Then finding the coefficients $(a_0, a_1,\cdots a_k)$ becomes a pure linear algebra problem
\begin{align}
(1+\Lambda)^n \begin{pmatrix} a_0\\ a_1\\ \cdots \\a_k\end{pmatrix}=\begin{pmatrix} 1\\ 0\\ \cdots \\0\end{pmatrix}
\end{align}
Since $(1+\Lambda)$ is invertible, the desired $\bm a$ exists and is given by the first column of  $(1+\Lambda)^{-n}$. The matrix $(1+\Lambda)^{-n}$ can be computed by a formal Taylor expansion and the Taylor series truncates because $\Lambda^{k+1}=0$:
\begin{align}\label{invLam}
(1+\Lambda)^{-n}=\sum_{\ell=0}^k (-)^\ell \binom{n+\ell-1}{\ell}\Lambda^\ell
\end{align}
Combining (\ref{invLam}) and $(\Lambda^\ell)_{i,1}=\delta_{i,\ell+1}$, we obtain
\begin{align}
a_\ell=(-)^\ell \binom{n+\ell-1}{\ell}
\end{align}
and this finishes our proof of (\ref{Pkg}).

Now we start proving the proposition \ref{Qpure}. Using lemma \ref{xtov}, we  rewrite $\hat P_k (x,v)$ as 
\begin{align}
\hat P_k (x,v)=\frac{(-)^{s-t+k}}{(s-t+k)!}P_k(x)(x^2 v^{i_1})\cdots (x^2 v^{i_{t-k}})(z\cdot\partial_x)^{s-t+k}\frac{x^{i_{t+1-k}}\cdots x^{i_s}}{x^2}
\end{align} 
where $x^2 v^{i_j}$ can be moved to the right of $(z\cdot\partial_x)^{s-t+k}$ since $z\cdot\partial_x (x^2 v^i)=0$
\begin{align}
\hat P_k (x,v)=P_k(x)(z\cdot\partial_x)^{s-t+k}g(x,v), \,\,\,\,\, g(x, v)=\frac{(-)^{s-t+k}}{(s-t+k)!}x^{2(t-k-1)}v^{i_1} v^{i_{t-k}}x^{i_{t+1-k}}\cdots x^{i_s}
\end{align}
Then proposition \ref{Qpure} immediately follows from lemma \ref{zxmove}.

\section{Computation of the decorated Harish-Chandra characters }\label{fullchar}
In section \ref{comchar}, we have computed the Harish-Chandra characters $\Theta_{\CF_{\Delta,s}}(g)=\Tr_{\CF_{\Delta,s}}g$ for $g$ sitting on the one-dimensional subgroup generated by the boost $D$. In this appendix, we will consider more general $g$ by also turning on angular momenta in the trace $\Tr_{\CF_{\Delta,s}}$, i.e. 
\begin{align}
\Theta_{\CF_{\Delta,s}}(q, \bm u)=\Tr_{\CF_{\Delta,s}} \left(q^D \prod_{i=1}^r e^{u_i J_i}\right)
\end{align}
where $r=\floor*{\frac{d}{2}}$,  and $J_i=M_{2i-1,i}$ are generators of the Cartan subalgebra of $\SO(d)$.

Let's start from the $s=0$ case. Following the derivation in section \ref{comchar}, we first conjugate $q^D \prod_{i=1}^r e^{u_i J_i}$ by a special conformal transformation
\begin{align}
\Theta_{\CF_{\Delta}}(q, \bm u)=\Tr_{\CF_{\Delta,s}} \left(e^{-b\cdot K} g_0 e^{b\cdot K}\right), \,\,\,\,\, g_0\equiv q^D \prod_{i=1}^r e^{u_i J_i}
\end{align}
where $b^i\in\mathbb R^d$ is an arbitrary auxiliary vector. Then we need to compute the action of $e^{-b\cdot K} g_0 e^{b\cdot K}$ on $|x\rangle$. With some straightforward but tedious computations, we obtain 
\begin{align}\label{conjg0}
e^{-b\cdot K} g_0 e^{b\cdot K} |x\rangle=\Omega(q,\bm u, x, b)^{\bar\Delta} |\tilde x(q, \bm u, x, b)\rangle
\end{align}
where 
\begin{align}
&\Omega(q,\bm u, x, b)=\frac{q}{1-2\,x\cdot b+b^2 x^2+2\,q(b\cdot x_{\bm u}-b\cdot b_{\bm u}x^2)+b^2 q^2 x^2}\nonumber\\
&\tilde x^i(q,\bm u, x, b)=\Omega(q,\bm u, x, b)(x^i_{\bm u}+x^2(q \,b^i-b^i_{\bm u}))
\end{align}
Here $x_{\bm u}^i$ ($b^i_{\bm u}$) is  related to  $x^i$ ($b^i$) by a rotation associated to $\prod_{i=1}^r e^{u_i J_i}$. More explicitly, $x_{\bm u}^i=R(\bm u)^i_{\,\,j} x^j$  where $R(\bm u)$ is a block diagonal matrix
\begin{align}
R(\bm u)=\begin{cases}\text{diag}\left\{\begin{pmatrix}\cos u_1 & \sin u_1\\ -\sin u_1& \cos u_1\end{pmatrix}, \begin{pmatrix}\cos u_2 & \sin u_2\\ -\sin u_2& \cos u_2\end{pmatrix},\cdots,\begin{pmatrix}\cos u_r & \sin u_r\\ -\sin u_r& \cos u_r\end{pmatrix}\right\}, &d=2r\\ \text{diag}\left\{\begin{pmatrix}\cos u_1 & \sin u_1\\ -\sin u_1& \cos u_1\end{pmatrix}, \begin{pmatrix}\cos u_2 & \sin u_2\\ -\sin u_2& \cos u_2\end{pmatrix},\cdots,\begin{pmatrix}\cos u_r & \sin u_r\\ -\sin u_r& \cos u_r\end{pmatrix},1\right\},  &d=2r+1\end{cases}
\end{align}
Given the action (\ref{conjg0}), the character $\Theta_{\CF_{\Delta}}(q, \bm u)$ can be expressed as 
\begin{align}
\Theta_{\CF_{\Delta}}(q, \bm u)=\int d^d x\, \Omega(q,\bm u, x, b)^{\bar\Delta}\delta^d(\tilde x(q,\bm u, x, b)-x)
\end{align}
This integral is localized to the two fixed points of the map $x\to \tilde x(q,\bm u, x, b)$, namely $x^i=0$ and $x^i=\frac{b^i}{b^2}$. At these points, the scaling factor $\Omega(q,\bm u, x, b)$ becomes 
\begin{align}
\Omega(q,\bm u, 0, b)=q, \,\,\,\,\,\Omega(q,\bm u, b^i/b^2, b)=q^{-1}
\end{align}
The remaining task is to figure out the Jacobian determinant associated to the map $x^i\to \tilde x^i-x^i$ at these fixed points. At $x^i=0$, the Jacobian matrix (denoted by $J_1$) is simply $qR(\bm u)^i_{\,\, j}-\delta^i_j$ and hence the corresponding Jacobian  is 
\begin{align}
|\det J_1|=\prod_{i=1}^r(1-2q\cos u_i+q^2)\times\begin{cases} 1, & d=2r\\ |1-q|, & d=2r+1\end{cases}
\end{align}
The Jacobian matrix at $x^i=\frac{b^i}{b^2}$ (denoted by $J_2$) takes a more complicated form 
\begin{align}\label{jac2}
J_2=\frac{1}{q} (1-2B)R(\bm u)(1-2B)-\mathbb{1}_{d}
\end{align}
where  $\mathbb{1}_{d}$ is the $d\times d$ identity matrix and $B^{i}_{\,\,j}=\frac{1}{b^2} b^i b_j$. Although $J_2$ depends on $b^i$ explicitly, its determinant does not. Using $B^2=B$, one can immediately see that $1-2B$ is an involutory matrix, which implies 
\begin{align}
|\det J_2|=|\det(q^{-1}R(\bm u)-\mathbb 1_d)|=\prod_{i=1}^r(1-2q^{-1}\cos u_i+q^{-2})\times\begin{cases} 1, & d=2r\\ |1-q^{-1}|, & d=2r+1\end{cases}
\end{align}
Altogether, the decorated character $\Theta_{\CF_{\Delta}}(q, \bm u)$ is 
\begin{align}
\Theta_{\CF_{\Delta}}(q, \bm u)&=\frac{q^{\bar\Delta}}{|\det J_1|}+\frac{q^{-\bar\Delta}}{|\det J_2|}=\frac{q^\Delta+q^{\bar\Delta}}{P_d(q, \bm u)}
\end{align}
where
\begin{align}
P_d(q, \bm u)=\prod_{i=1}^r(1-e^{iu_i}q)(1-e^{-iu_i}q)\times\begin{cases} 1, & d=2r\\ |1-q|, & d=2r+1\end{cases}
\end{align}
Compared to the undecorated character $\Theta_{\CF_{\Delta}}(q)$, c.f. eq. (\ref{chartheta1}), the only difference is that the denominator $|1-q|^d$ gets replaced by $P_d(q, \bm u)$. This result agrees with \cite{Basile:2016aen}. 

The derivation above can be straightforwardly generalized to spinning representations $\CF_{\Delta, s}$. We simply present the result here
\begin{align}
\Theta_{\CF_{\Delta,s}}(q, \bm u)=\Theta^{\SO(d)}_{\mY_s}(\bm u) \frac{q^\Delta+q^{\bar\Delta}}{P_d(q, \bm u)}
\end{align}
where $\Theta^{\SO(d)}_{\mY_s}(\bm u)$ is the $\SO(d)$ character corresponding to the spin-$s$ representation, defined as the trace of $\prod_{i=1}^r e^{ u_i J_i}$ over the spin-$s$ representation space (which, for example, is carried by $|0\rangle_{i_1\cdots i_s}$ in our settings). Some lower $d$ examples of $\Theta^{\SO(d)}_{\mY_s}(\bm u)$ are
\begin{align}
\Theta^{\SO(3)}_{\mY_s}(u_1)=\sum_{n=-s}^s e^{i n u_1}=\frac{\sin((s+\frac{1}{2})u_1)}{\sin(\frac{1}{2}u_1)}
\end{align}
\begin{align}
\Theta^{\SO(4)}_{\mY_s}(u_1,u_2)=\frac{\cos((s+1)u_1)-\cos((s+1)u_2)}{\cos (u_1)-\cos (u_2)}
\end{align}

\section{A new expression for certain $\text{SO}(d+2)$ characters}\label{SOd2}
In this appendix, we show an interesting property of the $\SO(d+2)$ characters $\Theta^{\SO(d+2)}_{\mY_{n,s}}(x)$ corresponding to  two-row representations $\mY_{n,s}$. Let's first give the statement.
\begin{claim}\label{claim1}
The character $\Theta^{\SO(d+2)}_{\mY_{n,s}}(x)$ has the same {\it polar} part (i.e. all terms of negative powers in $x$) and the same constant term as the following function 
\begin{align}
\boxed{P^d_{n,s}(x)\equiv\frac{D^d_s \,x^{-n}-D^d_{n+1}\,x^{1-s}}{(1-x)^d}}
\end{align} 
Since $\Theta^{\SO(d+2)}_{\mY_{n,s}}(x)$ is symmetric under $x\leftrightarrow x^{-1}$, it is then completely encoded in the function $P^d_{n,s}(x)$. In particular, when $s=0$, i.e. spin-$n$ representation, the  character $\Theta^{\SO(d+2)}_{\mY_n}(x)$ is encoded in 
\begin{align}
\boxed{Q^d_{n}(x)\equiv \frac{x^{-n}}{(1-x)^d}}
\end{align}
More explicitly, let $f(x)$ be a function with well-defined Laurent expansion around $x=0$. Denote the polar part of $f(x)$ by $[f(x)]_-$ and denote the polar part  together with the constant term of $f(x)$ by $[f(x)]_0$. With this notations, the claim implies that 
\begin{align}\label{newcharcomp}
\boxed{\Theta^{\SO(d+2)}_{\mY_{n,s}}(x)=\left[P^d_{n,s}(x)\right]_0+\left(\left.\left[P^d_{n,s}(x)\right]_-\right |_{x\to x^{-1}}\right)}
\end{align}
\end{claim}

The claim can be proved rigorously by induction, using  branching rules. Here we want to present a more intriguing but less rigorous argument, that shows  the similarity and difference between the $\mY_{n,s}$ representation of $\SO(d+2)$ and the infinite dimensional lowest-weight representations of $\SO(1, d+1)$. In view of this, we need to rewrite the $\mY_{n,s}$ representation  in a CFT-style language as follows. 
Denote the (antihermitian) generators of $\so(d+2)$ by $L_{AB}, 0\le A, B\le d+1$ which satisfy commutation relations 
\begin{align}
[L_{AB}, L_{CD}]=\delta_{BC} L_{AD}+\text{permutations}
\end{align}
In the differential operator realization, $L_{AB}=X_A\partial_B-X_B \partial_A$ where $X_A$ are coordinates in $\mathbb{R}^{d+2}$. 
Mimicking the conformal algebra, we define the following basis
\begin{align}
M_{ij}=L_{ij},\,\,\,\,\, H=iL_{0,d+1},\,\,\,\,\, P_i =L_{d+1,i}+i L_{0i},\,\,\,\,\,K_i =-L_{d+1,i}+i L_{0i}, 
\end{align}
where $1\le i\le d$. The new basis leads to some interesting commutators
\begin{align}
[M_{ij}, H]=0, \,\,\,\,\, [H, P_i]=P_i,\,\,\,\, [H, K_i]=-K_i\,\,\,\,\, [K_i, P_j]=2M_{ij}-2\delta_{ij} H
\end{align}
In particular, $P_i$ raises the eigenvalue of $H$ by 1 while $K_i$ lowers the eigenvalue of $H$ by 1. Define complex (lightcone) coordinate $z=X_{d+1}+i X_{0}, \bar z=X_{d+1}-i X_{0}$ and then the differential operator realization of  $H, P_i, K_i$ can be expressed as
\begin{align}
H=z\partial_z-\bar z \partial_{\bar{z}},\,\,\,\,\, P_i=z\partial_i -2\, X_i\partial_{\bar z},\,\,\,\,\, K_i=-\bar z\partial_i+2\, X_i \partial_z
\end{align}
First, let's consider spin-$n$ representation of $\SO(d+2)$ generated by the lowest weight state $|lw\rangle$ that satisfies
\begin{align}
H|lw\rangle=-n |lw\rangle,\,\,\,\,\, K_i |lw\rangle=0,\,\,\,\,\ M_{ij}|lw\rangle=0
\end{align}
A generic state in the Verma module generated by $|lw\rangle$ is a linear combination of the descendants $P_{i_1}\cdots P_{i_k}|lw\rangle$. However,  some of the $\so(d+2)$ descendants of $|lw\rangle$ are linearly dependent. To see this more explicitly, let's switch to the wavefunction picture, where the lowest weight state $|lw\rangle$ corresponds to  $\psi_{lw}(z,\bar z)=\bar z^n$ . One crucial observation is that all of the states $\psi_{i_1\cdots i_k}\equiv P_{i_1}\cdots P_{i_k}\psi_{lw}$ with $k\le n$ are linearly independent. To show this, it suffices to notice that the top component of $\psi_{i_1\cdots i_k}$ in $X_i$ is $(-2)^kX_{i_1}\cdots X_{i_k}\bar z^{n-k}$ for $k\le n$ which arises from the $-2X_i\partial_{\bar z}$ part of each $P_i$. It is clear that $X_{i_1}\cdots X_{i_k}\bar z^{n-k}$ are nonvanishing and linearly independent, and hence all $\psi_{i_1\cdots i_k}$ with $k\le n$ are linearly independent. Then it is almost trivial to construct the polar part and the constant term of the character $\Theta_{\mY _n}^{\SO(d+2)}(x)\equiv \Tr_{\mY_n} x^H$
\begin{align}
\left[\Theta_{\mY n}^{\SO(d+2)}(x)\right]_0=\sum_{k=0}^n \binom{d+k-1}{d-1} x^{k-n}=\left[\sum_{k=0}^\infty \binom{d+k-1}{d-1} x^{k-n}\right]_0=\left[Q^d_n(x)\right]_0
\end{align}
This is exactly the single-row case of the claim of this appendix. Then the full character can be recovered by using the $x\leftrightarrow x^{-1}$ symmetry. For example, due to this symmetry the coefficient of $x$ in $\Theta_{\mY_n}^{\SO(d+2)}(x)$ should be $\binom{d+n-2}{d-1}$ rather than the native counting $\binom{d+n}{d-1}$. Even more explicitly, we can find what are these $\binom{d+n-2}{d-1}$ states. The $H=0$ eigenspace $\CH^{(n)}_0$  is spanned by all $X_{i_1}\cdots X_{i_n}$. Acting $P_j$ on them yields basis of the $H=1$ eigenspace $\CH^{(n)}_1$
\begin{align}
P_j X_{i_1}\cdots X_{i_n}=z\delta_{j(i_1}X_{i_2}\cdots X_{i_n)}
\end{align}
Therefore $\CH^{(n)}_1$ is spanned by $z\,X_{i_1}\cdots X_{i_{n-1}}$ and has dimension $\binom{d+n-2}{d-1}$.

Next, we consider more complicated representations like $\mY_{n, s}$, for which we have a lowest weight state $|lw\rangle_{{i_1}\cdots i_s}$ that carries a spin-$s$ representation of $\SO(d)$. As before, $|lw\rangle_{{i_1}\cdots i_s}$ has quantum number $-n$ under $H$ and is annihilated by $K_i$. To represent the lowest weight state as a wavefunction, we need to introduce another copy of $\mathbb{R}^{d+2}$ with coordinate $Y^A$ such that $L_{AB}$ is realized as 
\begin{align}
L_{AB}=X_A\partial_{X_B}-X_B \partial_{X_A}+Y_A\partial_{Y_B}-Y_B \partial_{Y_A}
\end{align}
Define complex (lightcone) coordinate for the $Y$-space $w=Y_{d+1}+i Y_0, \bar w =Y_{d+1}-i Y_0$ and then $H,P_i,K_i$ can be expressed as 
\begin{align}
H=z\partial_z-\bar z \partial_{\bar{z}}&+w\partial_w-\bar w \partial_{\bar{w}},\,\,\,\,\, P_i=z\partial_{X_i} -2\, X_i\partial_{\bar z}+w\partial_{Y_i} -2\, Y_i\partial_{\bar w}\nonumber\\
 &K_i=-\bar z\partial_{X_i}+2\, X_i \partial_z-\bar w\partial_{Y_i}+2\, Y_i \partial_w
\end{align}
It is easy to check that the following wavefunction  actually corresponds to the lowest weight state
\begin{align}
\psi^{lw}_{i_1\cdots i_s}=\bar z^{n-s}(\bar z Y-\bar w X)_{i_1}\cdots (\bar z Y-\bar w X)_{i_s}-\text{trace}
\end{align}
Introduce a null $d$-vector $u_i$ and then $\psi^{lw}_{i_1\cdots i_s}$ can be more efficiently written as 
\begin{align}
\psi_{ns}^{lw}(X_A,Y_B;u_i)=\bar z^{n-s}(\bar z\, Y\cdot u-\bar w\, X\cdot u)^s
\end{align}
The claim $\left[\Theta^{\SO(d+2)}_{\mY_{n,s}}(x)\right]_0=\left[\frac{D^d_s x^{-n}-D^{d}_{n+1}x^{1-s}}{(1-x)^d}\right]_0$, in particular the $x^{1-s}$ part on the R.H.S, signals that there should be a constraint on the descendants $\psi^{lw}_{i_1\cdots i_s}$ at level $(n+1-s)$. To show this, let's focus on the  component of $\psi^{lw}_{i_1\cdots i_s}$ with the highest degree in $X_i$, which is roughly $\bar z^{n-s}\bar w^s X_{i_1}\cdots X_{i_s}$ up to pure trace component. The top $X_i$ components in the descendants of $\psi^{lw}_{i_1\cdots i_s}$ come from $-2X_i\partial_{\bar z}\in P_i$ with $\partial_{\bar z}$ acting on $\bar z^{n-s}$, that is,
\begin{align}
P_{i_1}\cdots P_{i_k} \psi^{lw}_{j_1\cdots j_s}\ni (-2)^k\left(\partial^k_{\bar z}\bar z^{n-s}\right)\,\bar w^s X_{i_1}\cdots X_{i_k}X_{j_1}\cdots X_{j_s}
\end{align}
Therefore for level $0\le k\le n-s$, all the naive descendants $P_{i_1}\cdots P_{i_k} \psi^{lw}_{j_1\cdots j_s}$ are nonvanishing and linearly independent. However at level $k=n+1-s$, one can easily check that $(u\cdot P)^k\psi_{ns}^{lw}$ vanishes identically, which means the following spin-$(n+1)$ wavefunction is actually zero
\begin{align}\label{constraintrace}
P_{(i_1}\cdots P_{i_{n+1-s}}\psi^{lw}_{j_1\cdots j_s)}-\text{trace}=0
\end{align}
To be more explicit,  let's take the two-form representation $\mY_{1,1}$ as an example.  In this case, the lowest-weight  wavefunction is 
\begin{align}
\psi^{lw}_i(X,Y)=\bar z\, Y_i-\bar w\, X_i
\end{align}
and the states at the next level 
\begin{align}
P_j \psi^{lw}_i(X,Y)=2 (X_i\, Y_j-X_j\, Y_i)+(w\,\bar z-z\,\bar w)\delta_{ij}
\end{align}
where we only have antisymmetric and pure trace components and hence $P_{(j} \psi^{lw}_{i)}-\frac{\delta_{ij}}{d}P_{k} \psi^{lw}_{k}=0$.
Without the constraint (\ref{constraintrace}),  the polar part of $\Theta^{\SO(d+2)}_{\mY_{n, s}}(x)$ would simply be encoded in $D^d_s \frac{x^{-n}}{(1-x)^d}$, i.e. the result of $s=0$ multiplied by a spin degeneracy. With this constraint, we should subtract the contribution of itself together with its descendants, which is $D^d_{n+1}\frac{x^{1-s}}{(1-x)^d}$ from a simple counting. Altogether, we get 
\begin{align}\left[\Theta^{\SO(d+2)}_{\mY_{n, s}}(x)\right]_0=\left[\frac{D^d_s x^{-n}-D^{d}_{n+1}x^{1-s}}{(1-x)^d}\right]_0=\left[P^d_{n,s}(x)\right]_0\end{align}

The claim \ref{claim1} has some important corollaries. We first rewrite $P^d_{n,s}$ as 
\begin{align}\label{PtoQQ}
P^d_{n,s}(x)=D^d_s Q^d_n(x)-D^d_{n+1} Q^d_{s-1}(x)
\end{align}
Since $P^d_{n,s}(x)$ encodes the character of $\mY_{n,s }$ and $ Q^d_n(x)$ encodes the character of $\mY_{n}$, the elementary equation (\ref{PtoQQ}) yields a highly nontrivial relation of characters
\begin{align}\label{charmagic}
\boxed{\Theta^{\SO(d+2)}_{\mY_{n, s}}(x)=D^d_s \Theta^{\SO(d+2)}_{\mY_{n}}(x)-D^d_{n+1}\Theta^{\SO(d+2)}_{\mY_{s-1}}(x)}
\end{align}
It allows us to construct the Weyl characters of two-row representations by only using the Weyl characters of single-row representations. Taking the limit $x\to 1$ on both sides of (\ref{charmagic}), we obtain a nontrivial relation involving the dimensions of different representations
\begin{align}\label{originalmagic}
\boxed{D^{d+2}_{n, s}=D^d_s D^{d+2}_{n}-D^d_{n+1} D^{d+2}_{s-1}}
\end{align}
This formula plays an important role in the derivation of the character integral representation of one-loop partition functions of spinning fields in $\text{dS}_{d+1}$ \cite{Anninos:2020hfj}.

\normalsize
\newpage

\bibliographystyle{JHEP}
\bibliography{Repref}

\end{document}